\documentclass[smallextended]{svjour3}

\usepackage[export]{adjustbox}
\usepackage[utf8]{inputenc}
\usepackage{mathpazo}
\usepackage{amsmath}
\usepackage[round]{natbib}
\usepackage[margin=1in]{geometry}
\usepackage{amssymb}
\usepackage{standalone}
\usepackage{graphicx}
\usepackage{overpic}
\usepackage{epstopdf}

\usepackage[normalem]{ulem}
\usepackage[rgb,dvipsnames]{xcolor}
\usepackage{tikz}
\usepackage{bm}
\usepackage[subrefformat=parens]{subcaption}
\usepackage{booktabs}
\usepackage[math]{cellspace}
\AtBeginDocument{
    \heavyrulewidth=.08em
    \lightrulewidth=.05em
    \cmidrulewidth=.03em
    \belowrulesep=.65ex
    \belowbottomsep=0pt
    \aboverulesep=.4ex
    \abovetopsep=0pt
    \cmidrulesep=\doublerulesep
    \cmidrulekern=.5em
    \defaultaddspace=.5em
}

\newcommand \beq{\begin{eqnarray}}
\newcommand \eeq{\end{eqnarray}}
\newcommand \beqno{\begin{eqnarray*}}
\newcommand \eeqno{\end{eqnarray*}}
\newcommand \bit{\begin{itemize}}
\newcommand \eit{\end{itemize}}

\graphicspath{{./figures/}{./Nonlocal2D1S/}{./final_figures/}}

\newcommand*\diff{\mathop{}\!\mathrm{d}}

\begin{document}
\def\mytitle{Patterning of nonlocal transport models in biology: the impact of spatial dimension}
\title{\mytitle}
\titlerunning{\mytitle}
\author{Thomas Jun Jewell\textsuperscript{a,}\footnote{Corresponding author\\Emails: jewell@maths.ox.ac.uk, andrew.krause@durham.ac.uk, maini@maths.ox.ac.uk, gaffney@maths.ox.ac.uk} \mbox{\and Andrew L. Krause\textsuperscript{b} \and Philip K. Maini\textsuperscript{a} \and Eamonn A. Gaffney\textsuperscript{a}}}
\authorrunning{T.J. Jewell, A. L. Krause, P. K. Maini, E. A. Gaffney}

\institute{
    \textsuperscript{a}
     Wolfson Centre for Mathematical Biology, Mathematical Institute, University of Oxford, Andrew Wiles Building, Radcliffe Observatory Quarter, Woodstock Road, Oxford, OX2 6GG, United Kingdom \\
    \textsuperscript{b}
     Mathematical Sciences Department, Durham University, Upper Mountjoy Campus, Stockton Rd, Durham DH1 3LE, United Kingdom \\
    }

\date{\today}

\maketitle

\begin{abstract}
Throughout developmental biology and ecology, transport can be driven by nonlocal interactions. Examples include cells that migrate based on contact with pseudopodia extended from other cells, and animals that move based on their vision of other animals. Nonlocal integro-PDE models have been used to investigate contact attraction and repulsion in cell populations in 1D. In this paper, we generalise the analysis of pattern formation in such a model from 1D to higher spatial dimensions. Numerical simulations in 2D demonstrate complex behaviour in the model, including spatio-temporal patterns, multi-stability, and the selection of spots or stripes heavily depending on interactions being attractive or repulsive. Through linear stability analysis in $N$ dimensions, we demonstrate how, unlike in local Turing reaction-diffusion models, the capacity for pattern formation fundamentally changes with dimensionality for this nonlocal model. Most notably, pattern formation is possible only in higher than one spatial dimension for both the single species system with repulsive interactions, and the two species system with `run-and-chase' interactions. The latter case may be relevant to zebrafish stripe formation, which has been shown to be driven by run-and-chase dynamics between melanophore and xanthophore pigment cells.
\end{abstract}


\maketitle

\section{Introduction}

Biological development primarily occurs in two or three spatial dimensions. Mathematical models of biology must therefore balance the analytical and numerical challenges associated with these higher dimensions, against the ability to capture the essential driving phenomena \citep{ModelingCells3D}. Most initial proof of concept papers for pattern formation models choose the simplicity of a 1D domain. In the context of canonical local reaction-diffusion models, this can be justified by linear theory, where dispersion relations are effectively the same for any number of dimensions \citep{KrauseModernTuring}. Of course, the nonlinear analysis and the possible shapes of patterns does change with dimensionality. A key example is the distinction between spots and stripes in 2D that does not exist in 1D \citep{ermentrout_spots_stripes}. Conversely, the impact of different spatial dimensions on nonlocal models of biological self-organisation, particularly in terms of linear stability and capacity for pattern formation, is understudied.

In this work, we look at the highly influential nonlocal attraction and repulsion model for cell populations introduced by \citet{Painter2015_nonlocal_rd_developmental_biology}, based on the earlier model by \citet{ARMSTRONG2006}. We generalise their analysis of pattern formation of cell aggregates from 1D to $N$ dimensions ($N$D), motivated by 2D and 3D biological applications. Most studies analysing the stability of related models only examine the 1D case, with the notable exception of \citet{Dyson_existence_N_dim_nonlocal}, which takes a functional analytical viewpoint and does not consider dispersion relations, for instance.

Distinct from models such as \citet{nonlocal_kinetics_britton, nonlocal_kinetics_2006}, in which the nonlocality features only in reaction-kinetics, the \citet{Painter2015_nonlocal_rd_developmental_biology} and \citet{ARMSTRONG2006} models incorporate nonlocality purely through cell-migration, in an advection term. These latter nonlocal models thus lie within the broader context of a shift in focus in the field of pattern formation from biochemical reaction-kinetics to transport and migration dynamics. Increasingly, an area of active interest is highlighting that many biological pattern forming processes and the predictions of corresponding theoretical models both heavily depend on the details of transport dynamics beyond simple diffusion \citep{Kondo_zebrafish_review}.

The \citet{Painter2015_nonlocal_rd_developmental_biology} model assumes that agents can interact nonlocally, which is primarily interpreted in a developmental biology context in terms of cells which can extend protrusions (pseudopodia) such as filopodia or lamellipodia that transmit signals or force to other cells which are relatively far from the original cell. In some contexts, such as for the pigment cells of newts, these filopodia can extend up to ten times the cell diameter \citep{Newt_stripes}. The nonlocal interaction can be repulsive, such as with fibroblast cells in the neural crest which move apart when their lamellipodia come into contact \citep{neural_crest_development}. It can also be attractive, for example representing adhesion between cells. In fact, \citet{nonlocal_Vasculogenesis_Villa2022} use such an approach to incorporate adhesion in their model of vasculogenesis, the de novo formation of blood vessels.

Nonlocal transport models are also used in the context of ecology, where animal populations form territorial patterns driven by nonlocal interactions \citep{Potts2019_spatial_memory_ecosystems, potts2019directionally}. Animal swarms, such as flocks of birds or swarms of locusts, are also modelled using a similar framework, with investigations by \citet{early_swarm_nonlocal_model} predating the work of \citet{ARMSTRONG2006}. For a recent example and a brief review of swarming models, see \citet{EdGreen_locust_swarming}. Nonlocality in ecology can take the form of direct long-ranged interaction mediated by sight or smell, or indirect interaction through leaving scent markings or through retained memory of previous encounters between animals.

A significant application of nonlocal transport models in the context of pattern formation is to the development of zebrafish stripes. Zebrafish are model species commonly used to study pattern formation in developmental biology due to the availability of their mutant types, the comprehensive understanding of their genome, and the ability for their stripes to regenerate in adults \citep{Kondo_zebrafish_review}. \citet{YamanakaKondo2014} proposed that these stripes are primarily formed through nonlocal `run-and-chase' interactions between two species of pigment cell: xanthophore cells extend pseudopodia to melanophore cells and are attracted towards melanophores, whilst melanophores are repelled away from the xanthophores. With this in mind, \citet{Painter2015_nonlocal_rd_developmental_biology} analysed their own model in 1D, but found that purely run-and-chase dynamics with two species could not drive pattern formation. In this work, we investigate whether this capacity for pattern formation changes with the introduction of two or more spatial dimensions. Furthermore, we examine not only run-and-chase dynamics, but all combinations of attractive and/or repulsive interactions with general interaction kernels for both the single species and two species cases.

We begin with the single species case in $N$D, defining the model in Section \ref{section:model}. In Section \ref{section:dispersion_relation_derivation} we perform linear stability analysis, making use of hyperspherical Bessel functions to derive the dispersion relation, which we then summarise and analyse in Section \ref{section:dispersion_relation_summary} to gain insight on the conditions for pattern formation. In Section \ref{section:physical_interpretation} we present a physical interpretation of the results to intuitively explain the effect of dimensionality on pattern formation. In Section \ref{section:numerics} we present numerical simulations of the 2D case to validate the predictions of linear instability and explore beyond the linear regime. Then, in Section \ref{section:two_species}, we apply all of these techniques to investigate the two species case. Finally, in Section \ref{section:discussion}, we discuss the significance of our findings.

\section{Model in $N$ Dimensions}
\label{section:model}

We nominally interpret the following model in a developmental biology context, in which the agents are cells with nonlocal terms representing pseudopodia induced interactions. However due to the generality of each term, the model could apply to a general class of systems in which agents diffuse and interact through nonlocal advection.

We study the continuum model introduced in \citet{Painter2015_nonlocal_rd_developmental_biology} describing the time evolution of the population density, $u(\boldsymbol{x},t)$, for a single species of cell at position $\boldsymbol{x}\in \mathbb{R}^N$ and time $t$. An infinite domain is chosen in order to focus on patterning due to nonlocal interactions, rather than geometry or boundary conditions. In the model, cells can diffuse, proliferate and die, and interact nonlocally. This nonlocal interaction induces a flux per unit cell density at $(\boldsymbol{x},t)$ given by
\begin{equation}
    \boldsymbol{F}(\boldsymbol{x},t)=p(u(\boldsymbol{x},t))\mu\int_{\mathbb{R}^N} \boldsymbol{\hat{s}}\,\frac{1}{\xi^N}\tilde{\Omega}\left(\frac{s}{\xi}\right)g(u(\boldsymbol{x}+\boldsymbol{s},t))\boldsymbol{\diff s^N}.
\label{eq:NonlocalForce}
\end{equation}
Here, the integral sums the flux per unit cell density at $\boldsymbol{x}$ induced by interactions from cells at every point $\boldsymbol{x}+\boldsymbol{s}$, where $\boldsymbol{s}\equiv |\boldsymbol{s}|\boldsymbol{\hat{s}}\,\equiv\,s\boldsymbol{\hat{s}}$. Accordingly, $\boldsymbol{\diff s^N}$ is the $N$ dimensional volume element. The magnitude of flux generated by cell density $u$ varies as some function $g(u)$.  This flux is assumed to be parallel to the direction of separation between cells, hence the $\boldsymbol{\hat{s}}$ term in the integrand. How this interaction varies with separation is characterised by the interaction kernel, $\tilde{\Omega}\left(\frac{s}{\xi}\right)$, which is normalised without loss such that
\begin{equation}
\int_{\mathbb{R}^N}\tilde{\Omega}\left(\frac{s}{\xi}\right)\boldsymbol{\diff s^N}=\xi^N,
\label{eq:omega_normalisation}
\end{equation}
where $\xi$ is a characteristic length scale called the signalling range. The overall magnitude of the generated flux scales with $\mu$, the interaction strength, which also dictates the direction: $\mu>0$ corresponds to attractive interactions, whilst $\mu<0$ corresponds to repulsive interactions. In line with \citet{Painter2015_nonlocal_rd_developmental_biology} we focus on the case where $\tilde{\Omega}\left(\frac{s}{\xi}\right)$ is a non-negative function, and so either all cells are attracting or all cells are repelling, for all separation distances. This choice does not affect the dispersion relation we later derive. It is important to note that we assume that the nonlocal interaction is isotropic in space, and so $\tilde{\Omega}\left(\frac{s}{\xi}\right)$ is a function of only the magnitude of separation, $s$. Finally, the flux is regulated by some packing function, $p(u)$, representing, for example, contact inhibition and more generally prevents infinite densities forming at a point. 

The total flux of cells at $(\boldsymbol{x},t)$ due to the nonlocal interaction is then taken as proportional to $u(\boldsymbol{x},t)\boldsymbol{F}(\boldsymbol{x},t)$, where $\boldsymbol{F}(\boldsymbol{x},t)$ is given by Eq. \eqref{eq:NonlocalForce}. The divergence, $\boldsymbol{\nabla}\cdot$, of this flux thus contributes to the time evolution of the density, given by
\begin{equation}
    \frac{\partial u}{\partial t} = \nabla^2 u + h(u) -\boldsymbol{\nabla}\cdot\left(up(u)\frac{\mu}{\xi^{N}} \int_{\mathbb{R}^N} \boldsymbol{\hat{s}}\,\tilde{\Omega}\left(\frac{s}{\xi}\right)g(u(\boldsymbol{x}+\boldsymbol{s},t))\boldsymbol{\diff s^N}\right),
    \label{eq:non_dim_PDE}
\end{equation}
where $u=u(\boldsymbol{x},t)$, unless explicitly specified. Diffusion is incorporated through the Laplacian. Local cell proliferation and death is included via the function $h(u)$. We specify that these proliferation-death kinetics have a positive equilibrium at $u=U$ where $h(U)=0$ and $\frac{\partial h(u)}{\partial u}\big\rvert_{U}<0$. Additionally we focus on the case where $p(U)\geq0$ and $\frac{\partial g(u)}{\partial u}\big\rvert_{U}>0$. These two inequalities are not required for the dispersion relation we derive in the stability analysis, however if $p(U)\frac{\partial g(u)}{\partial u}\big\rvert_{U}<0$ then some of the predicted behaviour of attractive and repulsive interactions would be swapped. The first inequality is the statement that (at density $U$) the packing forces, which are a reactive normal force, cannot be actively stronger than the nonlocal interaction against which they are reacting, and hence cannot reverse the direction of the flux of cells. The second inequality specifies that (at density $U$), the higher the density of cells in a region, the stronger the interaction from that region. Both inequalities are reasonable assumptions for most models.

For conciseness Eq. \eqref{eq:non_dim_PDE} is already non-dimensionalised, as in this work we are investigating the underlying dynamics and stability of such models, rather than strictly modelling a specific biological system. The population density $u$ should be thought of as relative to some packing density scale, whilst distances are relative to some reference length scale, and the nonlocal flux and proliferation-death kinetics have timescales relative to the timescale dictated by diffusion. For detailed discussion of the dimensionalisation, see \cite{Painter2015_nonlocal_rd_developmental_biology}.

In contrast to standard Turing patterning systems \citep{TuringOriginalPaper, KrauseModernTuring}, the model defined by Eq. \eqref{eq:non_dim_PDE} can support pattern formation even with just a single species. The conditions under which this pattern formation can occur, and the character of the patterns, was extensively explored in \cite{Painter2015_nonlocal_rd_developmental_biology} in 1D. Here we perform investigations in $N$D, which allows more concise analysis of the biologically relevant 2D and 3D cases as well as providing clearer insight into the underlying effects of changing the number of spatial dimensions.

We begin these investigations with linear stability analysis.

\section{Derivation of Dispersion Relation}
\label{section:dispersion_relation_derivation}

Following the standard procedure when determining the conditions for pattern formation \citep{Murray2002, KrauseModernTuring}, we linearise our integro-PDE about the positive spatially uniform steady state $u(\boldsymbol{x},t)=U$, and consider whether a small heterogeneous perturbation such that $u(\boldsymbol{x},t)=U+\tilde{u}(\boldsymbol{x},t)$, where $|\tilde{u}(\boldsymbol{x},t)| \ll 1$, will decay back to homogeneity or grow. From Eq. \eqref{eq:non_dim_PDE} we see that the homogeneous steady state is at the equilibrium of the proliferation-death kinetics, i.e. $h(U)=0$. Linearising about this state gives, to leading order,
\begin{equation}
    \frac{\partial \tilde{u}}{\partial t} = \nabla^2 \tilde{u} +h'(U)\tilde{u} -Up(U)g'(U)\frac{\mu}{\xi^N}\boldsymbol{\nabla}\cdot\left( \int_{\mathbb{R}^N} \boldsymbol{\hat{s}}\,\tilde{\Omega}\left(\frac{s}{\xi}\right)\tilde{u}(\boldsymbol{x}+\boldsymbol{s},t)\boldsymbol{\diff s^N}\right),
\label{eq:N-D_nonlocalPDE}
\end{equation} 
where $h'(U)\equiv \frac{\partial h(u)}{\partial u}\big\rvert_{U}$ and $g'(U)\equiv \frac{\partial g(u)}{\partial u}\big\rvert_{U}$.

We can see that for a spatially homogeneous solution to Eq.\,\eqref{eq:N-D_nonlocalPDE}, only the $h'(U)$ term is non-zero. For now, we consider a stable equilibrium of the proliferation-death kinetics, so that $h'(U)<0$, and thus the system is stable to spatially homogeneous perturbations, in analogue with the standard Turing instability. Later, in section \ref{section:transport_only}, we consider a system without proliferation and death, which is then no longer asymptotically stable to homogeneous perturbations.

To solve Eq. \eqref{eq:N-D_nonlocalPDE}, we use the standard technique of expanding into independent modes of the form $\tilde{u}=u_0 e^{i\boldsymbol{k}\cdot\boldsymbol{x}}e^{\lambda t}$, with wave-vector $\boldsymbol{k}$ and growth rate $\lambda$. This yields the dispersion relation

\begin{equation}
    \lambda = -k^2 +h'(U) - Up(U)g'(U)\frac{\mu}{\xi^N}\,ki\int_{\mathbb{R}^N} (\boldsymbol{\hat{k}}\cdot\boldsymbol{\hat{s}})\,\tilde{\Omega}\left(\frac{s}{\xi}\right)e^{i\boldsymbol{k}\cdot\boldsymbol{s}}\boldsymbol{\diff s^N},
\label{eq:ND1S_lambda_unevaluted}
\end{equation}
where $\boldsymbol{k}\equiv k\boldsymbol{\hat{k}}$.

It is important to note that Eq. \eqref{eq:ND1S_lambda_unevaluted} relies on the use of either: 1) an infinite domain, or 2) periodic boundary conditions on a finite rectangular domain with $\tilde{\Omega}\left(\frac{s}{\xi}\right)$ having compact support (with a small enough support compared to the domain lengths to prevent self interacting points). A finite domain without periodic boundary conditions would imply that the limits, and therefore value, of the integral in this equation would be dependent on $\boldsymbol{x}$, which invalidates the expansion in terms of spatial modes. The linear analysis of the model for different boundary conditions and geometries is left for future work. As previously noted, we assume an infinite domain, $\boldsymbol{x} \in \mathbb{R}^N$. The choice of periodic boundary conditions on a finite domain would yield the same dispersion relation, but with the added constraint that $\boldsymbol{k}$ can only take values that correspond to integer numbers of wavelengths spanning the domain, i.e. $k_i=\frac{2\pi}{L_i}n$, where $k_i$ is the component of $\boldsymbol{k}$ in the direction parallel to an axis of the domain, and $L_i$ is the length of the domain along this axis.

To evaluate the dispersion relation, Eq. \eqref{eq:ND1S_lambda_unevaluted}, we note that the nonlocal term is proportional to the $N$D Fourier transform of a radial vector function with isotropic argument, $\boldsymbol{\hat{r}}\,\tilde{\Omega}\left(\frac{r}{\xi}\right)$. To evaluate this, we first consider the simpler case of an isotropic scalar function, $f(r)$, and show that these Fourier transforms can be naturally described in terms of hyperspherical Bessel functions with hyperspherical Hankel transforms.

\subsection{$N$D Fourier Transform of $f(r)$}
\label{section:isotropic_scalar_function}
The $N$D Fourier transform of an isotropic function $f(r)$ is defined as
\begin{equation}
    \tilde{f}(k)=\int_{\mathbb{R}^N}f(r)e^{i\boldsymbol{k}\cdot\boldsymbol{r}}\boldsymbol{\diff r^N},
\label{eq:ND_fourier_transform}
\end{equation}
which is known \citep{Sneddon_transforms} to be equal to the Hankel transform given by
\begin{equation}
    \tilde{f}(k)=\left(\frac{1}{k}\right)^{\frac{N}{2}-1}(2\pi)^{\frac{N}{2}}\int_0^\infty r^{\frac{N}{2}}f(r)J_{\frac{N}{2}-1}(kr)\diff r,
\label{eq:ugly_hankel_transform}
\end{equation}
where $J_\alpha(x)$ is the $\alpha$ order Bessel function of the first kind.

We consider this transform in the context of hyperspherical Bessel functions, which are an $N$D generalisation of Bessel functions, defined by \citet{hyperspherical_Bessel} and \citet{ND_Bessel_functions} as 
\begin{equation}
    j_l^{(N)}(x)\equiv A_N \frac{J_{l+\frac{N}{2}-1}(x)}{x^{\frac{N}{2}-1}},
\label{eq:hyperspherical_bessel_definition}
\end{equation}
where $l\in \mathbb{N}\cup\{0\}$. For convenience, we will use normalisation $A_N=\Gamma\left(\frac{N}{2}\right)2^{\frac{N}{2}-1}$ so that $j_0^{(N)}(0)=1$, where $\Gamma$ is the gamma function.

In this case, Eq. \eqref{eq:ugly_hankel_transform} becomes
\begin{equation}
    \tilde{f}(k)=\frac{2 \pi^{\frac{N}{2}}}{\Gamma\left(\frac{N}{2}\right)}\int_0^\infty f(r)\, r^{N-1}\, j_0^{(N)}(kr) \diff r.
\label{eq:zeroth_order_HS_hankel}
\end{equation}
In other words, the $N$D Fourier transform of an isotropic scalar function is equivalent to a zeroth order hyperspherical Hankel transform.

In 1D, $j_0^{(1)}(x)=\text{cos}(x)$ and the hyperspherical Hankel transform reduces to a Fourier cosine transform. In 2D, $j_0^{(2)}(x)=J_0(x)$ and so the transform is the canonical zeroth order Hankel transform. Similarly in 3D, $j_0^{(3)}(x)=j_0(x)\equiv \frac{\text{sin}(x)}{x}$, which is the zeroth order spherical Bessel function, and so the transform is a zeroth order spherical Hankel transform.

With this result for the transform of $f(r)$, we can find the relevant transform for $\boldsymbol{\hat{r}}f(r)$, which matches the form in our dispersion relation.

\subsection{$N$D Fourier Transform of $\boldsymbol{\hat{r}}f(r)$}
\label{section:isotropic_vector_function}

Many of the relationships between standard Bessel functions of different orders $l$ also apply for hyperspherical Bessel functions. For example, the derivative of a zeroth order Bessel function is the negative of a first order Bessel function, $\frac{\diff J_0(x)}{\diff x}=-J_1(x)$, and this is also true for hyperspherical Bessel functions. We can prove this using a well documented recurrence relation for standard Bessel functions, $\frac{\diff}{\diff x}\left(\frac{J_\alpha(x)}{x^\alpha}\right)=-\frac{J_{\alpha+1}}{x^{\alpha}}$, with the definitions in Eq. \eqref{eq:hyperspherical_bessel_definition} such that
\begin{equation}
     \frac{\diff}{\diff x}j^{(N)}_0(x)=A_N\frac{\diff}{\diff x}\left(\frac{J_{\frac{N}{2}-1}(x)}{x^{\frac{N}{2}-1}}\right)=-A_N\frac{J_{\frac{N}{2}}(x)}{x^{{\frac{N}{2}-1}}}=-j^{(N)}_1(x).
\label{eq:deriv_j0_is_j1}
\end{equation}

This relation is useful for determining the $N$D Fourier transform, $\boldsymbol{\tilde{F}}(\boldsymbol{k})$, of a radial vector function with isotropic argument, $\boldsymbol{\hat{r}}f(r)$, which is given by
\begin{equation}
        \boldsymbol{\tilde{F}}(\boldsymbol{k})=\int_{\mathbb{R}^N} \boldsymbol{\hat{r}}f(r)e^{i\boldsymbol{k}\cdot\boldsymbol{r}}\boldsymbol{\diff r^N}.
\end{equation}

Implicit in the working presented below is sufficient smoothness of the integrands to allow the commutation of limiting processes. Then, by using the gradient in $k$-space, $\boldsymbol{\nabla}_k$, we can deduce
\begin{equation}
    \boldsymbol{\tilde{F}}(\boldsymbol{k})=-i\,\boldsymbol{\nabla}_k \int_{\mathbb{R}^N}\frac{f(r)}{r}e^{i\boldsymbol{k}\cdot\boldsymbol{r}}\boldsymbol{\diff r^N}.
\end{equation}    
Now as $\frac{f(r)}{r}$ is itself an isotropic function, we can use the result of Section \ref{section:isotropic_scalar_function} to write
\begin{equation}
\begin{split}
    \boldsymbol{\tilde{F}}(\boldsymbol{k})&=-i\,\boldsymbol{\nabla}_k\,\frac{2\pi^{\frac{N}{2}}}{\Gamma\left(\frac{N}{2}\right)}\int_0^\infty \frac{f(r)}{r}\, r^{N-1}\, j_0^{(N)}(kr) \diff r\\
    &=-\boldsymbol{\hat{k}} \,i \, \frac{2\pi^{\frac{N}{2}}}{\Gamma\left(\frac{N}{2}\right)}\int_0^\infty \frac{f(r)}{r}\, r^{N-1}\, \left[\frac{\partial}{\partial k}j_0^{(N)}(kr)\right] \diff r \\
    &= \boldsymbol{\hat{k}} \,i \, \frac{2\pi^{\frac{N}{2}}}{\Gamma\left(\frac{N}{2}\right)}\int_0^\infty f(r)\, r^{N-1}\,j_1^{(N)}(kr) \diff r.
\end{split}
\label{eq:1st_order_HS_hankel}
\end{equation}
Thus, the $N$D Fourier transform of a general  $\boldsymbol{\hat{r}}f(r)$, has direction parallel to $\boldsymbol{k}$ and has magnitude equal to the \textbf{first} order hyperspherical Hankel transform of $f(r)$.

In 1D, $j_1^{(1)}(x)=\text{sin}(x)$ and we have the Fourier sine transform. In 2D, $j_1^{(2)}(x)=J_1(x)$ and so the transform is the canonical first order Hankel transform. In 3D, $j_1^{(3)}(x)=j_1(x)\equiv \frac{\text{sin}(x)}{x^2}-\frac{\text{cos}(x)}{x}$, which is the first order spherical Bessel function, and corresponds to the first order spherical Hankel transform.

For both $f(r)$ in Eq. \eqref{eq:zeroth_order_HS_hankel} and $\boldsymbol{\hat{r}}f(r)$ in Eq. \eqref{eq:1st_order_HS_hankel} we have reduced an $N$D integral to a 1D integral. In this process, the integral kernels change from complex exponentials to hyperspherical Bessel functions. All angular information is then contained within these new integral kernels. This angular information is intrinsic to, and changes with, the particular $N$ number of dimensions, even when our original functions are isotropic. Section \ref{section:physical_interpretation} explores an intuitive physical picture for why this angular information is dimensionally dependent and why it affects pattern formation. 

With the above observations in mind, we should expect our biological system to behave differently in different numbers of dimensions. This is captured in the dispersion relation, which can now be determined using Eq. \eqref{eq:1st_order_HS_hankel}.

\subsection{Back to the dispersion relation}
From the results of Section \ref{section:isotropic_vector_function}, we can evaluate the nonlocal term in Eq. \eqref{eq:ND1S_lambda_unevaluted} as
\begin{equation}
    i\int_{\mathbb{R}^N} (\boldsymbol{\hat{k}}\cdot\boldsymbol{\hat{s}})\,\tilde{\Omega}\left(\frac{s}{\xi}\right)e^{i\boldsymbol{k}\cdot\boldsymbol{s}}\boldsymbol{\diff s^N} = -\frac{2\pi^{\frac{N}{2}}}{\Gamma\left(\frac{N}{2}\right)}\int_0^\infty \tilde{\Omega}\left(\frac{s}{\xi}\right)\, s^{N-1}\,j_1^{(N)}(ks) \diff s,
\end{equation}
and so the dispersion relation becomes
\begin{equation}
    \lambda = -k^2 +h'(U) + \frac{2\pi^{\frac{N}{2}}}{\Gamma\left(\frac{N}{2}\right)}\,Up(U)g'(U)\frac{\mu}{\xi^N}\,k\int_0^\infty \tilde{\Omega}\left(\frac{s}{\xi}\right)\, s^{N-1}\,j_1^{(N)}(ks) \diff s.
\label{eq:ND1S_lambda_EVALUATED}
\end{equation}

\subsection{Behaviour of $j_1^{(N)}(x)$}
\begin{figure}
    \centering
    \includegraphics[width=0.7\textwidth]{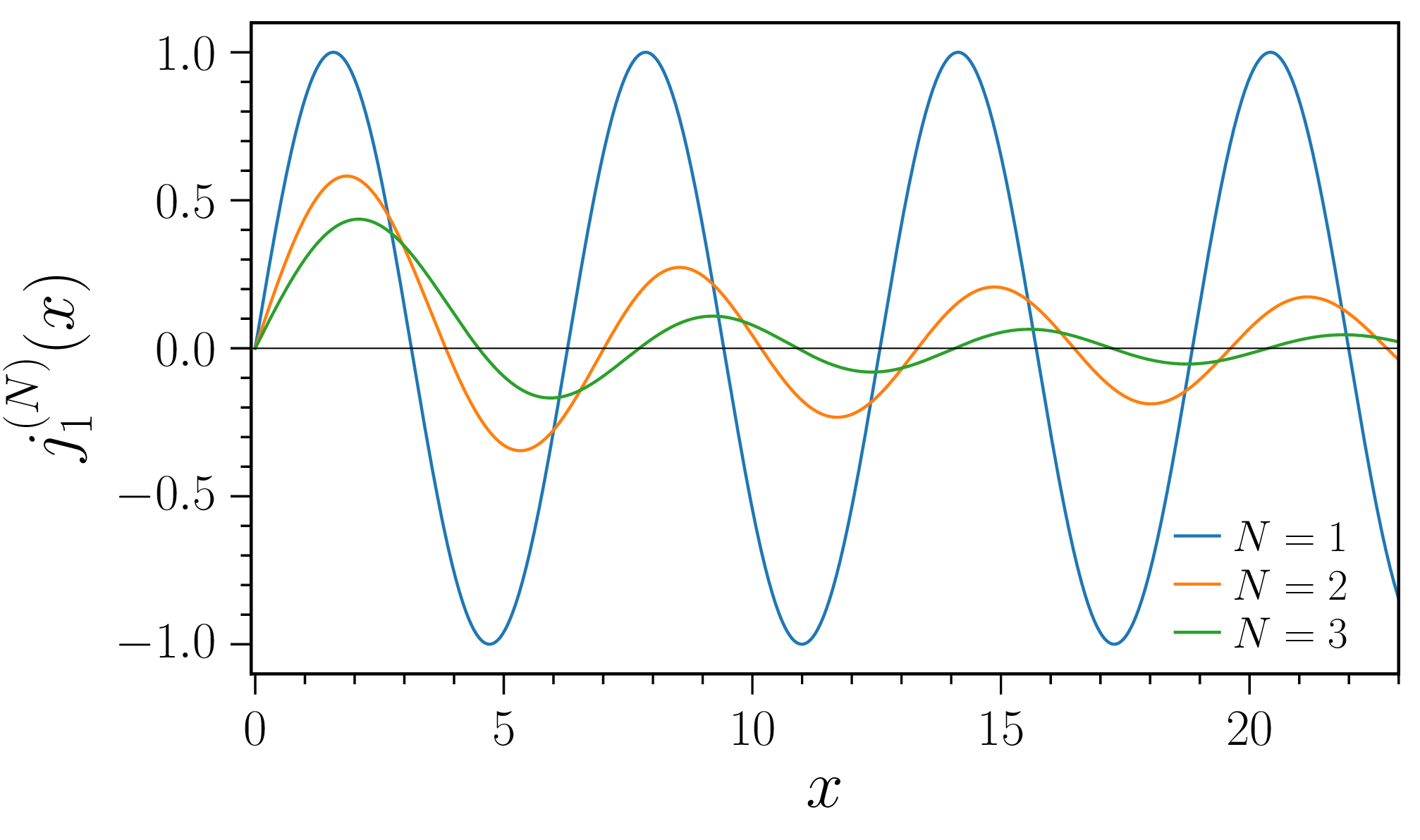}
    \caption{Plots of $j_1^{(N)}(x)$ for the $N=1,2,3$ cases. $j_1^{(1)}(x)=\text{sin}(x)$, $j_1^{(2)}(x)=J_1(x)$, and $j_1^{(3)}(x)=j_1(x)=\frac{1}{x^2}\text{sin}(x)-\frac{1}{x}\text{cos}(x)$.}
    \label{fig:hyperspherical_Bessels}
\end{figure}

In order to interpret the dispersion relation, it is useful to understand the overall behaviour of $j_1^{(N)}(x)$, of which the $N=1,2,3$ cases are plotted in Fig. \ref{fig:hyperspherical_Bessels}.

Firstly, as illustrated in Fig. \ref{fig:hyperspherical_Bessels}, $j_1^{(N)}(0)=0$ and $\frac{\diff j_1^{(N)}(x)}{\diff x}\big\rvert_{0}>0$ for all $N$. This can be seen by the function's small argument asymptotic form,
\begin{equation}
\begin{split}
    j_1^{(N)}(x) &\equiv \Gamma\left(\frac{N}{2}\right)2^{\frac{N}{2}-1} \frac{J_{\frac{N}{2}}(x)}{x^{\frac{N}{2}-1}} \sim \frac{\Gamma\left(\frac{N}{2}\right)2^{\frac{N}{2}-1}}{\Gamma({\frac{N}{2}+1})}\left(\frac{x}{2}\right)^{\frac{N}{2}}\frac{1}{x^{\frac{N}{2}-1}} \\
    &\sim \frac{x}{N} \,\,\,,
\end{split}
\end{equation}
in which we used the small argument asymptotic form for standard Bessel functions, $J_\alpha(x) \sim \frac{1}{\Gamma(\alpha+1)}\left(\frac{x}{2}\right)^\alpha$.

Secondly, considering the large argument approximation for Bessel functions, for which \\$J_\alpha(x)\propto \frac{1}{\sqrt{x}}\left[\text{sin}(x+\phi_\alpha) + \mathcal{O}\left(x^{-1}\right)\right]$ when $x \gg \alpha^2 - \frac{1}{4}$, it follows that all $j_1^{(N)}(x)$ are approximated by
\begin{equation}
    j_1^{(N)}(x) \propto \frac{1}{x^{\frac{N-1}{2}}}\left[\text{sin}(x+\phi_N) + \mathcal{O}\left(x^{-1}\right) \right],
\end{equation}
for $x \gg \frac{1}{4}(N^2 - 1)$. In essence, all $j_1^{(N)}(x)$ oscillate as a sine wave, with some phase $\phi_N$, enveloped by a negative power of $x$, with the power decreasing by a half with each added spatial dimension. Although these approximations have an unbounded relative error at the zeroes of the sine wave, due to the $\mathcal{O}(x^{-1})$ terms, the absolute error will be small and tends to zero as $x$ tends to infinity. For the purposes of this paper, in which we will approximate integrals of $j_1^{(N)}(x)$, only absolute errors are important. 

\subsection{Dependence on $\xi$}
Finally, we highlight the relationship between the nonlocal signalling range $\xi$ and the wavenumber $k$ in the dispersion relation. By substituting $s=q\xi$, we can reparameterise the nonlocal term in Eq. \eqref{eq:ND1S_lambda_EVALUATED} such that
\begin{equation}
\begin{split}
    \frac{1}{\xi^N}k\int_0^\infty \tilde{\Omega}\left(\frac{s}{\xi}\right)\, s^{N-1}\,j_1^{(N)}(ks) \diff s &= \frac{1}{\xi}k\xi \int_0^\infty \tilde{\Omega}\left(q\right)\, q^{N-1}\,j_1^{(N)}(k\xi q) \diff q \\
    &\equiv \frac{1}{\xi}G(k\xi).
\end{split}
\end{equation}
Here we see that all $k$ dependence takes the form of the product $k\xi$.

\section{Dispersion Relation}
\label{section:dispersion_relation_summary}

For our nonlocal integro-PDE model in $N$ dimensions, defined by Eq. \eqref{eq:non_dim_PDE},  the dispersion relation between linear growth rate $\lambda$ and spatial mode $\boldsymbol{k}$ is given by

\begin{alignat}{2}
&\lambda(k) = -k^2 +h'(U) + \frac{2\pi^{\frac{N}{2}}}{\Gamma\left(\frac{N}{2}\right)}\,Up(U)g'(U)\frac{\mu}{\xi^N}\,k\int_0^\infty \tilde{\Omega}\left(\frac{s}{\xi}\right)\, s^{N-1}\,j_1^{(N)}(ks) \diff s \label{eq:ND1S_lambda}, \quad &&N\in\mathbb{N}\\
&\phantom{\lambda(k)}=-k^2 +h'(U) + \frac{2\pi^{\frac{N}{2}}}{\Gamma\left(\frac{N}{2}\right)}\,Up(U)g'(U)\mu \frac{1}{\xi}G(k\xi), \quad &&N\in\mathbb{N} \\
\intertext{in which $j_1^{(N)}(x)$ are the first order, $N^{\text{th}}$ dimensional hyperspherical Bessel functions where}
&j_1^{(1)}(x) =  \text{sin}(x),\\
&j_1^{(2)}(x) =  J_1(x),\\
&j_1^{(3)}(x) =  j_1(x) = \frac{1}{x^2}\text{sin}(x)-\frac{1}{x}\text{cos}(x). \\
\intertext{Further, in summary, at $x=0$,}
&j_1^{(N)}(0) = 0 \label{eq:j=0_at_zero}, && N\in\mathbb{N}, \\
&\frac{\diff j_1^{(N)}(x)}{\diff x}\Bigg\rvert_{0}>0,  && N\in\mathbb{N}. \\
\intertext{For $x \gg \frac{1}{4}(N^2 - 1)$ we can approximate,}
&j_1^{(N)}(x) \propto \frac{1}{x^{\frac{N-1}{2}}}\,\text{sin}(x+\phi_N) \label{eq:j_large_x}, &&N\in\mathbb{N}, \\
\intertext{and so for $ks \gg \frac{1}{4}(N^2 - 1)$,}
&s^{N-1}j_1^{(N)}(ks) \propto \left(\frac{s}{k}\right)^{\frac{N-1}{2}}\,\text{sin}(ks+\phi_N) \label{eq:j_large_k}, &&N\in\mathbb{N},
\end{alignat}
to good approximation.

\subsection{Analysis of Dispersion Relation}
Here we have a dispersion relation that depends on diffusion through the $-k^2$ term, proliferation-death through $h'(U)$, and the nonlocal interaction appears as a hyperspherical Hankel transform of the nonlocal interaction kernel $\tilde{\Omega}\left(\frac{s}{\xi}\right)$. The form of this transform is controlled by the integral kernel, given by $s^{N-1}\,j_1^{(N)}(ks)$, and this depends on the dimensionality of the system. For example, in 1D the transform reduces to a Fourier sine transform, leading to the 1D dispersion relation derived in \citet{Painter2015_nonlocal_rd_developmental_biology}. In 2D and 3D, the integral kernels respectively are $s J_1(x)$ and $s^2 j_1(x)$, corresponding to the first order Hankel and first order spherical Hankel transforms. This dependence on dimensionality arises because the integral transforms represent summations of interaction signals originating from all directions in space. Summations over a disk, for example, will behave differently to summations over a line. Section \ref{section:physical_interpretation} uses this physical picture to intuitively explain some of the behaviour of our patterning system in different dimensions. However, to precisely determine this behaviour, we only have to study the properties of the integral transforms and their corresponding hyperspherical Bessel functions.

Firstly, for all dimensions, $\lambda(k)$ is always real, and so wavemodes are purely growing or decaying - we should expect to have no temporal oscillations in the linear regime. Another notable feature is that for all $N$ the homogeneous steady state's stability to homogeneous perturbations depends only on the stability of the proliferation-death kinetics. This is evident from $j_1^{(N)}(0)=0$ which implies that from Eq. \eqref{eq:ND1S_lambda} we have $\lambda(k=0)=h'(U)$. We are interested in proliferation-death kinetics with a stable equilibrium $U$, and so $\lambda(k=0)=h'(U)<0$.

Next, we can show that heterogeneous perturbations with infinitely high frequency will decay, which is a requirement for a physically realistic Turing patterning system. Eqs. \eqref{eq:j_large_k} and \eqref{eq:ND1S_lambda} tell us that as $k\to\infty$, the nonlocal interaction term in the dispersion relation cannot grow faster than $\mathcal{O}\left(k^{\left(\frac{3}{2}-\frac{N}{2}\right)}\right)$. For all $N$, this grows slower than the diffusion term, $-k^2$. Thus, as $k\to\infty$, \,$\lambda(k)\to -\infty$.

Additionally, this system can always have emerging spatial patterns, provided the interaction strength $\mu$ is of sufficiently large magnitude and appropriately positive or negative. For finite $k$, there will in general always be some value of $k$ for which the nonlocal interaction term in the dispersion relation is non-zero (for a non-zero interaction kernel). This means there will always be some value of $\mu$ for which this term is positive and outweighs the negative diffusion and proliferation terms. In this case, $\lambda(k)$ would be positive, and so a heterogeneous perturbation would grow.

Together, the previous three observations tell us that this system can have a Turing bifurcation for any number of dimensions.
Whether this occurs for attractive interactions ($\mu>0$) or repulsive interactions ($\mu<0$) requires more analysis. 

The fastest growing mode is simple to identify for parameters in which the nonlocal interaction is dominant over diffusion, as in this case we can approximate the growth rate as $\lambda(k)\approx Up(U)g'(U)\mu\frac{1}{\xi}G_N(k\xi)$. As this is a function of the product $k\xi$, its maximum is at $k\approx\frac{1}{\xi}\operatorname*{arg\,max}_x G_N(x)$ for attractive interactions, and $k\approx\frac{1}{\xi}\operatorname*{arg\,min}_x G_N(x)$ for repulsive interactions. Here, $\operatorname*{arg\,max}_x G_N(x)$ and $\operatorname*{arg\,min}_x G_N(x)$ are positive real numbers that only depend on the dimensionality. The wavelength of the initially emergent pattern is therefore directly proportional to the nonlocal signalling range $\xi$ and insensitive to changes in any other parameter. This is consistent with what we might intuitively expect and is also observed in 1D in \citet{Painter2015_nonlocal_rd_developmental_biology}.

\subsection{Attractive Nonlocal Interactions}
To have patterning for attractive nonlocal interactions ($\mu>0$) the hyperspherical Hankel transform in Eq. \eqref{eq:ND1S_lambda} must be positive for some $k>0$ as this would allow $\lambda(k)$ to be positive for a sufficiently large $|\mu|$. We can show that this transform is indeed always positive for small non-zero values of $k$ by looking at its derivative with respect to $k$:

\begin{equation}
    \begin{split}
        \frac{\partial}{\partial k}\left[\int_{0}^{\infty}\diff s\,\tilde{\Omega}\left(\frac{s}{\xi}\right)s^{N-1}\,j_1^{(N)}(ks)\right]\Bigg\rvert_{k=0} = \int_{0}^{\infty}\diff s\,\tilde{\Omega}\left(\frac{s}{\xi}\right)s^{N-1}\,\frac{\partial j_1^{(N)}(ks)}{\partial k}\Bigg\rvert_{k=0}>0,
    \end{split}
\end{equation}
where positivity is guaranteed by all terms in the integrand being positive.

We know that $j_1^{(N)}(0)=0$, which implies that the transform is equal to zero at $k=0$ and thus positive for small positive $k$. Therefore, for all $N$, attractive interactions can support Turing patterning for a sufficiently large interaction strength.

\subsection{Repulsive Nonlocal Interactions}
\label{section:repulsive_interactions_theory}
In contrast to attractive interactions, patterning can only occur for repulsive interactions ($\mu<0$) when the hyperspherical Hankel transform in Eq. \eqref{eq:ND1S_lambda} is negative. We know from the above section that for small $k$ the transform is positive and has positive gradient with respect to $k$. For $\mu<0$ this implies that $\lambda$ is negative and decreasing for small $k$. Whether it will ever increase and be above zero (as $k$ increases) depends on the interaction kernel.

\begin{figure}
    \centering
    \includegraphics[width=0.7\textwidth]{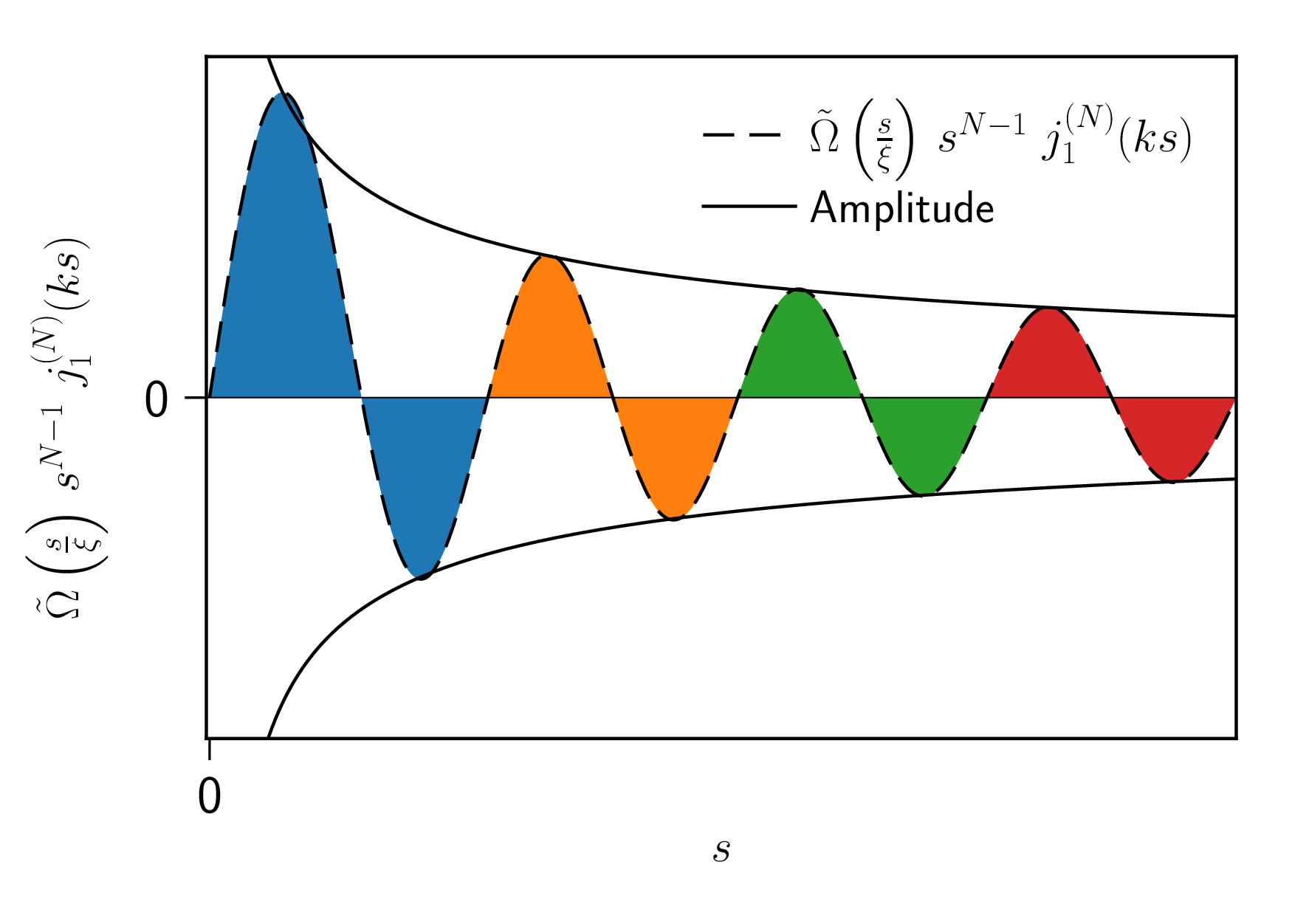}
    \caption{A schematic of the integrand of the hyperspherical Hankel transform, $\tilde{\Omega}\left(\frac{s}{\xi}\right)s^{N-1}\,j_1^{(N)}(ks)$, is plotted by the dashed line, for some decaying kernel. By pairing adjacent positive regions to negative regions (colour matched), we see that the total area is positive and so the integral is positive, when the amplitude of $\tilde{\Omega}\left(\frac{s}{\xi}\right)s^{N-1}\,j_1^{(N)}(ks)$ is non-increasing. However, if $\tilde{\Omega}\left(\frac{s}{\xi}\right)$ were to decay sufficiently slowly (or even increase) over some region, then the amplitude of $\tilde{\Omega}\left(\frac{s}{\xi}\right)s^{N-1}\,j_1^{(N)}(ks)$ could increase over this region, potentially allowing negative regions of the integrand to outweigh positive regions, leading to a negative integral.}
    \label{fig:Hankel_transform_sign}
\end{figure}

From Fig. \ref{fig:hyperspherical_Bessels} we see that $s^{N-1}j_1^{(N)}(ks)$ is  oscillatory with $s$, and we also know that the first half of its initial oscillation is positive. Therefore for the specific case of a monotonic interaction kernel, we can determine the sign of the transform by comparing the contributions from the positive part of each oscillation of $s^{N-1}j_1^{(N)}(ks)$ against the subsequent negative part. If the product $\tilde{\Omega}\left(\frac{s}{\xi}\right)\, s^{N-1}\,j_1^{(N)}(ks)$ decreases in amplitude as $s$ increases, then the contribution from the positive part of each oscillation will be larger than its subsequent negative part, and thus the integral transform will be positive for all $k$. See Fig. \ref{fig:Hankel_transform_sign} for a visual explanation. Similar arguments are used for the Fourier transform in \citet{sign_of_transform_argument}. If the above condition is broken, and instead the amplitude increased over some region, then the integral transform is not guaranteed to be positive for all $k$, which is significant because if \textbf{any} value of $k$ leads to a negative integral transform, then pattern formation can occur (for sufficiently high $|\mu|$).  
 
With the above argument, and the $\mathcal{O}\left(s^{\frac{N-1}{2}}\right)$ amplitude scaling from Eq. \eqref{eq:j_large_k}, it follows that: \\if $s^{\frac{N-1}{2}}\tilde{\Omega}\left(\frac{s}{\xi}\right)$ is a non-increasing function, then pattern formation with repulsive interactions is not possible. For example, in 1D, non-increasing interaction kernels can never support patterning for repulsive interactions. In contrast, in 2D, patterning with repulsive interactions can occur, but only for interaction kernels that decay slower than  $\mathcal{O}(\frac{1}{\sqrt{s}})$  over some region. For the boundary case where $s^{\frac{N-1}{2}}\tilde{\Omega}\left(\frac{s}{\xi}\right)$ is constant, the transform is non-negative because the first peak in $j_1^{(N)}(ks)$ is proportionally larger than the first trough for $N>1$ and equal to it for $N=1$. Hence the above wording of `non-increasing' rather than `decreasing'. It is worth noting that all of these arguments also apply when the interaction kernel has compact support, such as the `top-hat' kernel.

Compared to attractive interactions, which can enable patterning for small $k$, repulsive interactions can only enable patterning at larger values of $k$, because the hyperspherical Hankel transform is always positive for small $k$. Thus, assuming all other parameters are the same, repulsive interactions can form patterns with a shorter wavelength than attractive interactions. A further corollary is that repulsive interactions require a greater magnitude of interaction strength, $|\mu|$, to overcome the larger $-k^2$ diffusion term, in order to form patterns. Physically, this corresponds to the stabilising effect of diffusion being stronger at shorter length scales.

Most importantly, the inability for the 1D system to form patterns with repulsive interactions and non-increasing interaction kernels marks a fundamental difference between it and higher dimensional systems. The mathematical source of this difference is that the first order hyperspherical Hankel transform can only be negative for $N>1$, assuming a non-increasing interaction kernel. In general, for a non-increasing interaction kernel, increasing the number of dimensions enables faster decaying interaction kernels to support patterning with repulsive interactions. However, we must be careful not to assume that this always implies that increasing dimensionality promotes pattern formation with repulsive interactions. For specific model applications, the realistic form of an interaction kernel may also change with dimensionality, and potentially counter the change in the integral kernel.

\subsection{Transport-Only System}
\label{section:transport_only}

We now consider the specific case of $h(u)=0$, also studied in \citet{Painter2015_nonlocal_rd_developmental_biology}. In this case, cells do not proliferate or die, instead only undergoing diffusive and nonlocal transport, thus conserving total mass. \citet{Painter2015_nonlocal_rd_developmental_biology} showed that, unlike in classical local reaction-diffusion systems, transport-only pattern formation is possible for this nonlocal model.

Without proliferation-death kinetics, there is no longer a unique stable homogeneous steady state. Instead, the density $U$ of a homogeneous steady state can take any value. In practice, $U$ will be dictated by initial conditions. 

The dispersion relation, Eq. \eqref{eq:ND1S_lambda} with $h(u)=0$, implies that it is still possible to have $\lambda(k)>0$ for some $k>0$. Therefore, the transport-only system still exhibits Turing instabilities, in which the homogeneous state $U$ can undergo a transport-driven instability with a \textit{heterogeneous} perturbation to form a pattern. However, it is important to note that the homogeneous steady state is no longer asymptotically stable, 
but instead only Lyapunov stable, to \textit{homogeneous} perturbations. The zeroth mode growth rate, $\lambda(k=0)$, is equal to zero, as seen by the dispersion relation or equivalently by mass conservation. As such, a homogeneous perturbation $\delta U$ will move the system at homogeneous steady state $U$ to homogeneous steady state $U+\delta U$.

Through this process, an alternate route to instability, not considered in \citet{Painter2015_nonlocal_rd_developmental_biology}, can occur in which a homogeneous perturbation moves the system from a regime that was previously stable to heterogeneous perturbations to one that is unstable to heterogeneous perturbations. We see this from the dispersion relation, which is dependent on $U$, a quantity that is no longer fixed but instead subject to change by any homogeneous perturbation. We have verified through numerical simulations that this alternate route to instability can indeed occur for transport-only systems (results not shown, see Section \ref{section:numerics} for details of the numerical scheme).

\section{Physical Interpretation}
\label{section:physical_interpretation}
The dispersion relation can be understood more intuitively in a physical sense by comparing the magnitude and direction of the fluxes induced by the nonlocal interactions at a single point. In order for a heterogeneous perturbation to be stable, there cannot be a net flux of cells away from regions of high density.

We can analyse these fluxes by first considering only the nonlocal interaction with cell packing. In the original integro-PDE (Eq. \eqref{eq:non_dim_PDE}), the effects of diffusion, proliferation, and nonlocal interaction with packing contribute additively. In the linear dispersion relation, Eq. \eqref{eq:ND1S_lambda}, these processes similarly contribute additively and independently, and so we can consider each process separately to investigate its effect on stability. 

\subsection{1D}

For illustrative purposes, we first consider the simplest case of a 1D system with a top-hat function for the nonlocal interaction kernel, which has the form
\begin{equation}
    \tilde{\Omega}\left(\frac{s}{\xi}\right)=\begin{cases}
      \Omega_0, &s \leq \xi\\
      0, &s > \xi.
    \end{cases}
\end{equation}
This allows constant interaction within a distance $\xi$ and zero interaction beyond. 

\begin{figure}
    \begin{center}
    \subfloat[]{    \label{fig:Physical_forces_argument_1D_small}
    \includegraphics[width=0.3\textwidth]{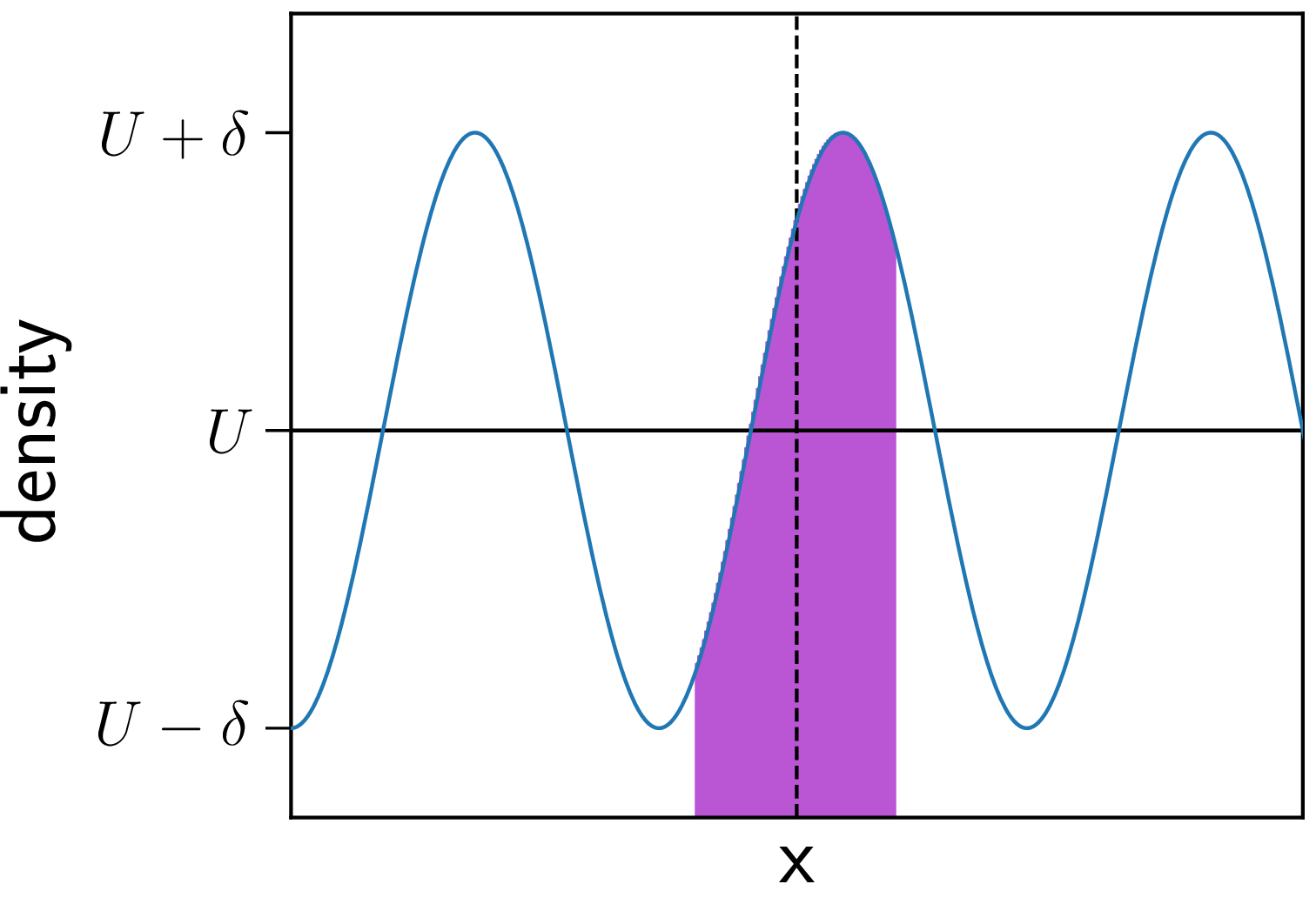}}    
    \quad
    \subfloat[]{    \label{fig:Physical_forces_argument_1D_med}
    \includegraphics[width=0.3\textwidth]{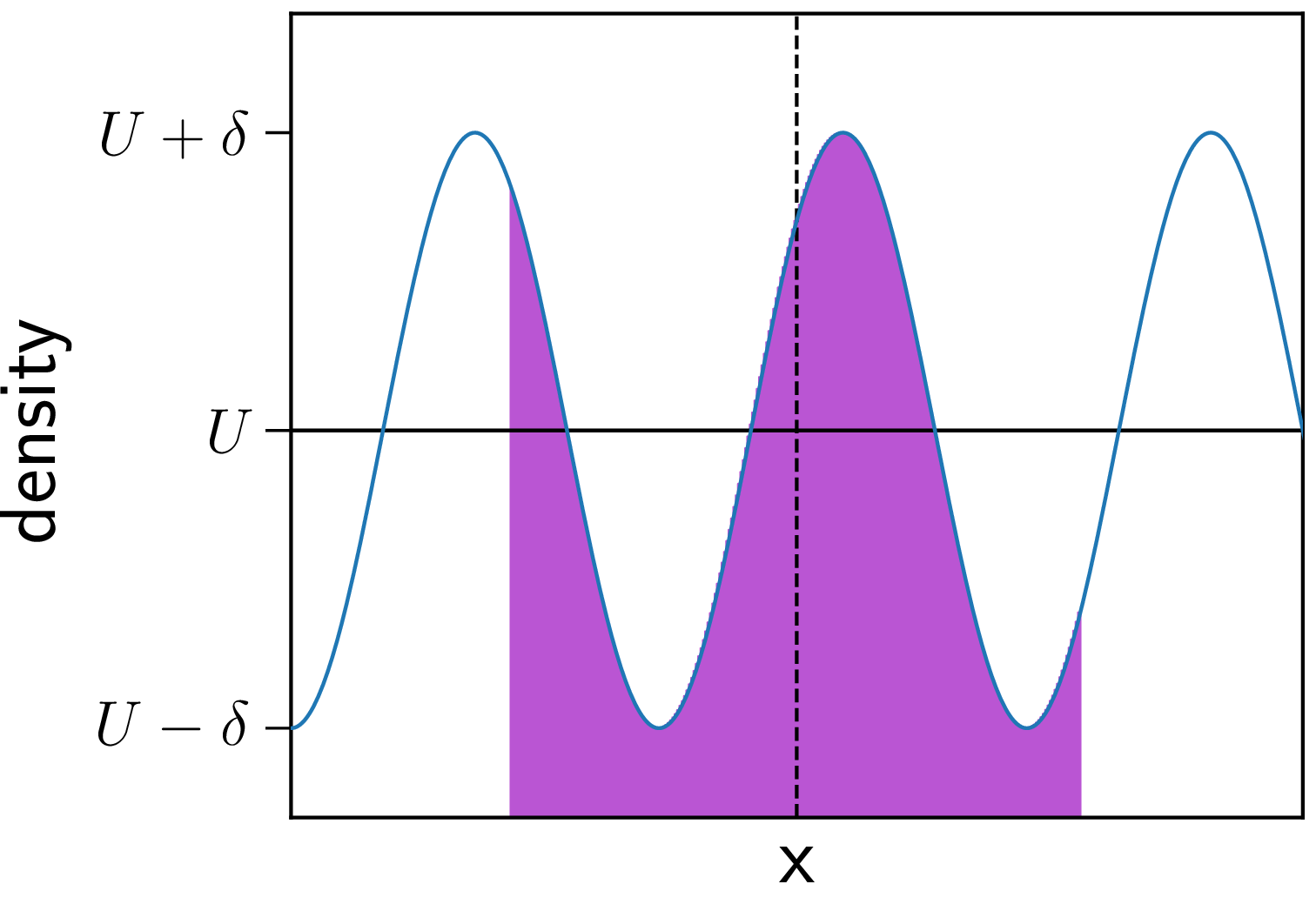}}    
    \quad
    \subfloat[]{    \label{fig:Physical_forces_argument_1D_big}
    \includegraphics[width=0.3\textwidth]{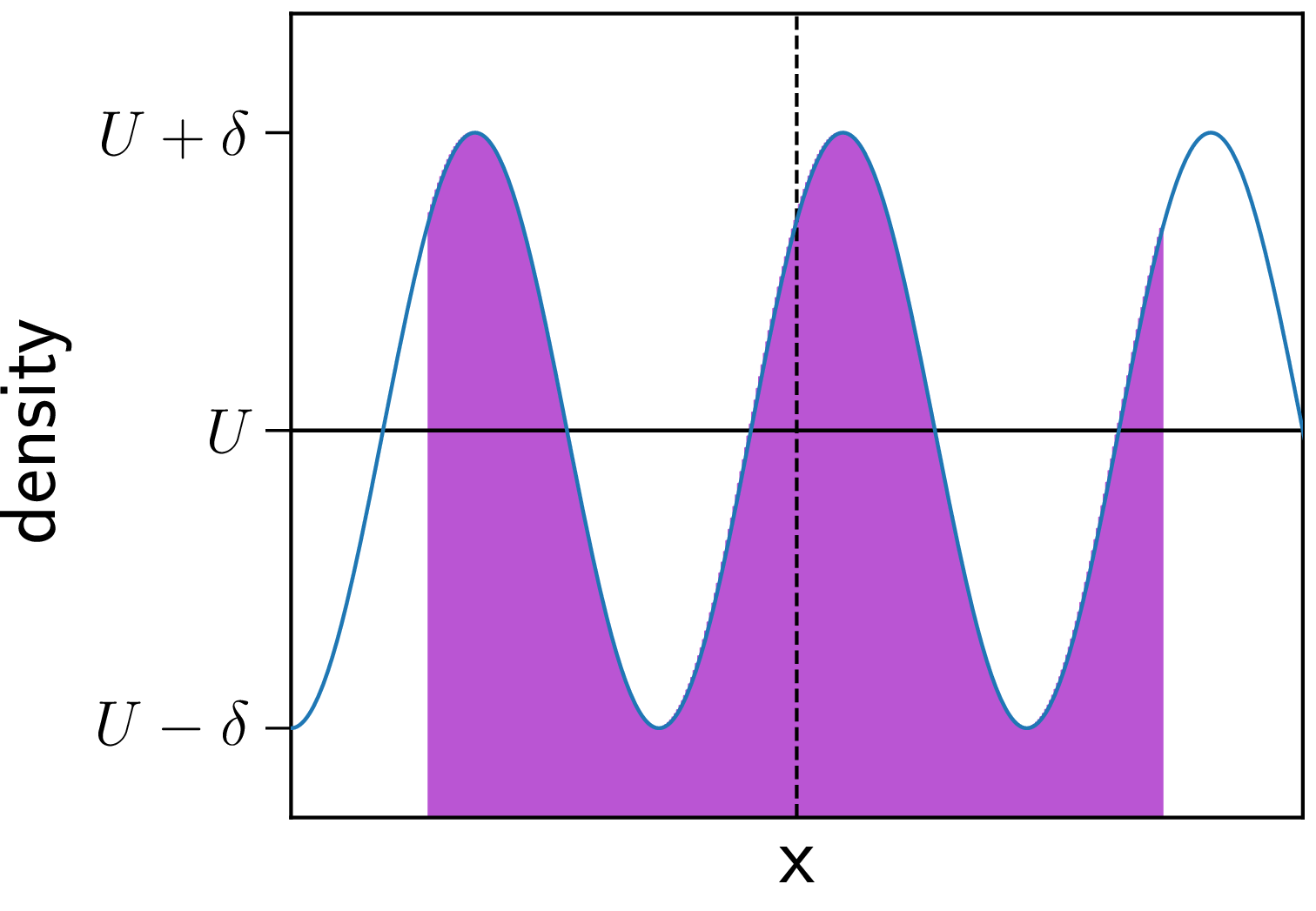}}   
  \\
  \subfloat[]{\label{fig:Physical_forces_argument_2D_small}%
    \includegraphics[width=0.4\textwidth]{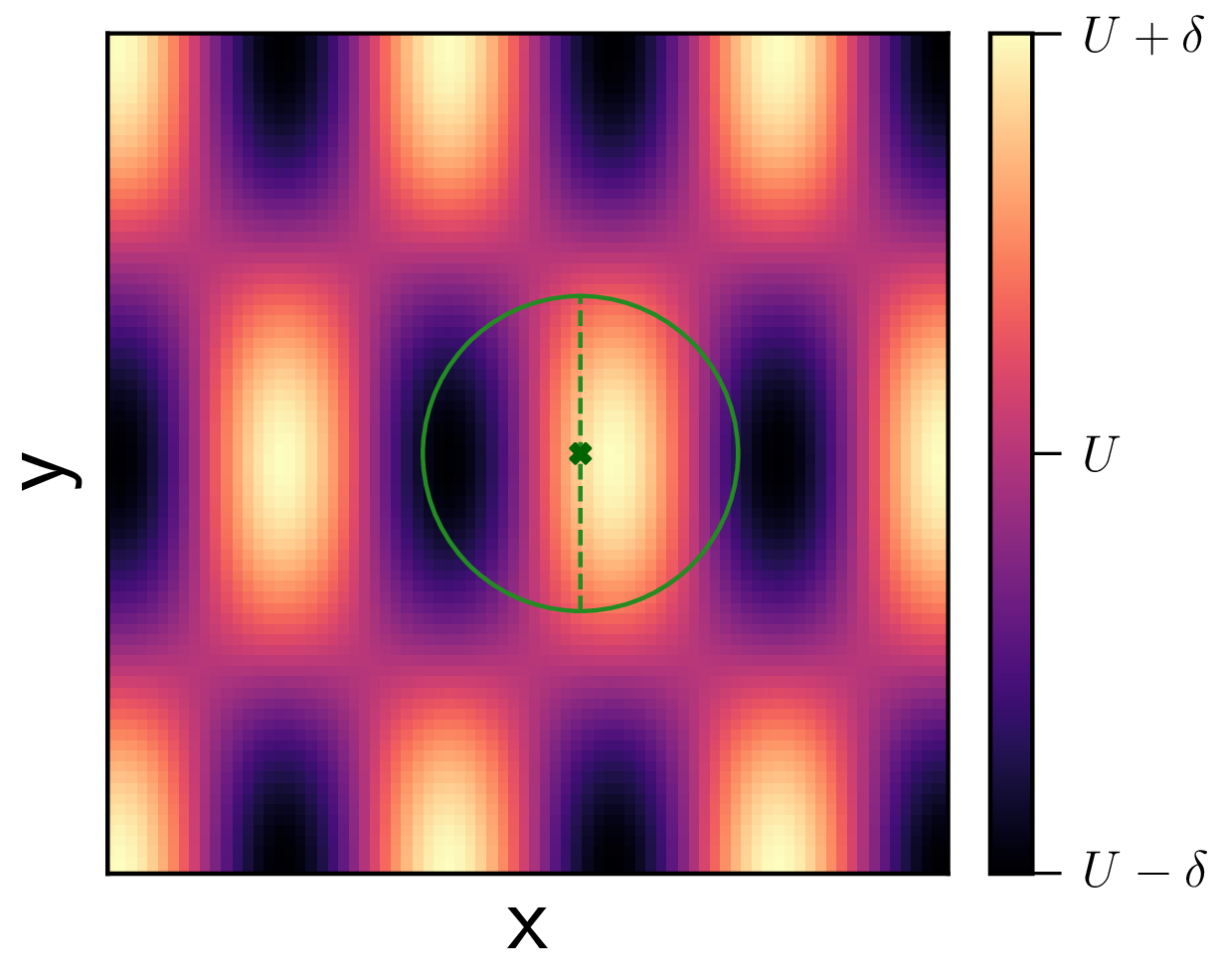}}
    \quad
    \subfloat[]{\label{fig:Physical_forces_argument_2D_big}%
    \includegraphics[width=0.4\textwidth]{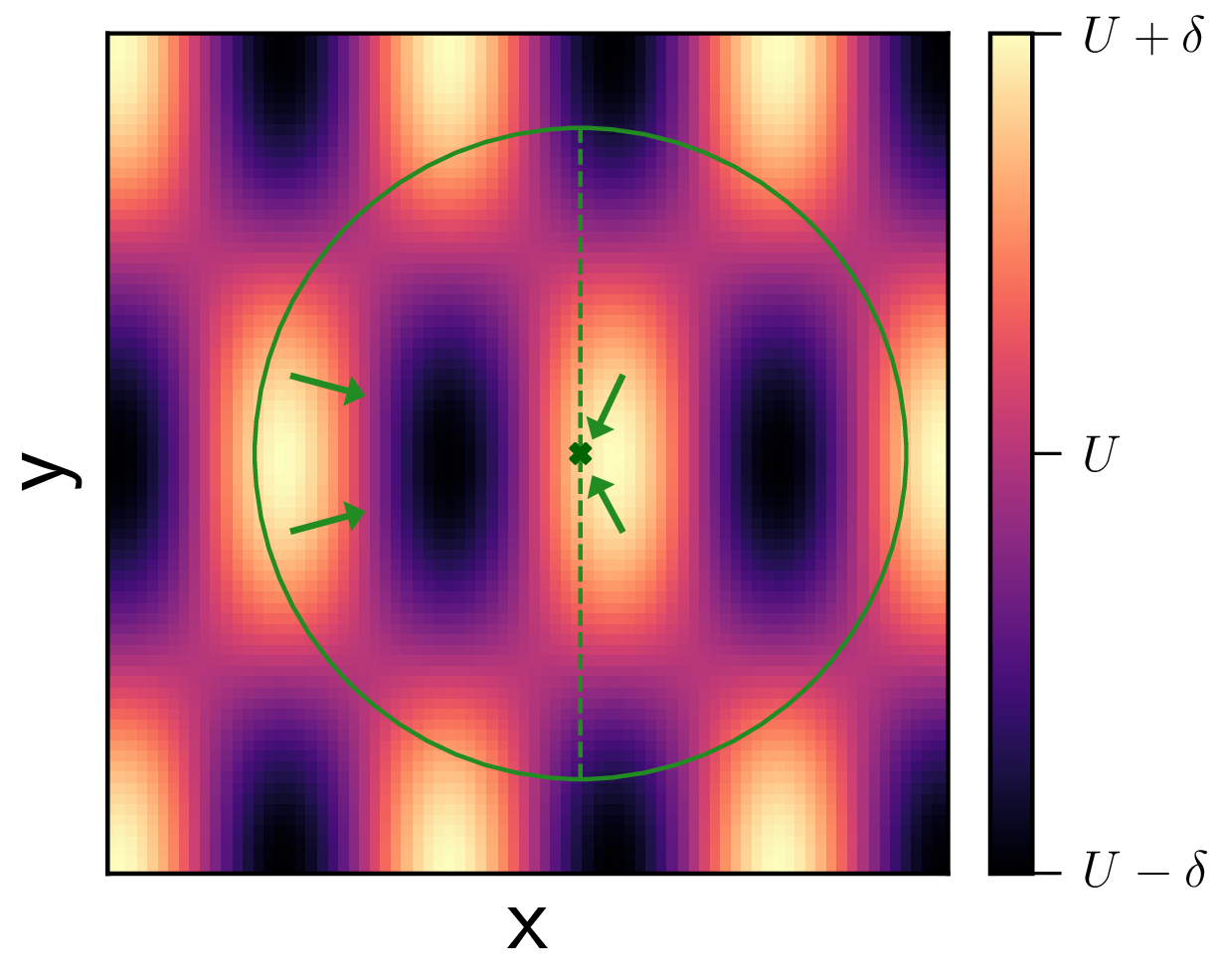}}
  \end{center}

    \caption{Schematics illustrating the influence of cells within a signaling range on an arbitrary point in 1D and 2D. \subref{fig:Physical_forces_argument_1D_small}-\subref{fig:Physical_forces_argument_1D_big} show an example perturbation of form $\delta\text{sin}(kx)$, with wavenumber $k$ and amplitude $\delta$, applied to a homogeneous state of density $U$ in a 1D system with nonlocal interactions. We consider the fluxes of cells at a point $p$ (dotted line) induced by cells within a distance $\xi$ (purple shaded region). In these examples, $\xi$ is smallest in \subref{fig:Physical_forces_argument_1D_small}, larger in \subref{fig:Physical_forces_argument_1D_med}, and largest in \subref{fig:Physical_forces_argument_1D_big}, where it equals a whole period of the perturbation. Thus, for \subref{fig:Physical_forces_argument_1D_small}, \subref{fig:Physical_forces_argument_1D_med} the shaded area is larger to the right of $p$ than to the left, and for \subref{fig:Physical_forces_argument_1D_big} the areas are equal. \subref{fig:Physical_forces_argument_2D_small}, \subref{fig:Physical_forces_argument_2D_big} show a heatmap of the density profile in the analogous 2D system with perturbation $\delta\,\text{sin}(k_x x)\,\text{sin}(k_y y)$ applied to homogeneous density $U$. We consider fluxes of cells at a point $\boldsymbol{p}$ (crossed marker) induced by cells within distance $\xi$ (green circle). $\xi$ is smaller in \subref{fig:Physical_forces_argument_2D_small} than in \subref{fig:Physical_forces_argument_2D_big}. Additionally, \subref{fig:Physical_forces_argument_2D_big} includes arrows demonstrating the direction of fluxes of cells at $\boldsymbol{p}$ induced by cells at different points (for a repulsive interaction). Note: the arrows are displayed at the points where the signals originate but each flux contribution occurs at $\boldsymbol{p}$.} 
    \label{fig:Physical_forces_argument} 
\end{figure}

With this in mind, for a sinusoidal perturbation to the homogeneous state with frequency $k$, we can compare the induced fluxes at an arbitrary point $p$. Fig. \ref{fig:Physical_forces_argument}\subref{fig:Physical_forces_argument_1D_small}-\subref{fig:Physical_forces_argument_1D_big} show such a point with the nearest high density peak on its right (by symmetry the following argument will also hold if the nearest peak was on its left). If the total flux to the left is larger than the total flux to the right, then the net movement of cells is from high density to low density and the pattern will decay to homogeneity. If the opposite is true, the amplitude of the perturbation will continue to grow, with some upper limit due to cell packing. 

To determine the induced flux from each direction we simply sum the number of cells to the left and right of point $p$ within  distance $\xi$. Fig. \ref{fig:Physical_forces_argument}\subref{fig:Physical_forces_argument_1D_small}-\subref{fig:Physical_forces_argument_1D_big} illustrate how the number of cells on the right is always larger than the number of cells on the left. The only exception is when $\xi$ equals a whole number of periods of $k$, in which case they are equal, and thus the net flux is zero, but any inclusion of diffusion and/or proliferation-death will destabilise the pattern. This is a fine-tuning observation; however, with all other values of $\xi$, for an attractive interaction, the greater number of cells to the right leads to a net induced flux to the right, and so the perturbation can always grow. Conversely, for repulsive interactions, the net flux at $p$ is to the left and the pattern decays to homogeneity.

The above argument can be expressed mathematically via:
\begin{equation}
\begin{split}
        \boldsymbol{F} &\propto \mu \boldsymbol{\hat{x}} \left[ \int_0^\infty \tilde{\Omega}\left(\frac{x}{\xi}\right)\text{sin}(k(x+p))\diff x - \int_{-\infty}^0 \tilde{\Omega}\left(\frac{x}{\xi}\right)\text{sin}(k(x+p))\diff x \right] \\
        &\propto \mu \boldsymbol{\hat{x}} \left[ \int_0^\xi \text{sin}(k(x+p))\diff x - \int_{-\xi}^0 \text{sin}(k(x+p))\diff x \right] \\
        &\propto \mu \boldsymbol{\hat{x}} \,2\,\text{cos}(kp)\left[1-\text{cos}(k\xi)\right].
\end{split}
\label{eq:forces_integral}
\end{equation}
As $p$ is defined to the left of the nearest peak, $-\frac{\pi}{2}<kp\leq \frac{\pi}{2}$ and thus $2\,\text{cos}(kp)\left[1-\text{cos}(k\xi)\right]\geq 0$. This shows that the contribution from the right is larger than from the left, in this case with the equality applying when $\xi=\frac{2\pi n}{k}$ for $n\in \mathbb{N}$, as stated above.

If we generalise to decaying interaction kernels in which cells further from $p$ induce less flux at $p$, the same argument holds - the contribution from the right of $p$ is always larger than the contribution from the left, even if we allow the interaction to act over an infinite range. Therefore in the general 1D case, attractive nonlocal interactions promote pattern formation whilst repulsive nonlocal interactions inhibit it. Including diffusion and/or proliferation-death only further dissipates any heterogeneity\footnote{This is true in the single species system. However with multiple species, diffusion can promote heterogeneity through the classic Turing mechanism \citep{TuringOriginalPaper}.}, and so cannot enable pattern formation with repulsive interactions.

Furthermore, we see from Fig. \ref{fig:Physical_forces_argument} that increasing $\xi$ has an identical effect to increasing $k$ on the proportion of the pattern contained within the scale of the interaction range, and therefore has the same effect on stability, provided nonlocal transport dominates over diffusion and proliferation-death kinetics. This is consistent with how the nonlocal interaction term in the dispersion relation is a function of the product $k\xi$ and how, when nonlocal transport dominates, the final pattern wavelength is directly proportional to $\xi$.

All of the above conclusions are also predicted by the dispersion relation. In fact, the integral in Eq.\,\eqref{eq:forces_integral} is essentially the 1D hyperspherical Hankel transform. However, this physical interpretation also allows us to more intuitively see why stability changes in higher dimensions.

\subsection{Higher dimensions}

To extend this argument to 2D, consider the induced fluxes in the $x$ direction on point $\boldsymbol{p}$ in Fig.\,\ref{fig:Physical_forces_argument}\subref{fig:Physical_forces_argument_2D_small},\,\subref{fig:Physical_forces_argument_2D_big}. In \subref{fig:Physical_forces_argument_2D_small}, there are many more cells in the semicircle to the right of $\boldsymbol{p}$ than in the semicircle to the left. Therefore this scenario, in which $\xi$ is small, is similar to the 1D case: attractive interactions will move cells at $\boldsymbol{p}$ from lower density to higher density, thereby promoting pattern formation. Conversely, repulsive interactions will move cells at $\boldsymbol{p}$ from higher density to lower density, inhibiting any pattern formation.

However, as $\xi$ increases it becomes important to consider a fundamental difference between $2$D and $1$D: the induced fluxes are not all parallel but instead point at different angles.

For the $\xi$ used in Fig.\,\ref{fig:Physical_forces_argument}\subref{fig:Physical_forces_argument_2D_big}, the difference between the number of cells on the left versus the right of $\boldsymbol{p}$ is fairly small, and so although there are still more cells to the right, the effect of the induced fluxes pointing at different angles is sufficient to ensure that the total flux induced by cells from the left is larger than by cells from the right. This is because, as illustrated by the arrows in Fig.\,\ref{fig:Physical_forces_argument}\subref{fig:Physical_forces_argument_2D_big}, fluxes induced by cells that are further away horizontally from $\boldsymbol{p}$ have larger horizontal components (compared to their magnitude) than fluxes induced by closer cells. This angular effect can reverse the previous results such that patterns can be, instead, dissipated by attractive interactions and promoted by repulsive interactions.

The same logic and conclusions apply to decaying interaction kernels with infinite range (i.e. without compact support) by considering $\xi$ as a scale length instead of a cut off. Decaying kernels correspond to cells that are further away from each other having weaker interactions. If this decay is sufficiently fast, it can dominate the effect of changing angle with distance, thereby prohibiting any pattern formation with repulsive interactions. This is predicted in Section \ref{section:repulsive_interactions_theory} through the statement that any kernel decaying faster than $s^{\frac{1-N}{2}}$ cannot support pattern formation with repulsive interactions.

In general, the angular effect described here is present in all dimensions higher than 1, implying that patterning with repulsive interactions is possible for all $N>1$. The exact form and extent of this angular effect will also differ with dimensions, i.e. the sum of induced fluxes from within a sphere on a 3D domain will not behave identically to the above 2D example. All of this information is fully captured in the dispersion relation through the integral kernel, $s^{N-1}j_1^{(N)}(ks)$.

\section{Numerical Validation of Instability \& Exploration Beyond Instability}
\label{section:numerics}

\subsection{Numerical Methods}

\subsubsection{Numerical Integration Scheme}
To validate our dispersion relation and explore pattern forming behaviour beyond linear instability, we numerically integrate the full integro-PDE (Eq. \eqref{eq:non_dim_PDE}) in 2D. As in \citet{Painter2015_nonlocal_rd_developmental_biology} we choose logistic growth for the proliferation-death kinetics, $h(u)=\rho u(1-\frac{u}{U})$, and a linear packing function, $p(u)=1-u$, which restricts $U<1$. We assume that the induced flux increases linearly with cell density, $g(u)=u$. As such, we integrate
\begin{equation}
    \frac{\partial u}{\partial t} = \nabla^2 u + \rho u(1-\frac{u}{U}) -\boldsymbol{\nabla}\cdot\left(u(1-u)\frac{\mu}{\xi^{2}} \int_{-\frac{L}{2}}^{\frac{L}{2}}\int_{-\frac{L}{2}}^{\frac{L}{2}} \boldsymbol{\hat{s}}\,\tilde{\Omega}\left(\frac{s}{\xi}\right)u(\boldsymbol{x}+\boldsymbol{s},t)\diff s_x \diff s_y\right).
\label{eq:2D_PDE}
\end{equation}

As a proxy for an infinite domain we choose a square domain, $[0,L]^2$, with periodic boundary conditions, as these are less restrictive on the possible allowed wavemodes than Neumann or Dirichlet conditions, and thus minimise any effects of the boundary on the final pattern. \citet{Painter2015_nonlocal_rd_developmental_biology} make the same choice, so this also allows for a more direct comparison. For initial conditions, we choose the homogeneous state $u=U$ with a perturbation of Gaussian white noise with zero mean and standard deviation $10^{-3}$. This choice of perturbation ensures every possible wavemode is excited.

We use the method-of-lines, first discretising in space and then integrating the resulting ODEs with backwards differentiation formulae. The latter is implemented  through the $\texttt{integrate.solve\_ivp}$
function from Python's SciPy library \citep{Scipy}. For errors in the integration over time, we choose $10^{-11}$ for both the relative and absolute tolerances. For the spatial discretisation, we use $100\times 100$ mesh points, having verified that this captures the same behaviour as with any further increases in precision. Notably, the nonlocal term in Eq. \eqref{eq:2D_PDE} is equivalent to a convolution between $\boldsymbol{\hat{s}}\,\tilde{\Omega}\left(\frac{s}{\xi}\right)$ and $u(\boldsymbol{x})$, and so we efficiently compute this using fast Fourier transform methods with the convolution theorem. The diffusion term is discretised using the standard centred 5-point stencil. All code and associated documentation can be found at the repository \citep{My_GitHub}.

\subsubsection{Estimating the Dispersion Relation from Simulation}
In order to test the dispersion relation, we use a method of computing the growth rate $\lambda_{\boldsymbol{k}}$ of a particular spatial mode $\boldsymbol{k}$ directly from the simulation. This is achieved by decomposing the density $u(\boldsymbol{x},t)$ during the linear regime (i.e. at early time) into modes as 
\begin{equation}
    u(\boldsymbol{x},t) = \sum_{\boldsymbol{k}} M_{\boldsymbol{k}}(t) e^{i\boldsymbol{k}\cdot\boldsymbol{x}},
\end{equation}
where $M_{\boldsymbol{k}}(t)$ is the amplitude of each spatial mode $\boldsymbol{k}$, and thus evolves as
\begin{equation}
    M_{\boldsymbol{k}}(t) = M_{\boldsymbol{k}}(0)e^{\lambda_{\boldsymbol{k}}t},
\end{equation}
while in the linear regime. We can rearrange this to give
\begin{equation}
    \lambda_{\boldsymbol{k}}=\frac{1}{t}\text{log}\left(\frac{M_{\boldsymbol{k}}(t)}{M_{\boldsymbol{k}}(0)}\right),
\label{eq:log_normalised_growth_mode}
\end{equation}
which can be calculated directly from simulation - each $M_{\boldsymbol{k}}(t)$, including $M_{\boldsymbol{k}}(0)$, is calculated through a discrete Fourier transform of the spatially discretised $u(\boldsymbol{x},t)$ at each time step. 

In practise, we choose the time $t=10^{-6}$ for this calculation to ensure the amplitude of the perturbation is still small and the dynamics are approximately linear. We have verified that the specific choice of $t$ does not affect the results as long as it is sufficiently small (any $t<0.01$ is sufficient for any of the sets of parameters used in this paper, and ensures the simulation is in the linear regime).

\subsubsection{Example Interaction Kernels}
For consistency with \citet{Painter2015_nonlocal_rd_developmental_biology}, we use the same three examples of nonlocal interaction kernels; these are displayed in Table \ref{table:interaction_kernels}. The \textbf{O1} kernel is the `top-hat' function, representing uniform signalling within a given area and zero signalling outside that area. The \textbf{O2} kernel is the exponential decay function, representing a rapid decrease in interaction rate with increasing distance from the cell. Finally, the \textbf{O3} kernel, unlike the other functions, is not monotonic and instead has a maximum at a distance $\xi$ from the cell and a smooth drop off to zero interaction at shorter or further distances. A kernel with this form was also used in the animal swarming model of \citet{early_swarm_nonlocal_model}. Each kernel is appropriately normalised for the 2D case, according to Eq. \eqref{eq:omega_normalisation}.

Specialising from Eq. \eqref{eq:ND1S_lambda}, the dispersion relation in 2D with our chosen $p(u)$, $h(u)$, and $g(u)$ is given by
\begin{equation}
    \lambda(k)=-k^2 - \rho + 2\pi U(1-U)\mu\frac{k}{\xi^2}H_1(k;\xi),
    \label{eq:2D_dispersion_relation}
\end{equation}
where $H_1(k; \xi)$ is the first order canonical Hankel transform of the interaction kernel. This transform for each example kernel is shown in Table \ref{table:interaction_kernels}.

\begin{table}
\centering
\normalsize
\begin{tabular}{ |c |c| c| c| }
 \hline
    & \textbf{O1} & \textbf{O2} & \textbf{O3}\\
 \hline
 $\tilde{\Omega}\left(\frac{s}{\xi}\right)=$ & $\begin{cases}
      \frac{1}{\pi}, &s \leq \xi\\
      0, &s>\xi \end{cases}$ & $\frac{1}{2\pi} e^{-\frac{s}{\xi}}$ & $\frac{1}{\sqrt{2} \pi^{3/2}} \,\frac{s}{\xi} \, \text{exp}\left({-\frac{1}{2}\left(\frac{s}{\xi}\right)^2}\right)$ \\ 
 $H_1(k;\xi) =$ & $-\frac{1}{\pi}\frac{\xi}{k}\left[J_0(k\xi) -\frac{1}{k\xi}\int_{0}^{k\xi}{J_0(p)dp}\right]$ & $\frac{1}{2\pi} \frac{k \xi^3}{(1+k^2\xi^2)^{3/2}}$ & $\frac{1}{\sqrt{2} \pi^{3/2}}\, k\xi^3 \text{exp}\left({-\frac{1}{2} k^2 \xi^2}\right)$\\ 
 \hline
\end{tabular}
\caption{Table showing the \textbf{O1}-\textbf{O3} interaction kernels, $\tilde{\Omega}\left(\frac{s}{\xi}\right)$, and their respective first order (standard) Hankel transforms, $H_1(k;\xi)$.}
\label{table:interaction_kernels}
\end{table}

\begin{figure}
    \begin{center}
  \subfloat[]{\label{fig:heatmap_tophat_attractive}%
    \includegraphics[height=0.35\textwidth]{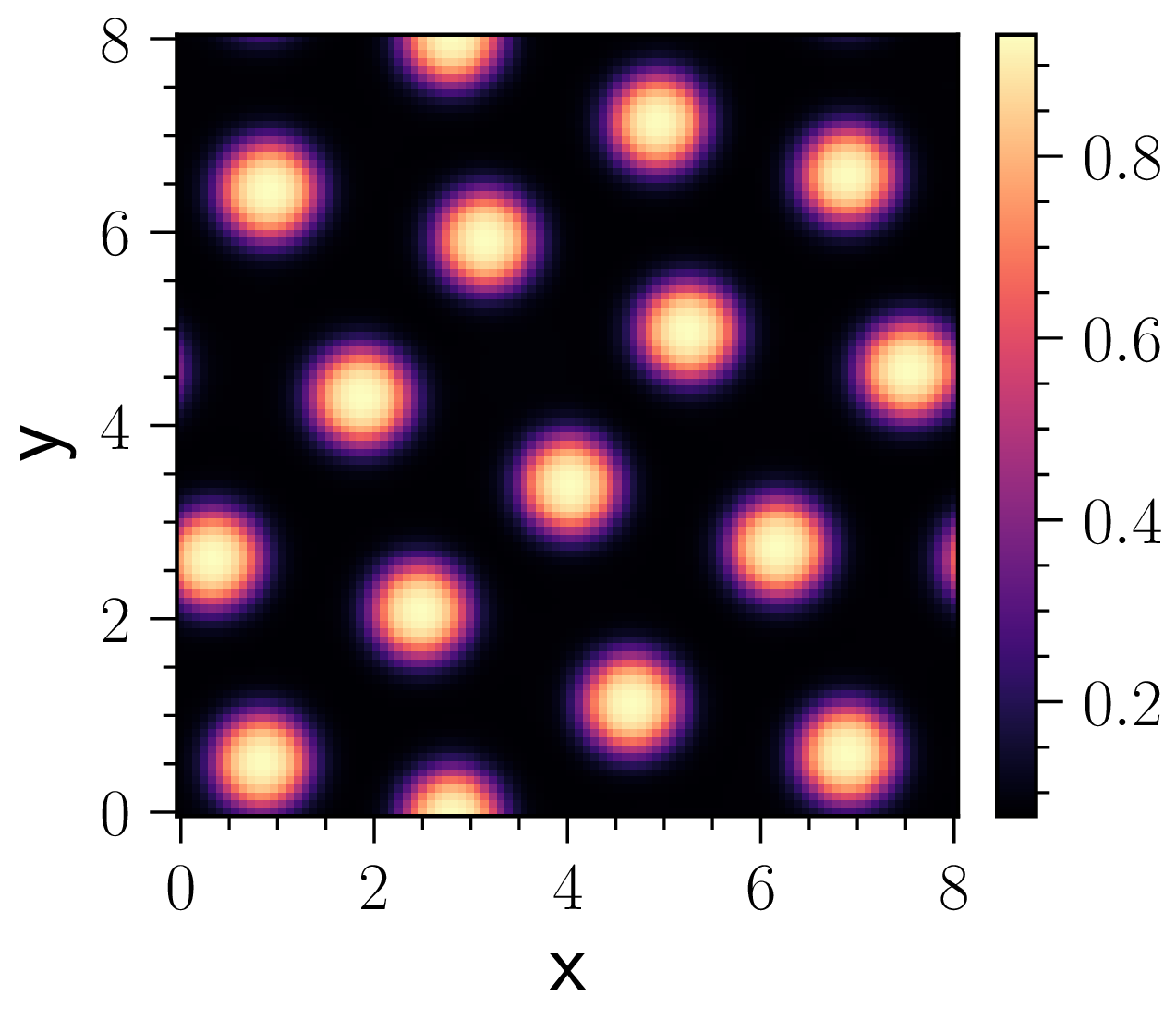}}
    \quad
    \subfloat[]{\label{fig:dispersion_tophat_attractive}%
    \includegraphics[height=0.35\textwidth]{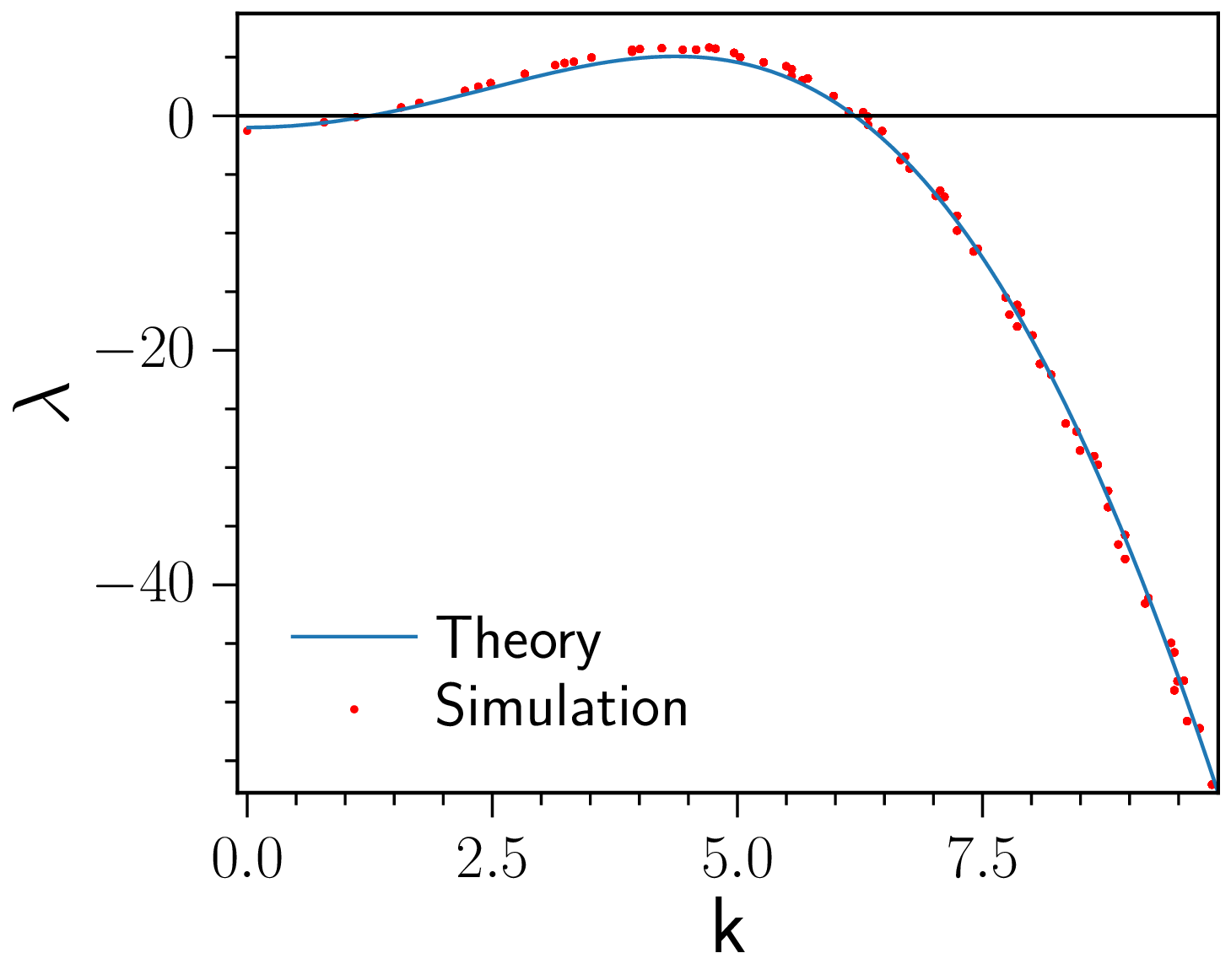}}
    \\
    \subfloat[]{\label{fig:heatmap_exp}%
    \includegraphics[height=0.35\textwidth]{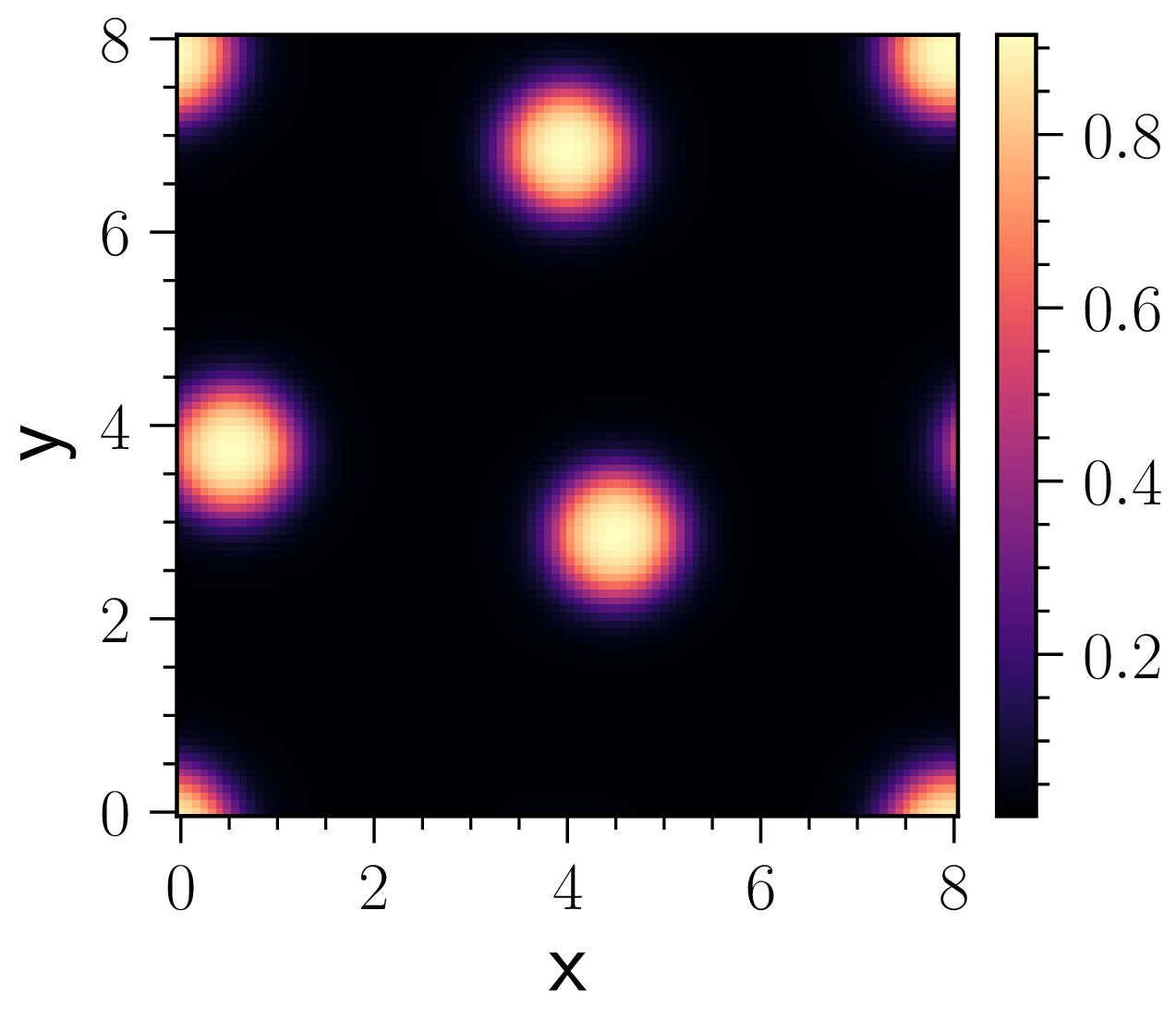}}
    \quad
    \subfloat[]{\label{fig:dispersion_exp}%
    \includegraphics[height=0.35\textwidth]{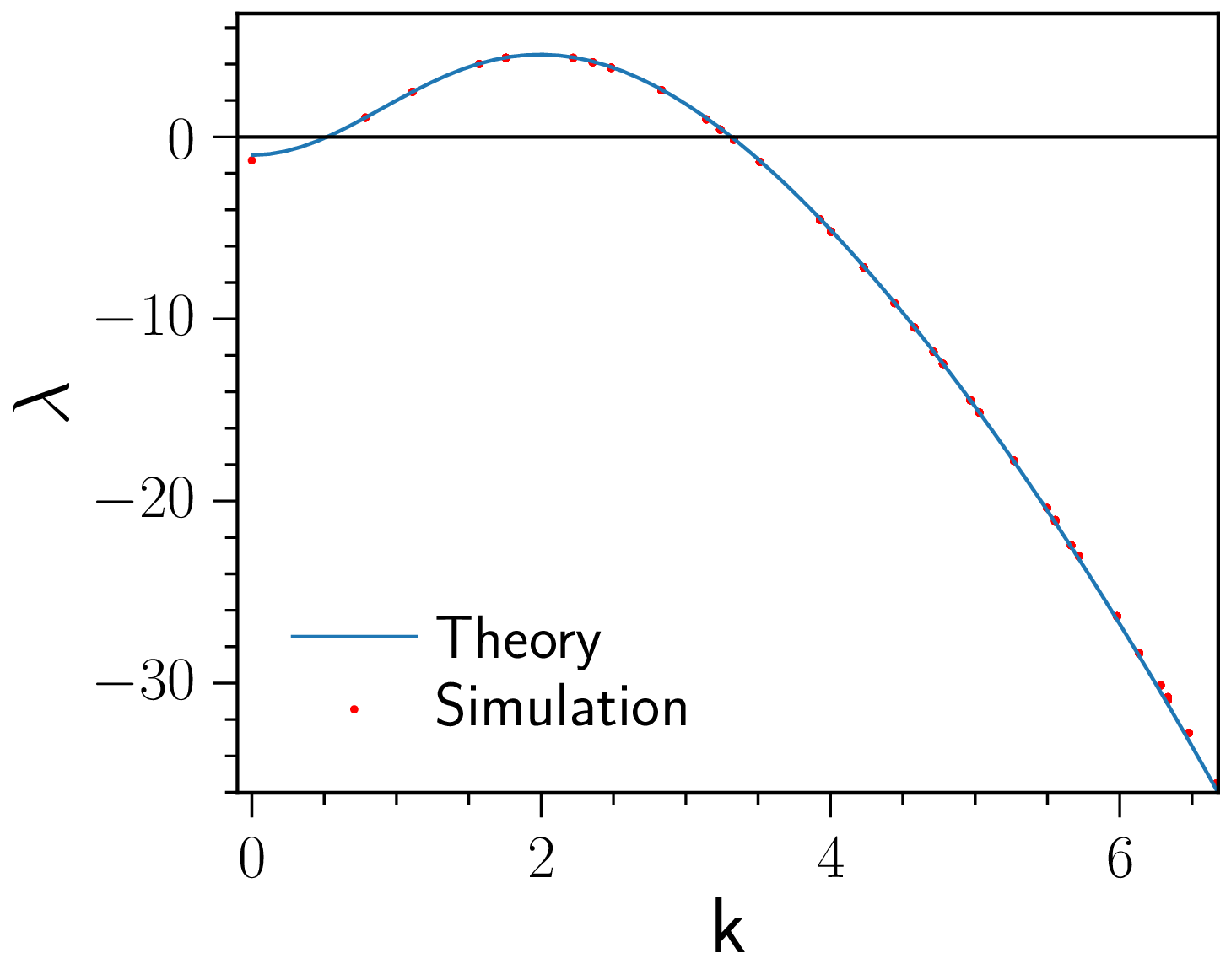}}
    \\
    \subfloat[]{\label{fig:heatmap_o3}%
    \includegraphics[height=0.35\textwidth]{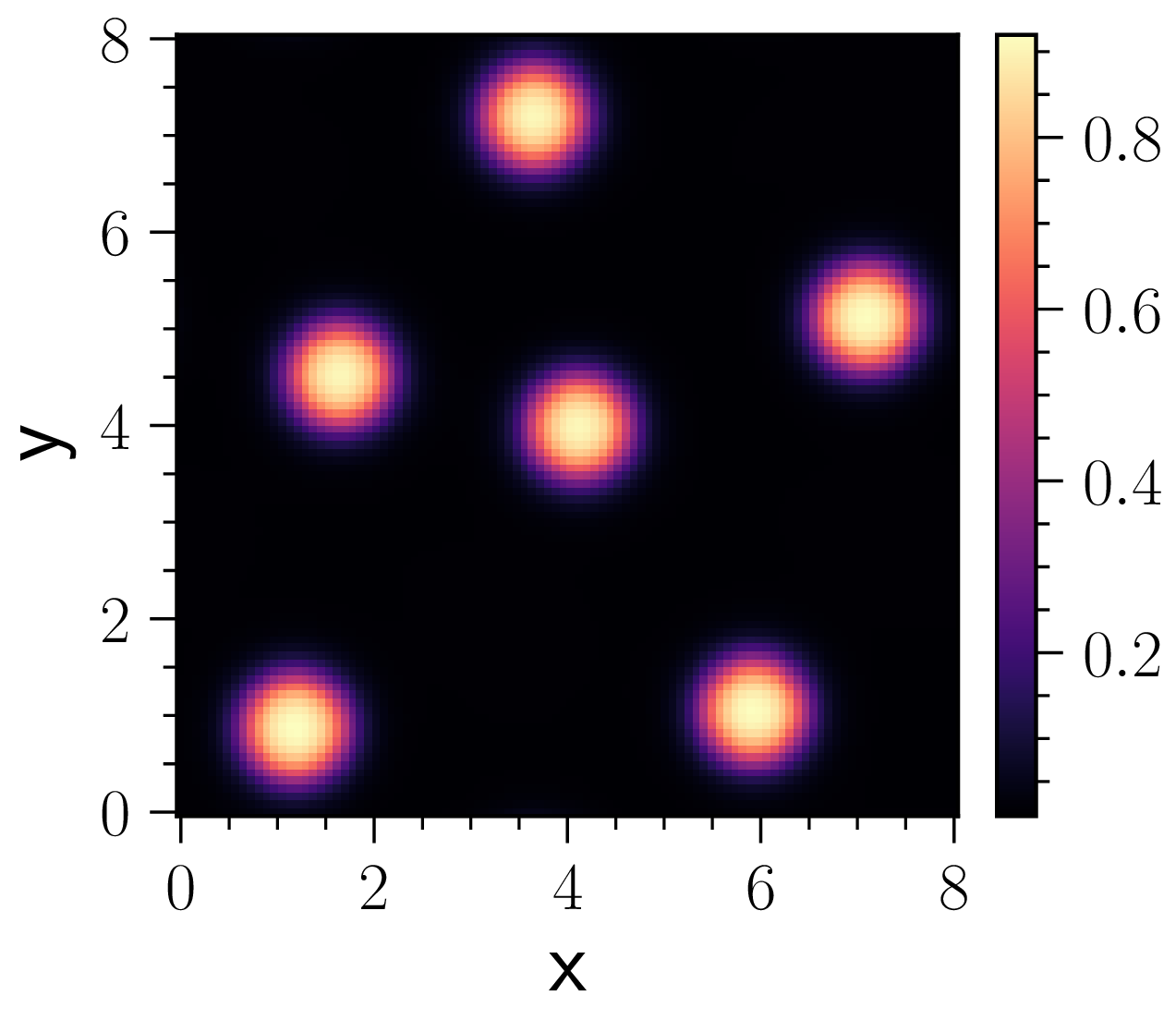}}
    \quad
    \subfloat[]{\label{fig:dispersion_o3}%
    \includegraphics[height=0.35\textwidth]{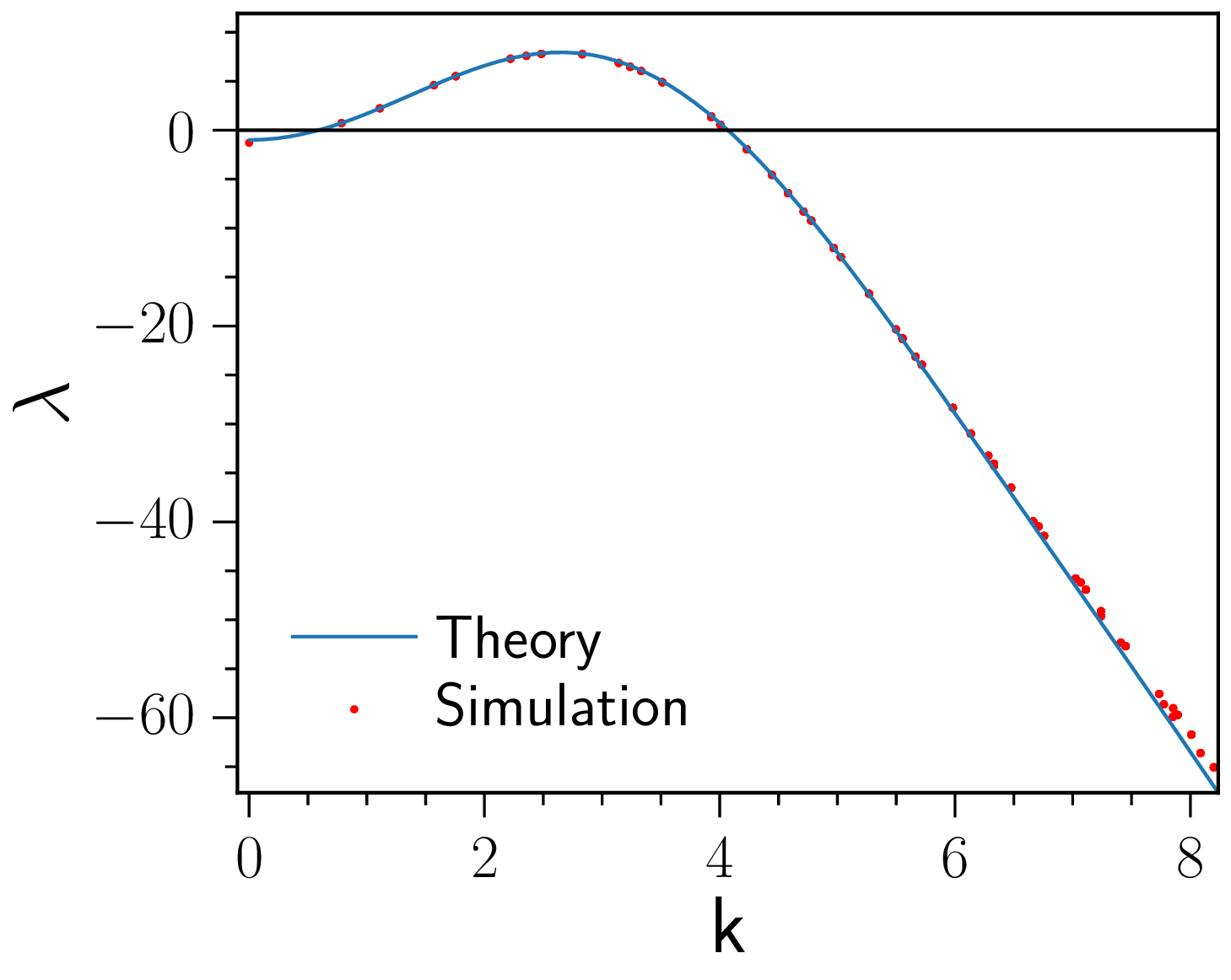}}
  \end{center}

    \caption{Simulation results for the 2D model, Eq. \eqref{eq:2D_PDE}, with \textbf{attractive interactions} and relatively low proliferation, for each example interaction kernel. \subref{fig:heatmap_tophat_attractive}, \subref{fig:heatmap_exp}, \subref{fig:heatmap_o3} heatmaps of the final cell density $u$ at steady state for a single simulation. \subref{fig:dispersion_tophat_attractive}, \subref{fig:dispersion_exp}, \subref{fig:dispersion_o3} the corresponding dispersion relations, with the analytical prediction (blue line) from Eq. \eqref{eq:2D_dispersion_relation} compared with estimates from the simulation (red dots) calculated using Eq. \eqref{eq:log_normalised_growth_mode} for each spatial mode $k$ supported on the finite domain. \subref{fig:heatmap_tophat_attractive}, \subref{fig:dispersion_tophat_attractive} uses the \textbf{O1} kernel; \subref{fig:heatmap_exp}, \subref{fig:dispersion_exp} uses the \textbf{O2} kernel; \subref{fig:heatmap_o3}, \subref{fig:dispersion_o3} uses the \textbf{O3} kernel. All simulations ran to $t=40$ and used parameters: $U=0.5$, $\mu=50$, $\rho=1$, $\xi=0.4$, $L=8$.}
    \label{fig:simulation_attractive} 
\end{figure}

\begin{figure}
    \begin{center}
  \subfloat[]{\label{fig:heatmap_repulsive_low_prolif}%
    \includegraphics[height=0.35\textwidth]{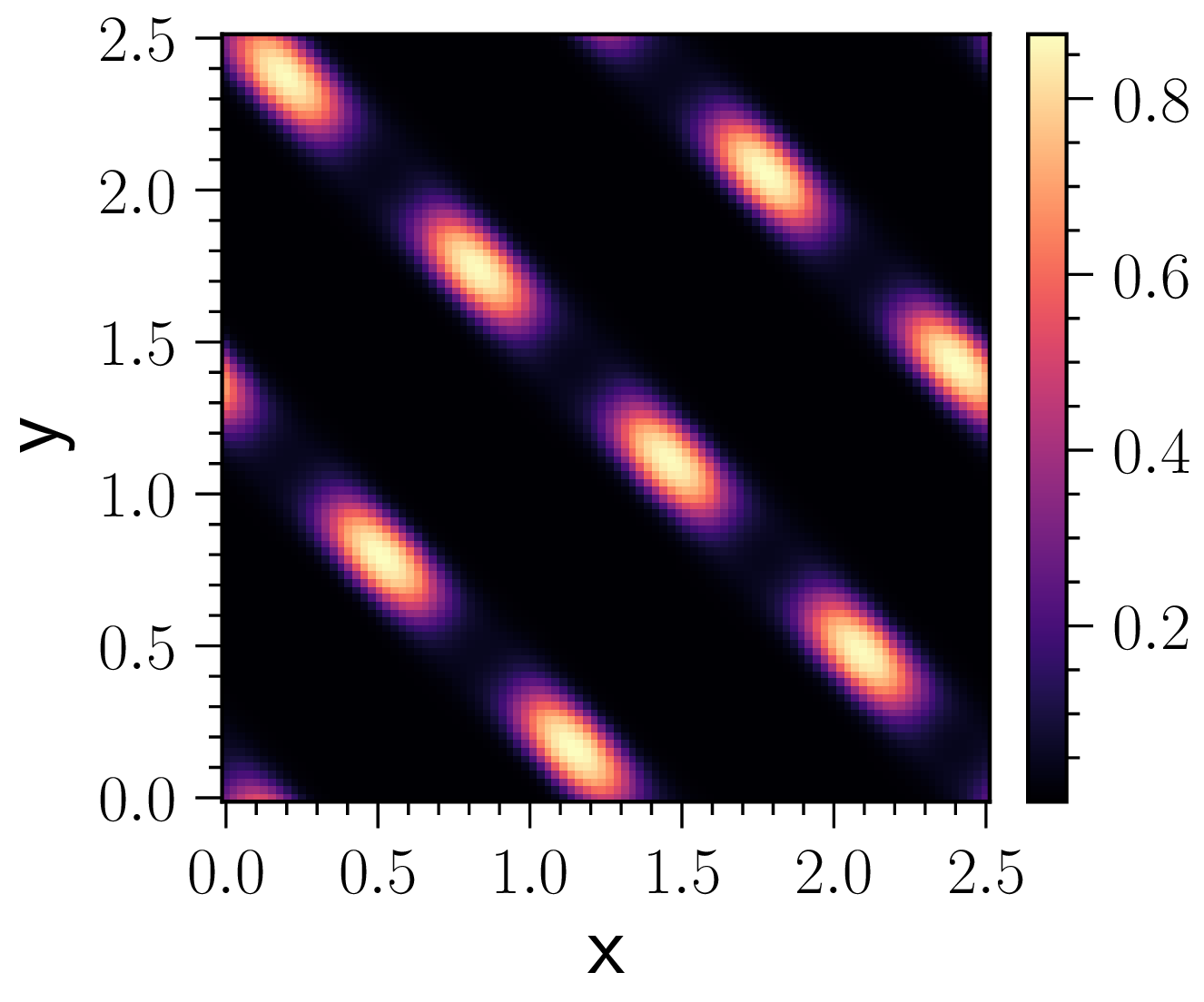}}
    \quad
    \subfloat[]{\label{fig:dispersion_repulsive_low_prolif}%
    \includegraphics[height=0.35\textwidth]{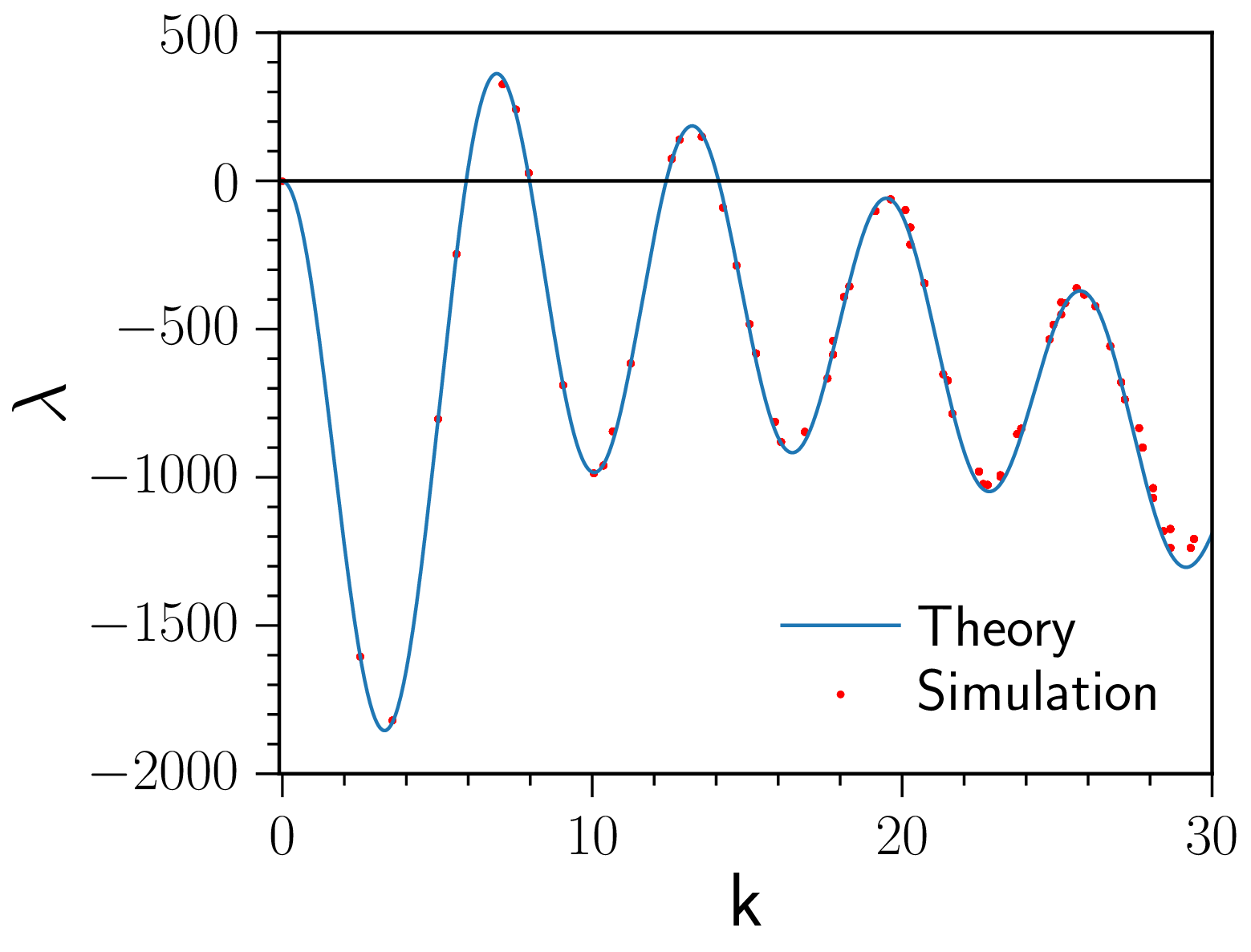}}
    \\
    \subfloat[]{\label{fig:heatmap_repulsive_med_prolif}%
    \includegraphics[height=0.35\textwidth]{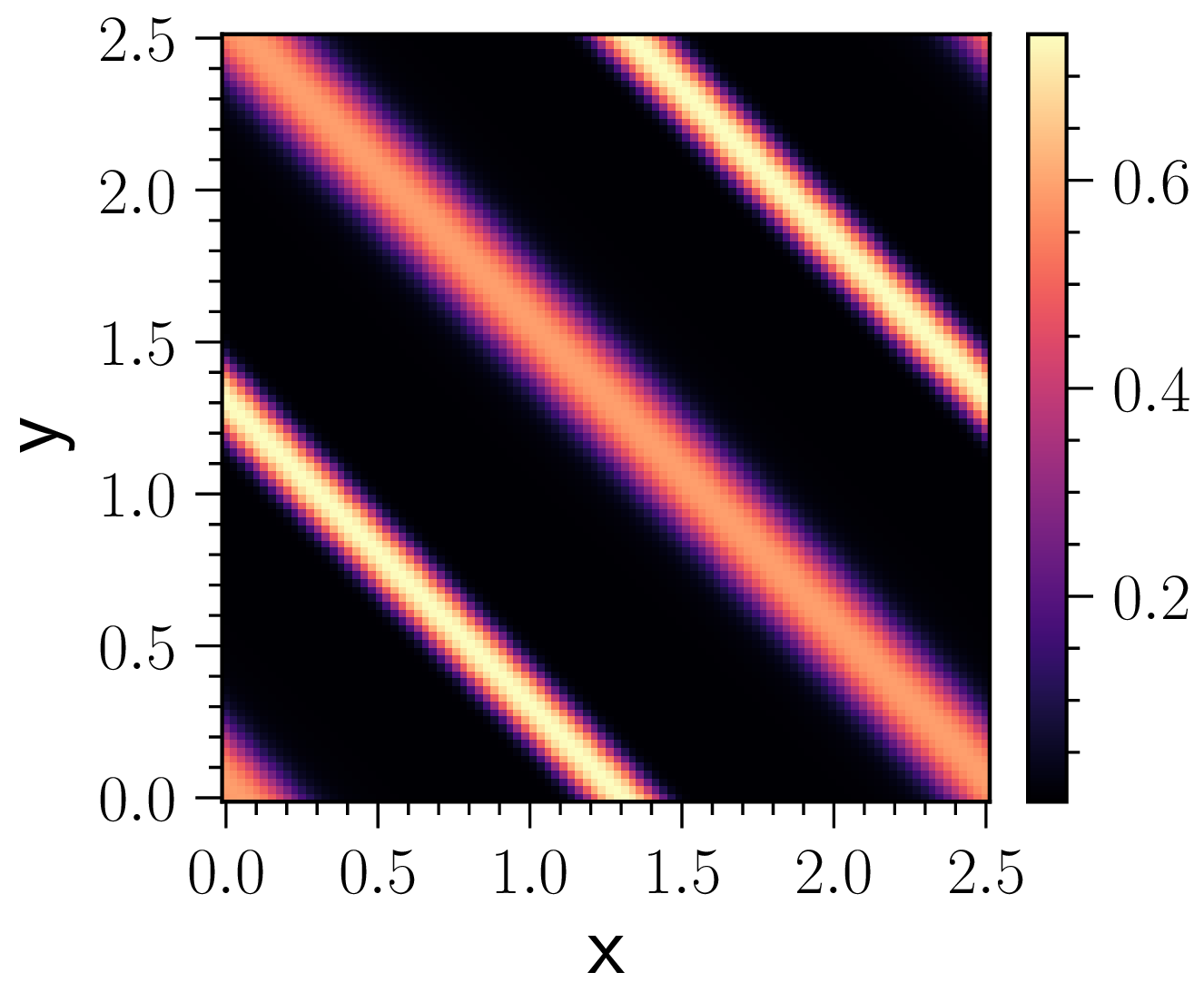}}
    \quad
    \subfloat[]{\label{fig:dispersion_repulsive_med_prolif}%
    \includegraphics[height=0.35\textwidth]{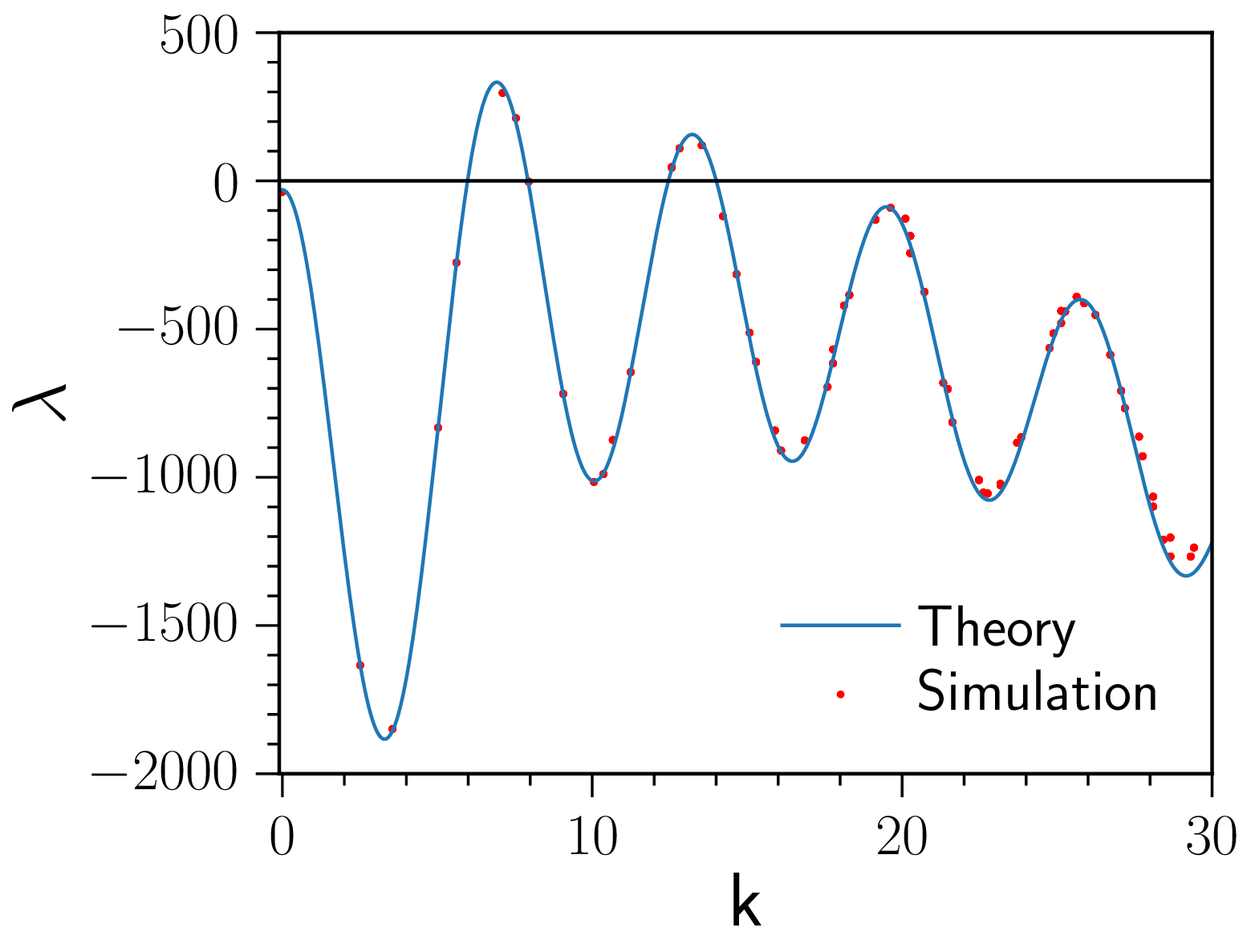}}
    \\
    \subfloat[]{\label{fig:heatmap_repulsive_high_prolif}%
    \includegraphics[height=0.35\textwidth]{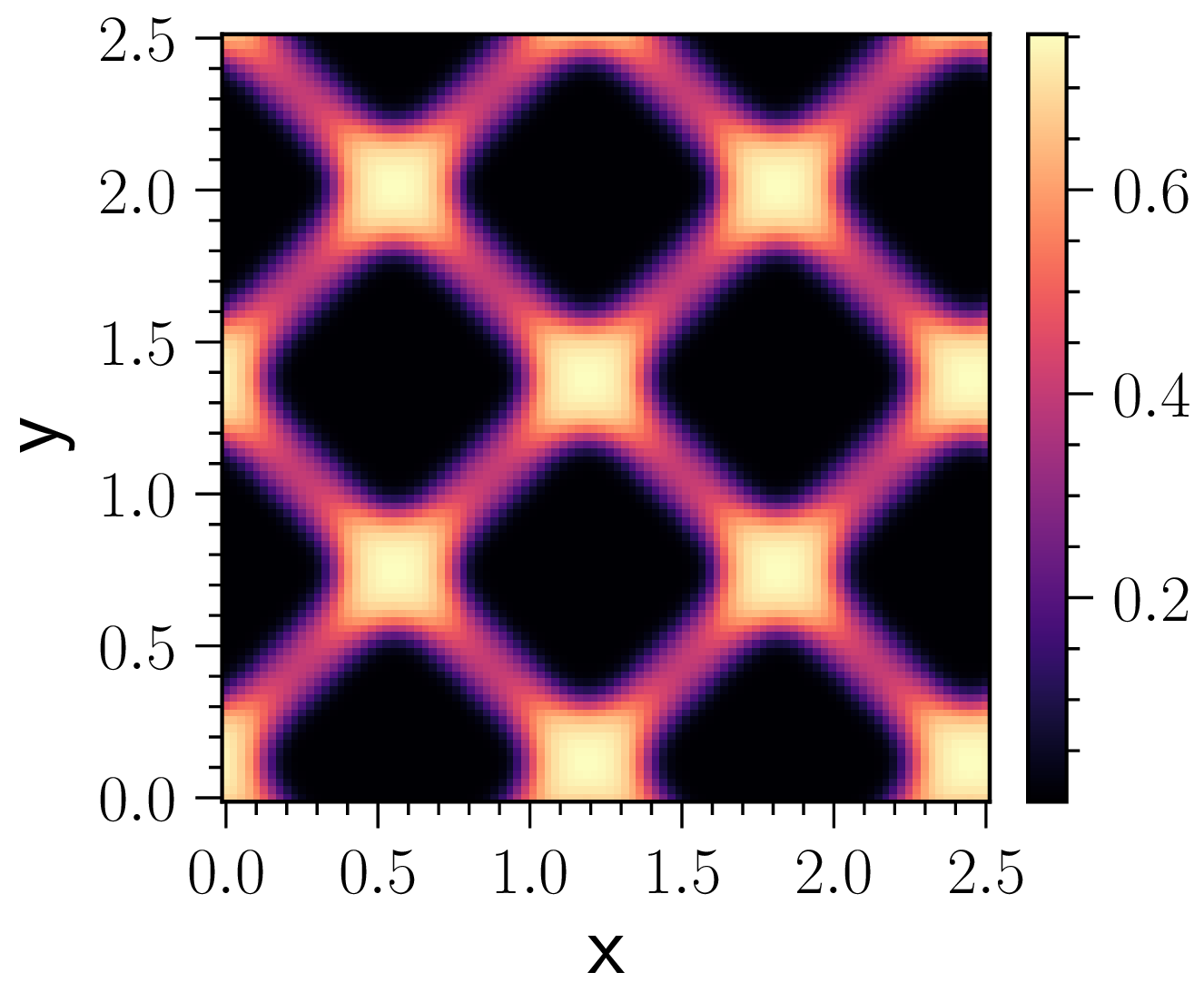}}
    \quad
    \subfloat[]{\label{fig:dispersion_repulsive_high_prolif}%
    \includegraphics[height=0.35\textwidth]{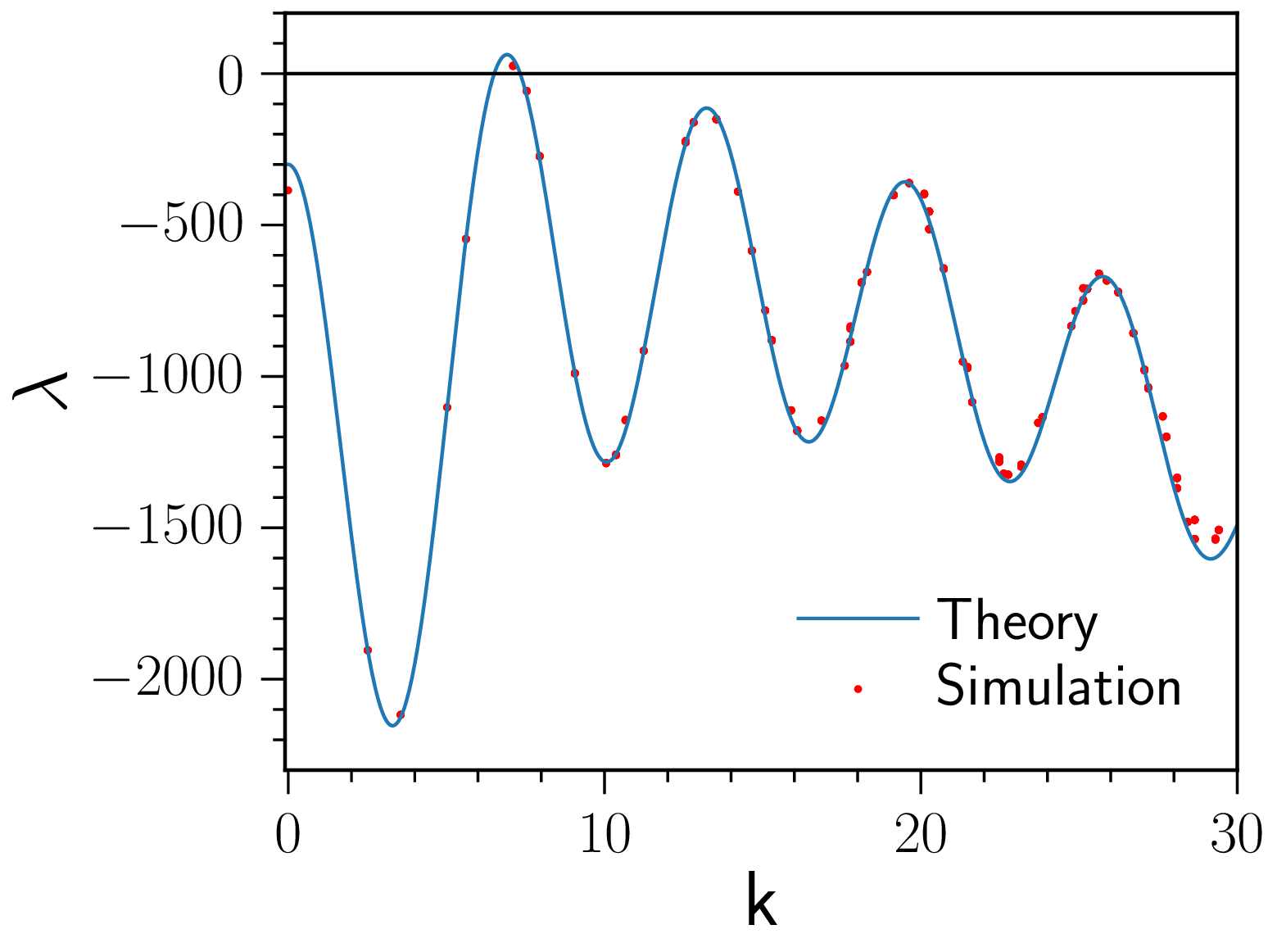}}
  \end{center}

    \caption{Simulation results for the 2D model, Eq. \eqref{eq:2D_PDE}, with \textbf{repulsive interactions} with the \textbf{O1} kernel, for a range of proliferation rates. \subref{fig:heatmap_repulsive_low_prolif}, \subref{fig:heatmap_repulsive_med_prolif}, \subref{fig:heatmap_repulsive_high_prolif} heatmaps of the final cell density $u$ at steady state for a single simulation. \subref{fig:dispersion_repulsive_low_prolif}, \subref{fig:dispersion_repulsive_med_prolif}, \subref{fig:dispersion_repulsive_high_prolif} the corresponding dispersion relations, with the analytical prediction (blue line) from Eq. \eqref{eq:2D_dispersion_relation} compared with estimates from the simulation (red dots) calculated using Eq. \eqref{eq:log_normalised_growth_mode} for each spatial mode $k$ supported on the finite domain. The proliferation-death rates are \subref{fig:heatmap_repulsive_low_prolif}, \subref{fig:dispersion_repulsive_low_prolif} $\rho=1$;  \subref{fig:heatmap_repulsive_med_prolif}, \subref{fig:dispersion_repulsive_med_prolif} $\rho=30$; \subref{fig:heatmap_repulsive_high_prolif}, \subref{fig:dispersion_repulsive_high_prolif} $\rho=300$. All simulations ran to $t=100$ and used parameters: $U=0.5$, $\mu=-5000$, $\xi=1$, $L=2.5$.
    }
    \label{fig:simulation_repulsive} 
\end{figure}

\begin{figure}
    \begin{center}
    \subfloat[$t=0$]{    \label{fig:breakup_1}    \includegraphics[height=0.25\textwidth]{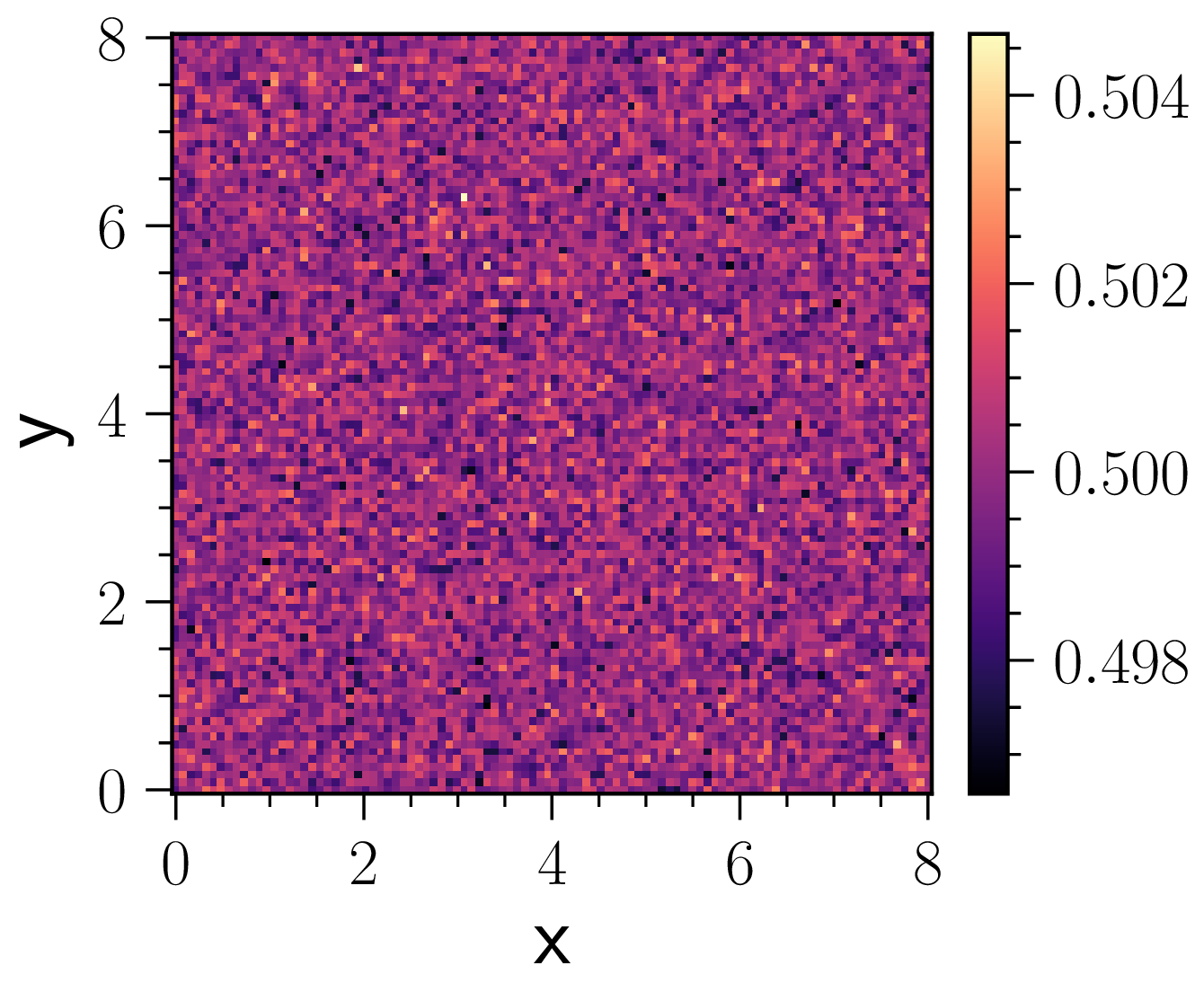}}
    \subfloat[$t=1.52$]{    \label{fig:breakup_2}    \includegraphics[height=0.25\textwidth]{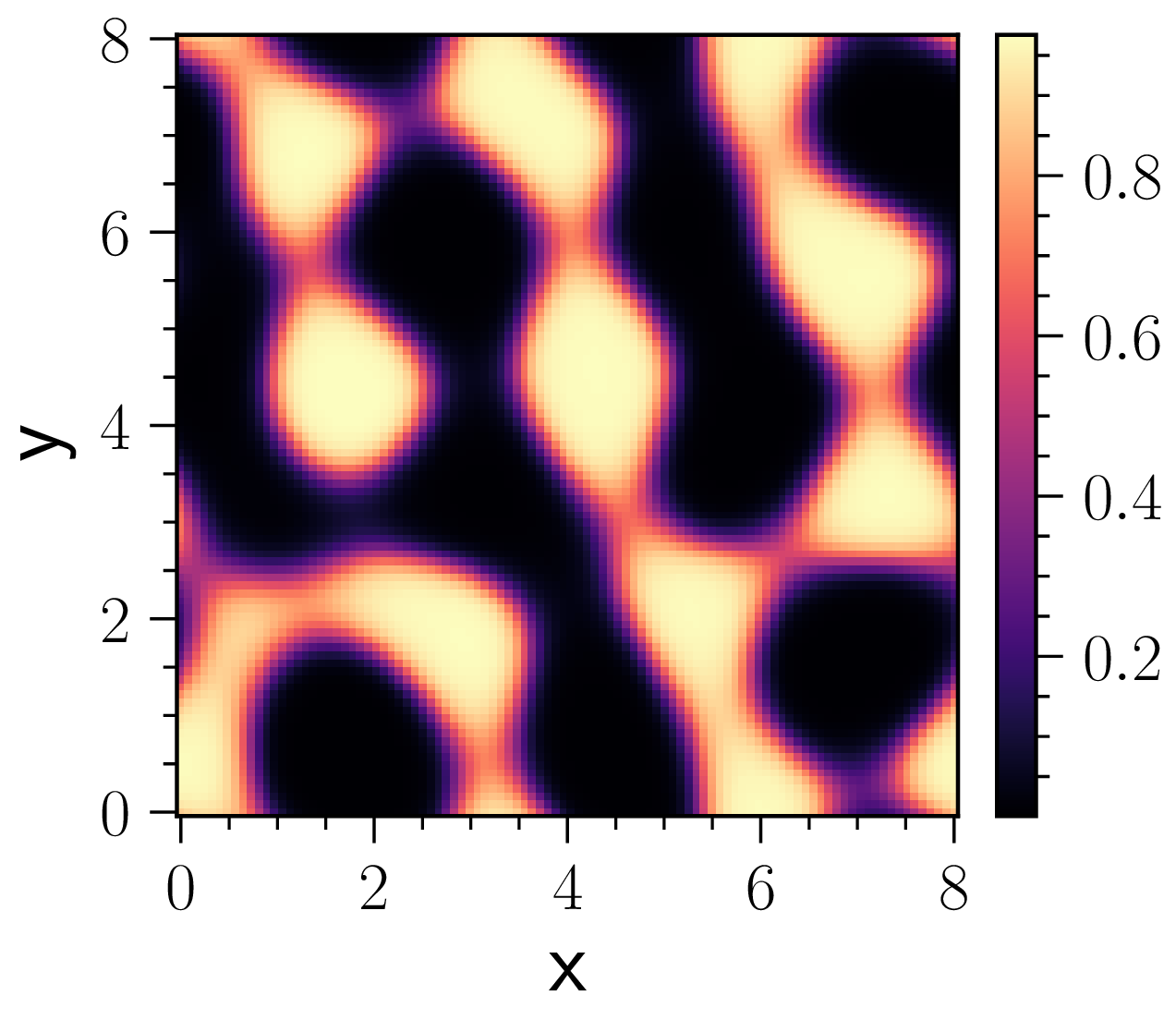}} 
    \subfloat[$t=3.60$]{    \label{fig:breakup_3}    \includegraphics[height=0.25\textwidth]{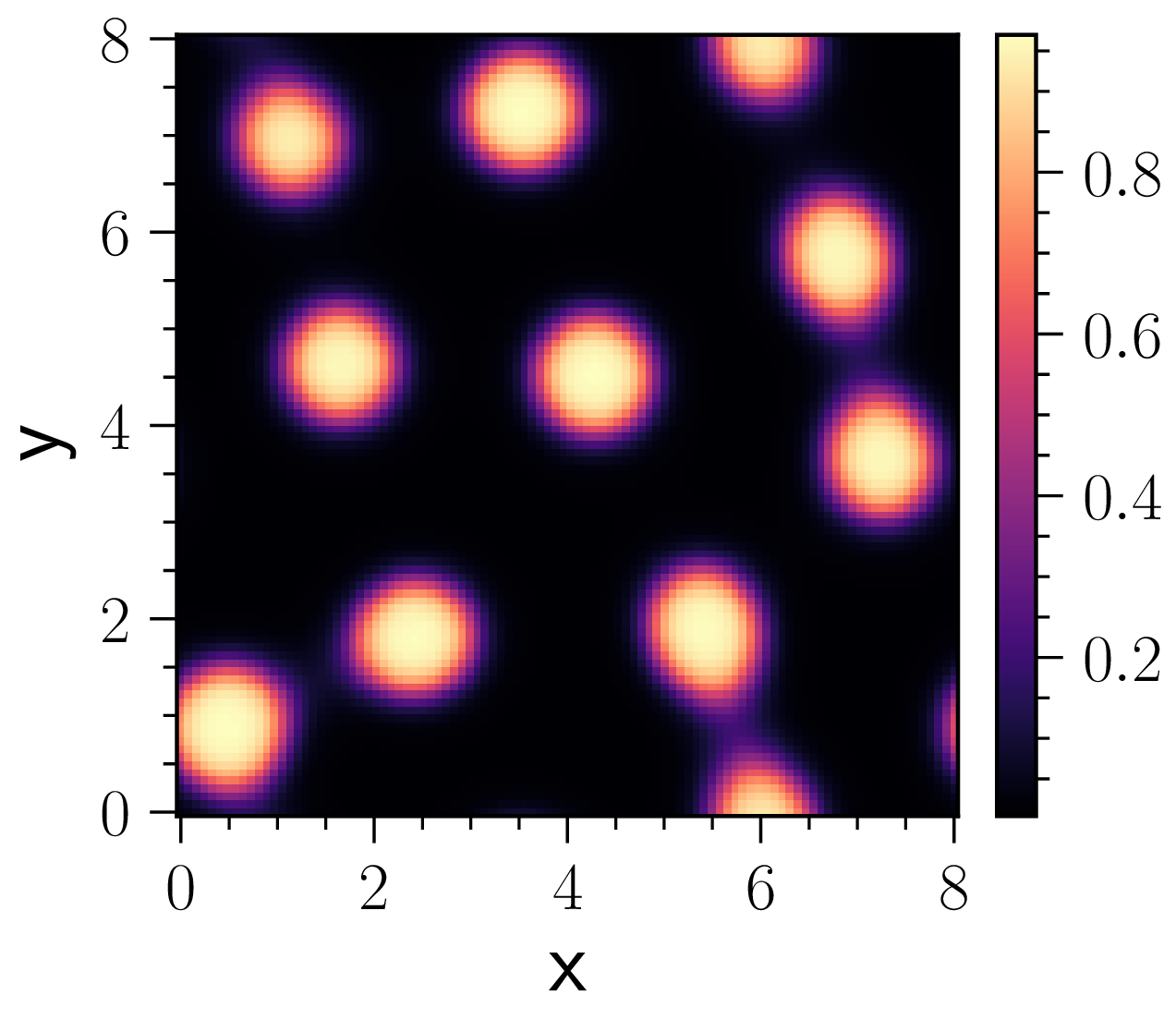}} 
    \\
    \subfloat[$t=4.96$]{    \label{fig:breakup_4}    \includegraphics[height=0.25\textwidth]{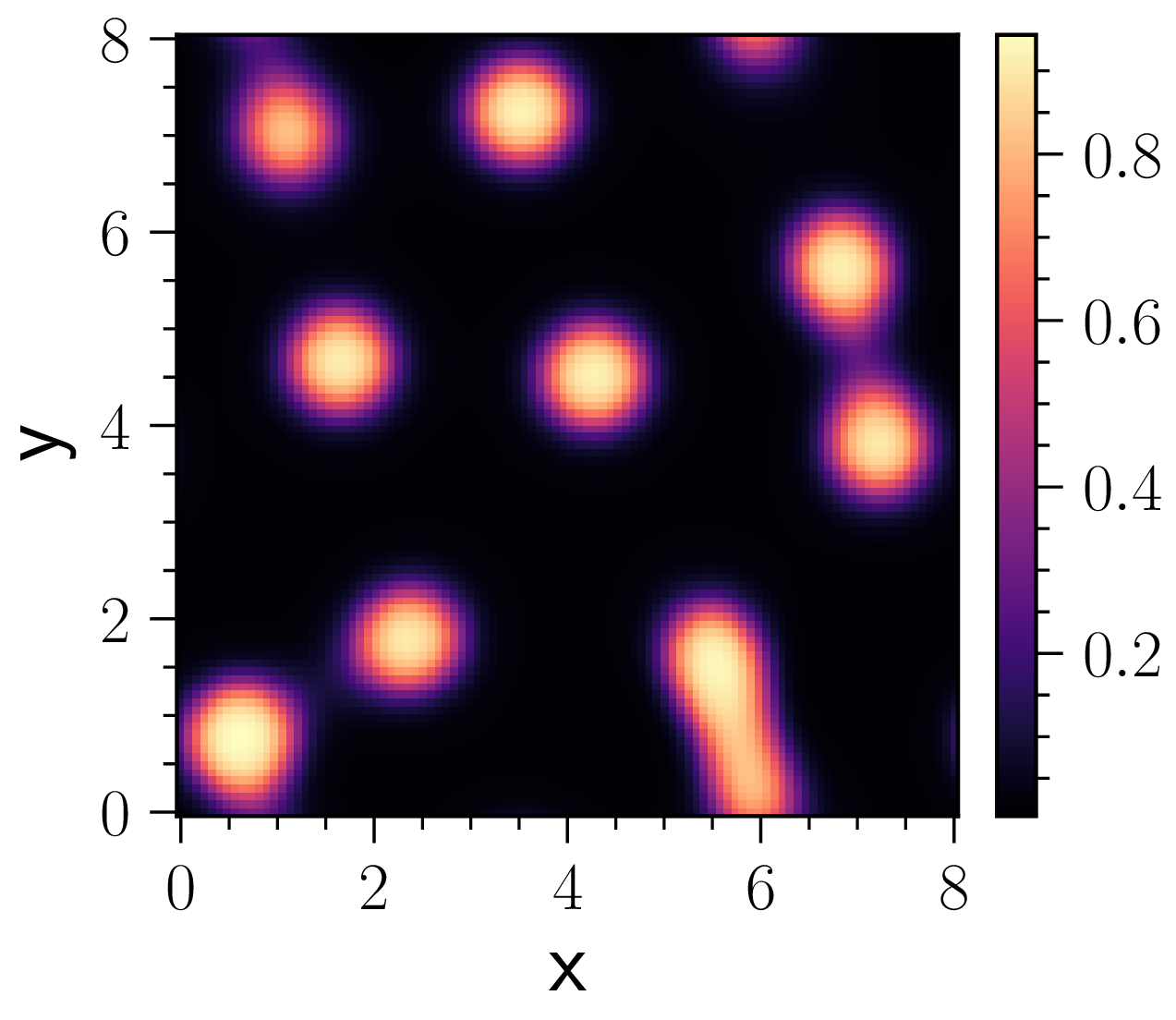}}
    \subfloat[$t=40.00$]{    \label{fig:breakup_5}    \includegraphics[height=0.25\textwidth]{attractive_o3_heatmap_t=40.00.eps}}
    
  \end{center}
    \caption{Cell density heatmaps at subsequent times for the same simulation as shown in Fig. \ref{fig:simulation_attractive}\subref{fig:heatmap_exp},\subref{fig:dispersion_exp}. Perturbations about the homogeneous state initially grow into a labyrinthine pattern, which then collapses into spots. Spots which are close together then coalesce, leaving a final steady state with fewer spots. Parameters: attractive interactions, \textbf{O3} kernel, $U=0.5$, $\mu=50$, $\rho=1$, $\xi=0.4$, $L=8$.}
    \label{fig:breakup_instability} 
\end{figure}

\subsection{Simulation Results}

\subsubsection{Steady State Patterns}
Figs. \ref{fig:simulation_attractive} and \ref{fig:simulation_repulsive} show typical examples of the long time steady state and the corresponding dispersion relation, for systems with a sufficiently high interaction strength $\mu$ to form heterogeneous patterns. We see that the dispersion relation derived from the linear stability analysis is reproduced in the simulations and accurately predicts the conditions for pattern formation. Notably, pattern formation is indeed possible with repulsive interactions, as demonstrated in Fig. \ref{fig:simulation_repulsive}, which uses the \textbf{O1} kernel. As predicted, the \textbf{O2} kernel does not support such patterns as it decays faster than $\mathcal{O}\left(\frac{1}{\sqrt{s}}\right)$ over all regions.

For all three example interaction kernels, we find that attractive interactions lead to spot patterns, as shown in Fig. \ref{fig:simulation_attractive}. Intuitively, this makes sense as any non-uniformity along a stripe would lead to the attractive interaction collapsing the stripe towards the higher density points, thereby forming spots. In fact, for most parameter choices, we indeed observe initial stripe or labyrinthine patterns which then collapse into spots, as demonstrated in Fig. \ref{fig:breakup_instability}. This mechanism resembles the ‘breakup’ or ‘varicose’ instabilities of homoclinic stripes studied in \citet{breakup_homoclinic_stripe}. Next, over a longer timescale, pairs of spots can then coalesce into single spots, until eventually a steady state is reached, often with fewer spots than during the initial collapse.

In contrast, we find that repulsion interactions lead to a steady state of stripes or perforated stripes, as shown in Fig. \ref{fig:simulation_repulsive}. For some parameter choices, perforated stripes can collapse to a form a regular lattice of symmetric spots - this is distinct from the spot patterns formed by attractive interactions which do not necessarily have a regular repeated structure. In other cases with repulsive interactions, such as in Fig. \ref{fig:heatmap_repulsive_low_prolif}, the perforated stripes remain as a steady state. To confirm that this particular simulation has indeed reached a steady state, we simulated to $t=100$, where we measured that $|\frac{\partial u(\boldsymbol{x,t})}{\partial t}|<2\times10^{-10}$ at every spatial point. Furthermore, for relatively high proliferation-death rates, such as in Fig. \ref{fig:simulation_repulsive}\subref{fig:heatmap_repulsive_med_prolif},\subref{fig:heatmap_repulsive_high_prolif} we observe that stripes no longer necessarily have equal width, and perpendicular stripes can stably coexist.

Figs. \ref{fig:simulation_attractive} and \ref{fig:simulation_repulsive} are examples of dynamics that have all asymptoted to their long time steady state. However, for other parameter choices, this model also supports spatio-temporal patterns.

\subsubsection{Spatio-Temporal Patterns}
\begin{figure}
    \begin{center}
    \subfloat[$t=5.64$]{    \label{fig:spatio_temporal_A_1}    \includegraphics[width=0.25\textwidth]{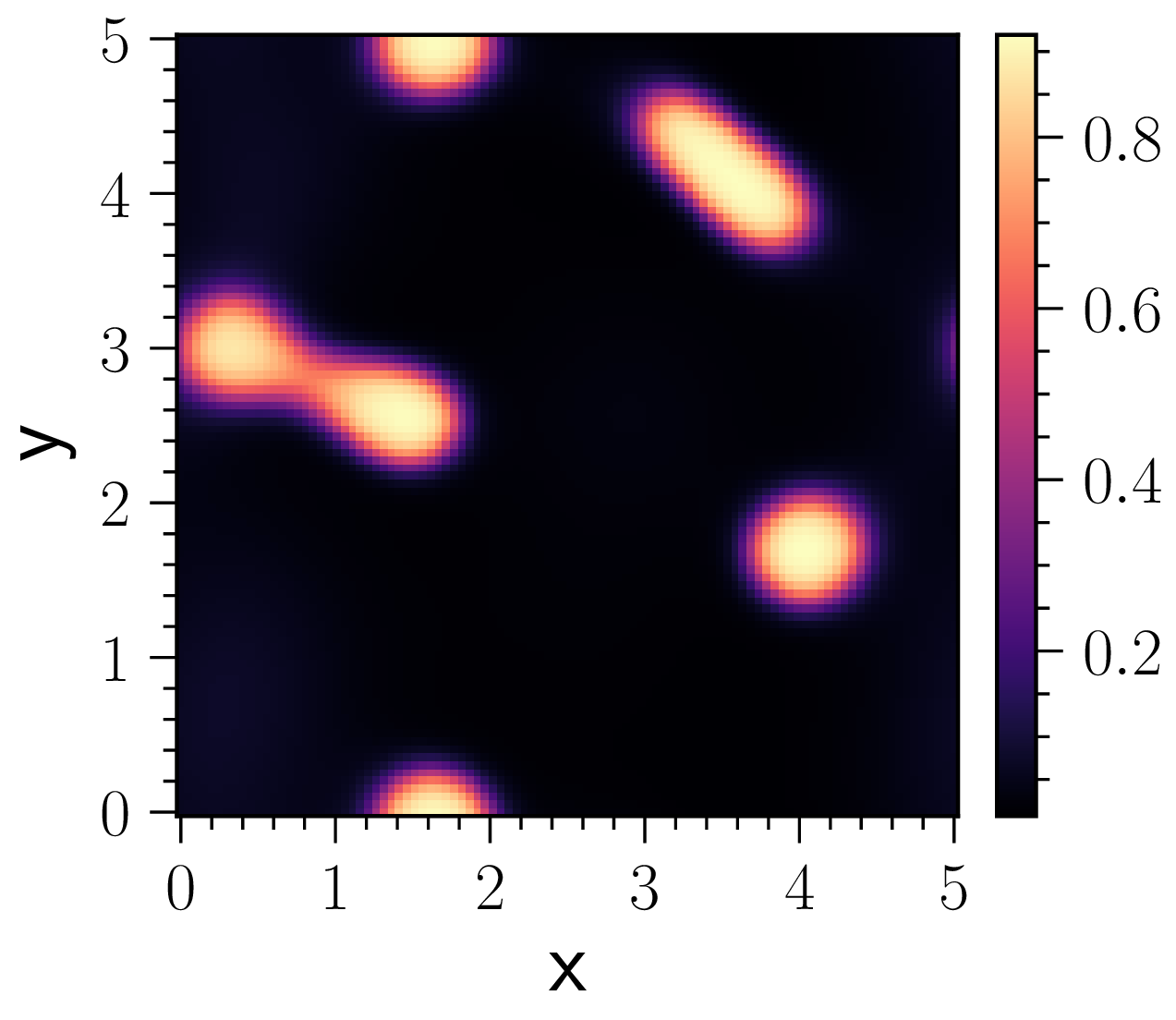}}    
    \subfloat[$t=6.08$]{    \label{fig:spatio_temporal_A_2}    \includegraphics[width=0.25\textwidth]{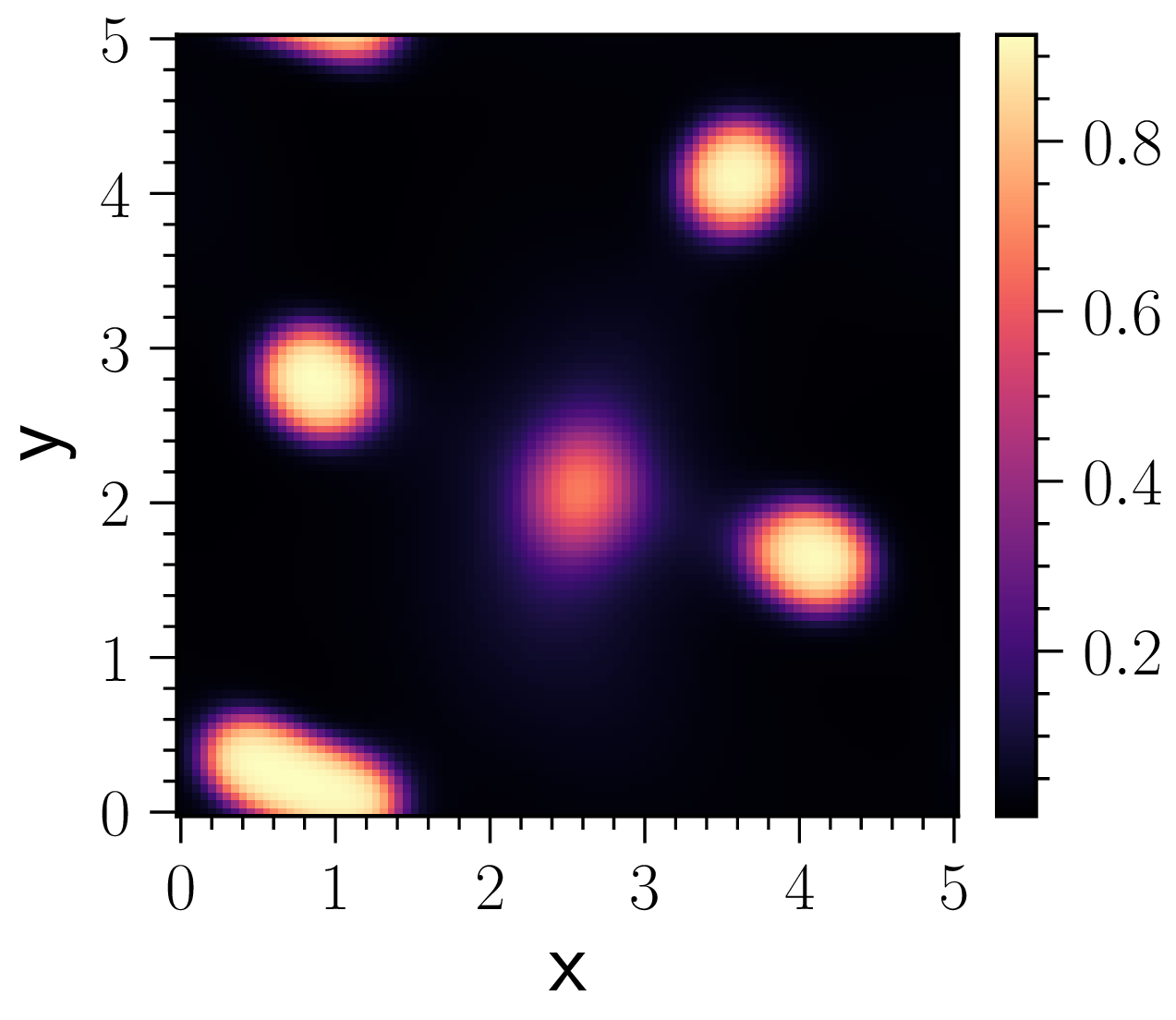}}    
    \subfloat[$t=6.22$]{    \label{fig:spatio_temporal_A_3}    \includegraphics[width=0.25\textwidth]{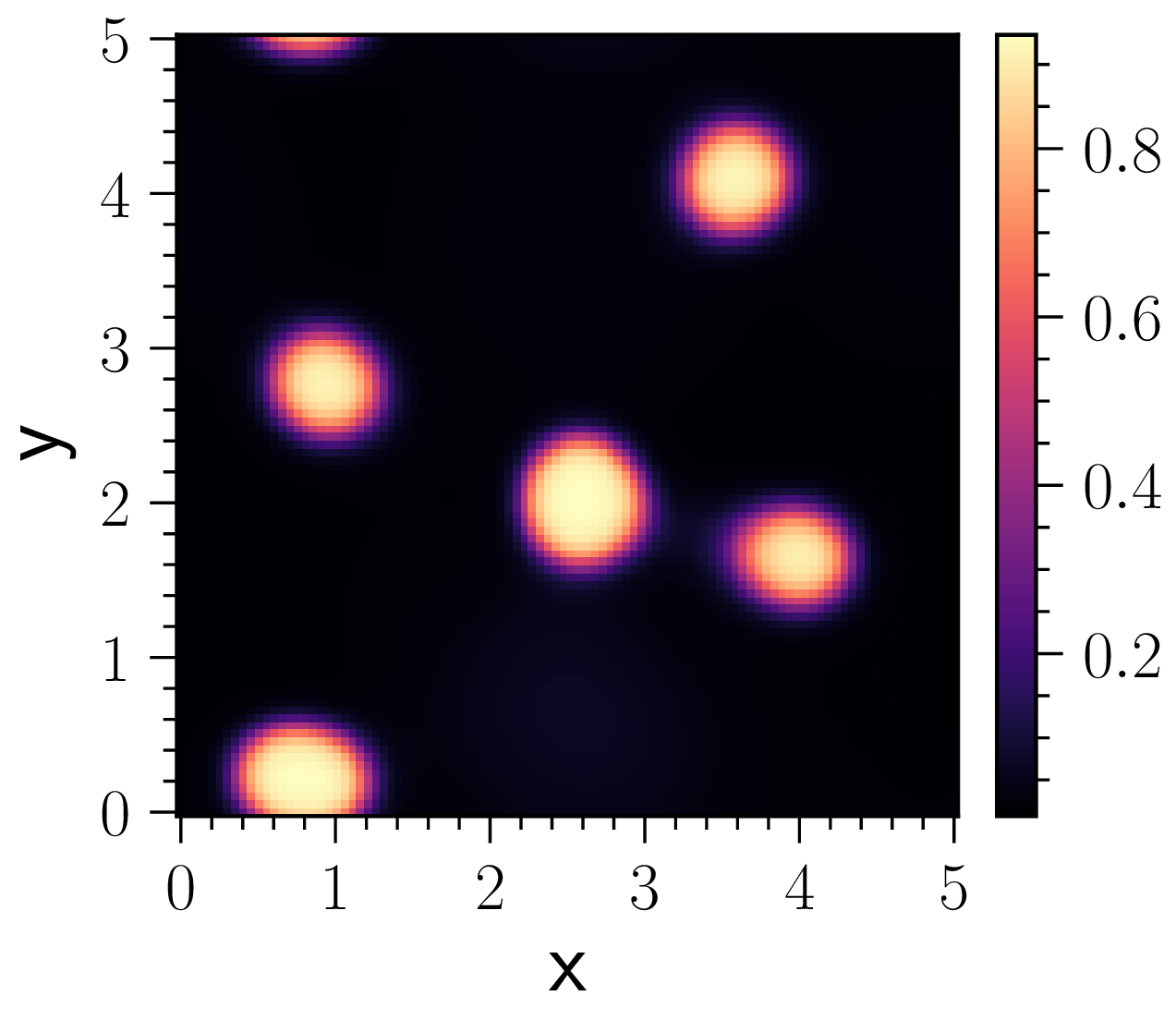}}    
    \subfloat[$t=6.48$]{    \label{fig:spatio_temporal_A_4}    \includegraphics[width=0.25\textwidth]{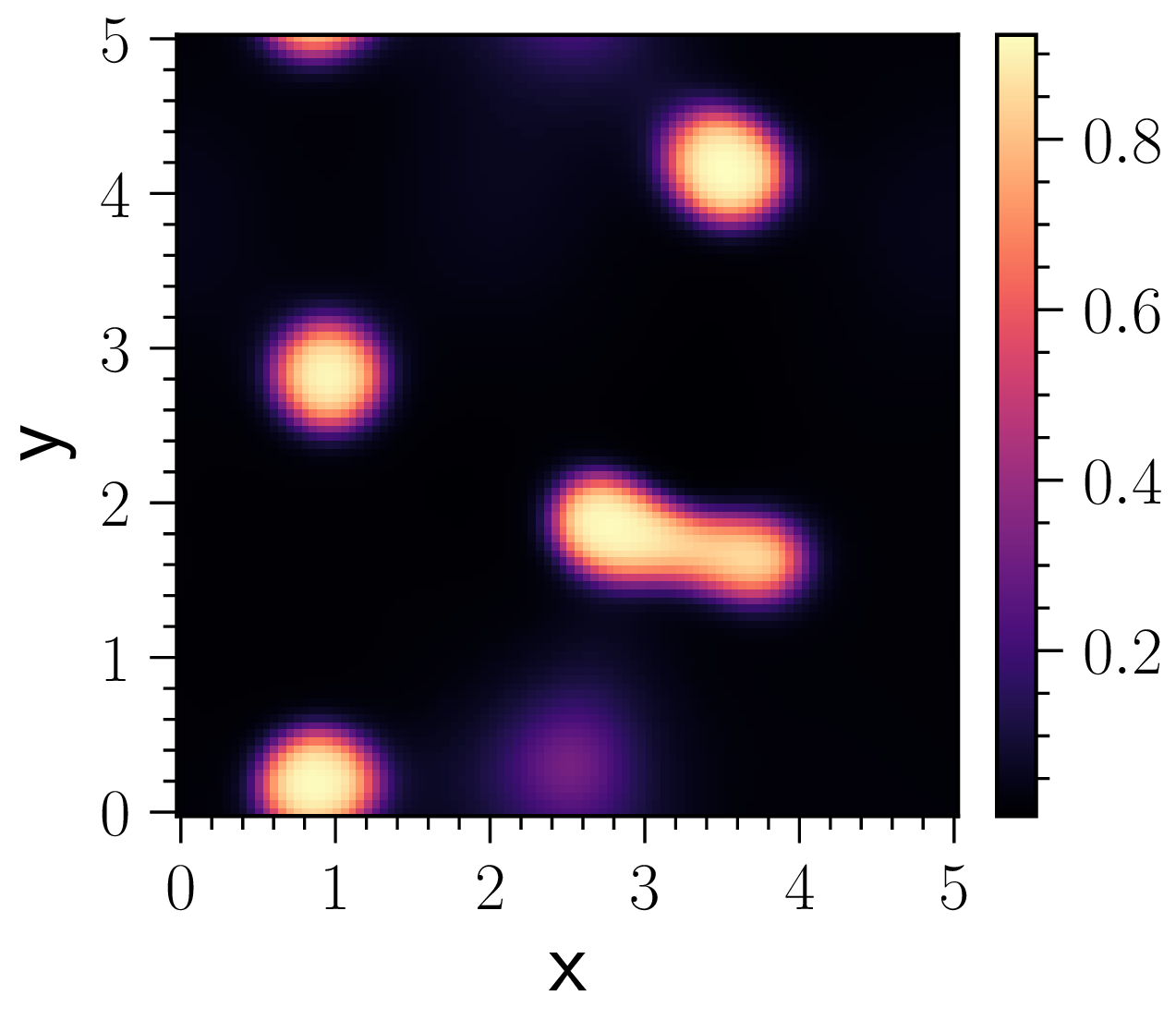}}
    \\
    \subfloat[$t=5.30$]{    \label{fig:spatio_temporal_B_1}    \includegraphics[width=0.25\textwidth]{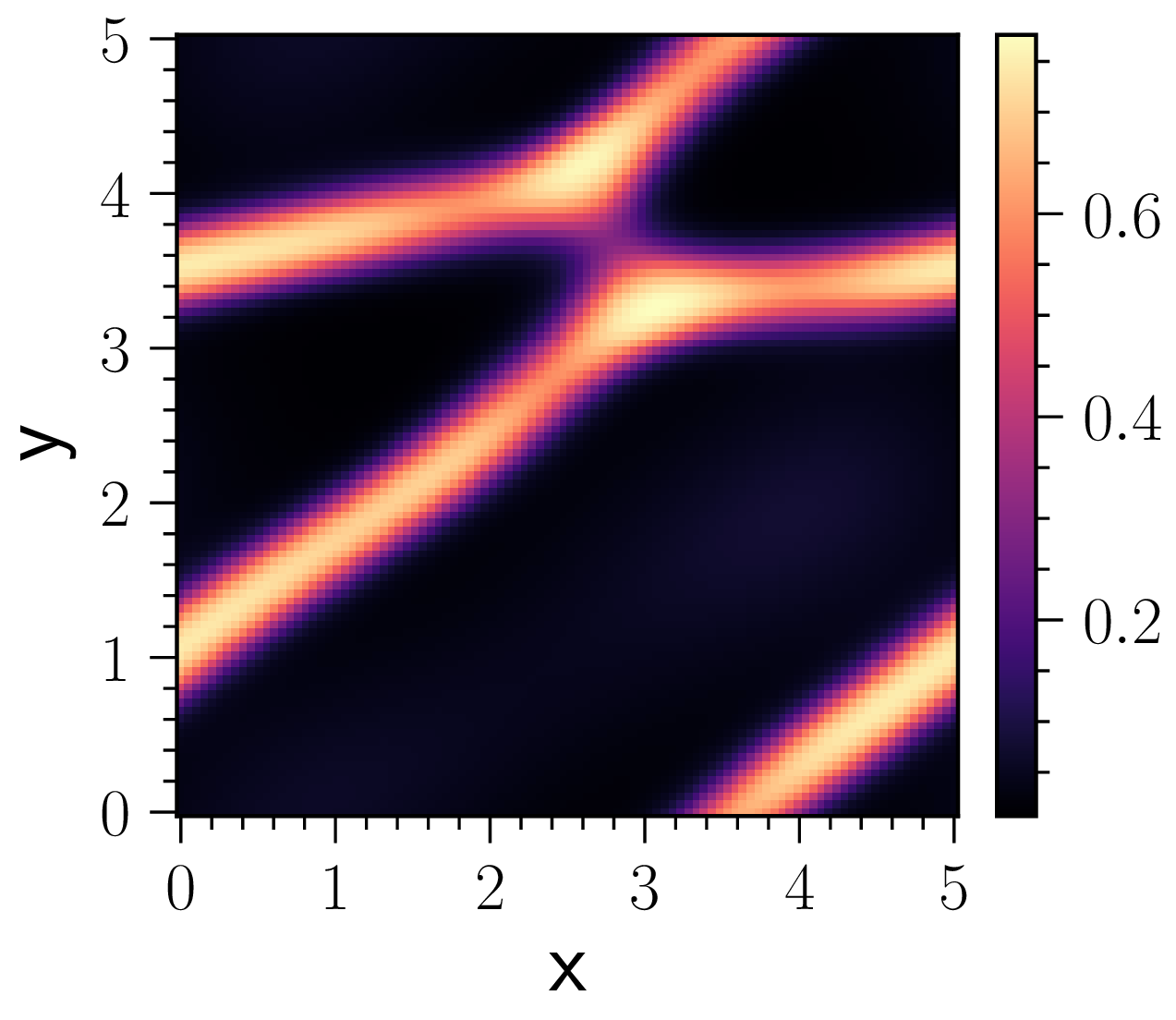}}    
    \subfloat[$t=5.40$]{    \label{fig:spatio_temporal_B_2}    \includegraphics[width=0.25\textwidth]{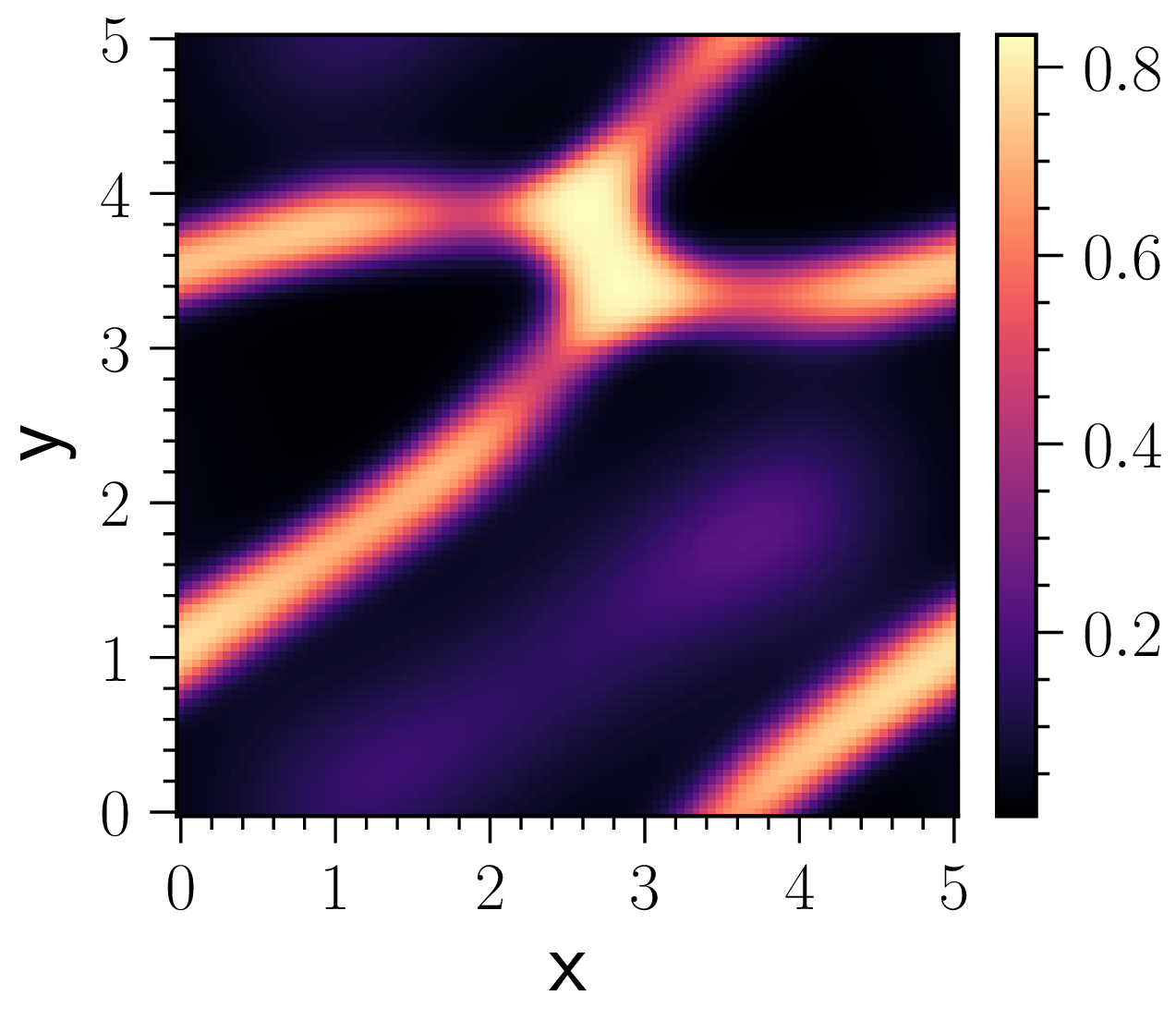}}    
    \subfloat[$t=5.50$]{    \label{fig:spatio_temporal_B_3}    \includegraphics[width=0.25\textwidth]{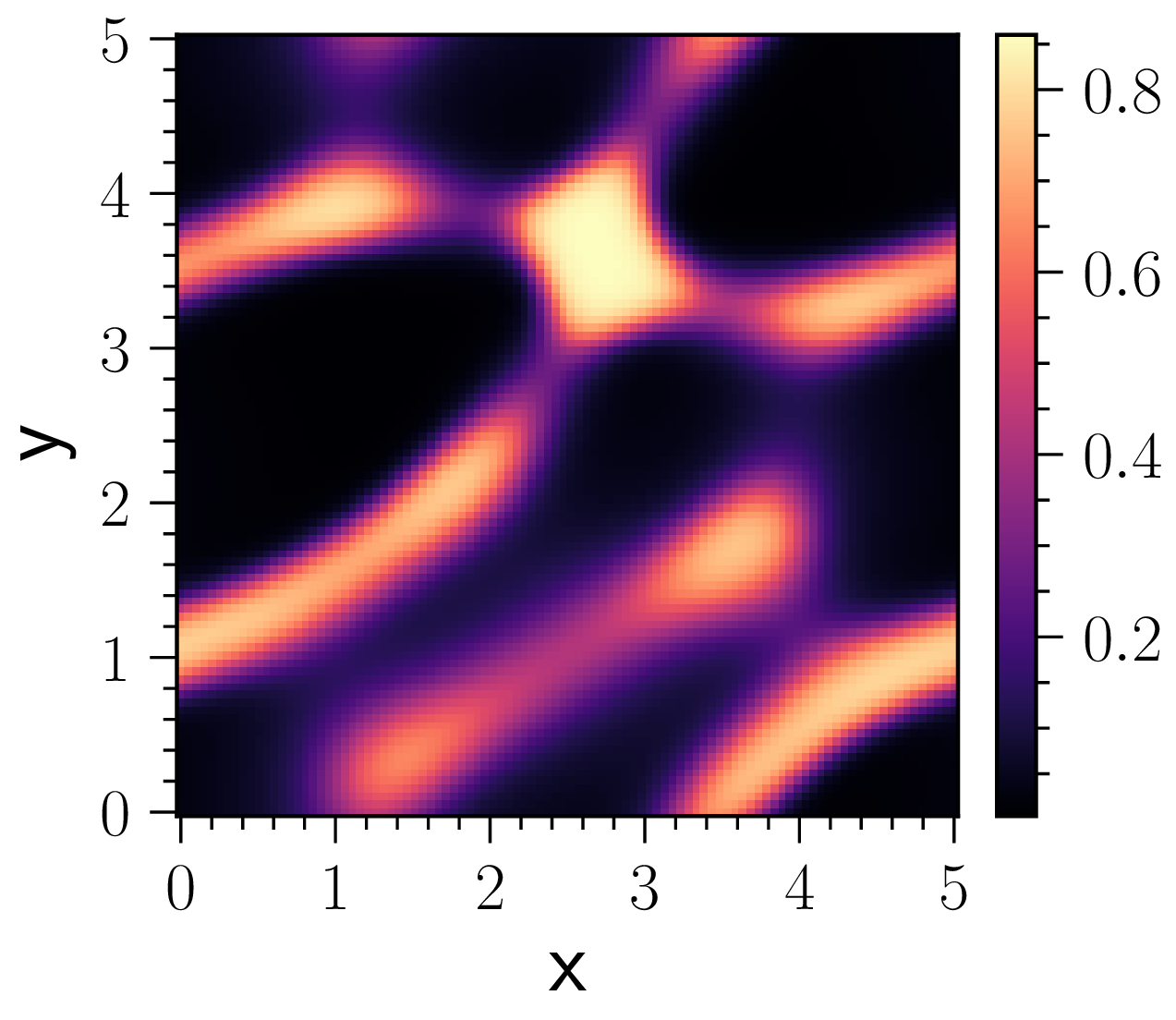}}    
    \subfloat[$t=5.60$]{    \label{fig:spatio_temporal_B_4}    \includegraphics[width=0.25\textwidth]{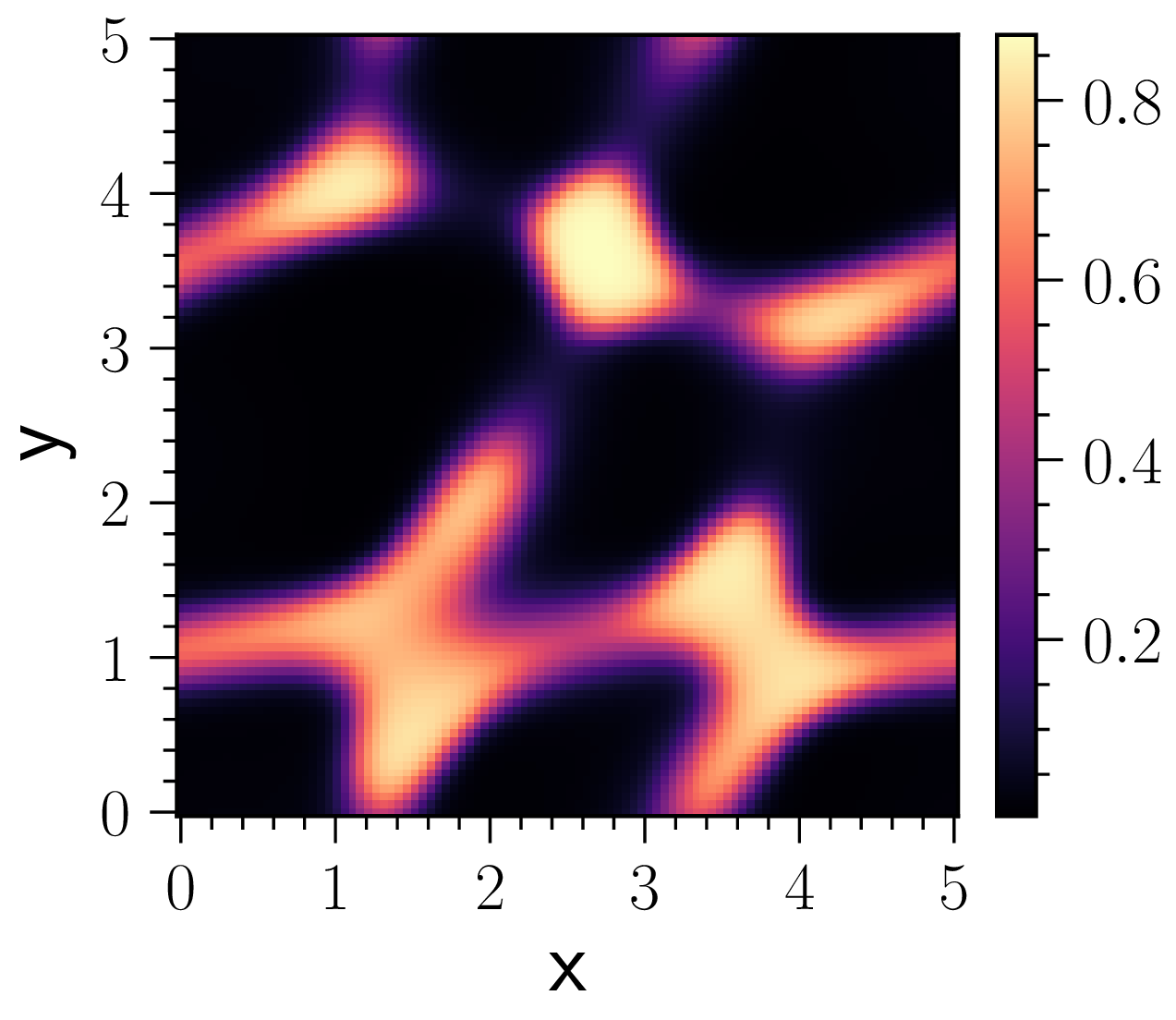}}

  \end{center}
    \caption{Examples of spatio-temporal patterning in the 2D model, Eq. \eqref{eq:2D_PDE} with \textbf{attractive interactions} and high proliferation. Each row of subfigures shows results for a different set of parameters. Within each row, each subfigure shows a heatmap of the cell density $u$ at a subsequent point in time. The top row simulation has lower proliferation, $\rho=10$, than the bottom row simulation, $\rho=20$. Both simulations use parameters: \textbf{O2} kernel, $U=0.5$, $\mu=150$, $\xi=0.3$,  $L=5$. Animations of both simulations can be found at the repository \citep{My_GitHub}.}
    \label{fig:spatio_temporal} 
\end{figure}

For the specific case of attractive interactions with a high proliferation, we observe evolving spatio-temporal patterns. Fig. \ref{fig:spatio_temporal} shows two such examples. We have confirmed numerically that spatio-temporal patterns are possible and have qualitatively similar behaviour with all three example interaction kernels.

As both the analytical and numerical results agree that linear growth rates $\lambda$ are purely real, these spatio-temporal patterns must be a fundamentally nonlinear effect. This is consistent with the fact that such behaviour in the simulation occurs at times significantly after the initial instability of the homogeneous state.

The mechanism of formation for these patterns is the same as that identified by \citet{Painter2015_nonlocal_rd_developmental_biology} for the 1D case: when the proliferation rate is high, new aggregates will form inside any regions of low density, and then attract and coalesce with existing aggregates, thereby moving and leaving behind new regions of low density in which new aggregates form, continuing the dynamics indefinitely. A similar `emerging and merging' mechanism was previously shown to drive spatio-temporal patterning in local Keller-Segel chemotaxis models in 1D \citep{chemotaxis_spatio-temporal_1D}.

In 2D, we can further categorise these spatio-temporal dynamics into two qualitatively distinct types. The first, shown in Fig. \ref{fig:spatio_temporal}\subref{fig:spatio_temporal_A_1}-\subref{fig:spatio_temporal_A_4}, features spontaneously forming spots that coalesce, forming new spots. The second, shown in Fig. \ref{fig:spatio_temporal}\subref{fig:spatio_temporal_B_1}-\subref{fig:spatio_temporal_B_4}, occurs with an even higher proliferation rate such that aggregates do not have time to fully coalesce into distinct spots before they are pulled towards and connected with new aggregates. The resulting dynamics then feature temporal labyrinthine patterns that constantly move, connect, and break apart. Notably, both of these patterns, as well as that shown in Fig.\,\ref{fig:multi_stability}, bear a striking resemblance to the spatio-temporal patterns generated by local chemotaxis models in 2D, as seen in \citet{chemotaxis_spatio-temporal_2D} and in our own simulations of Kegel-Segel dynamics.

\subsubsection{Multi-stability}
\begin{figure}
    \begin{center}
    \subfloat[]{    \label{fig:multi_stability_dispersion}
    \includegraphics[height=0.25\textwidth]{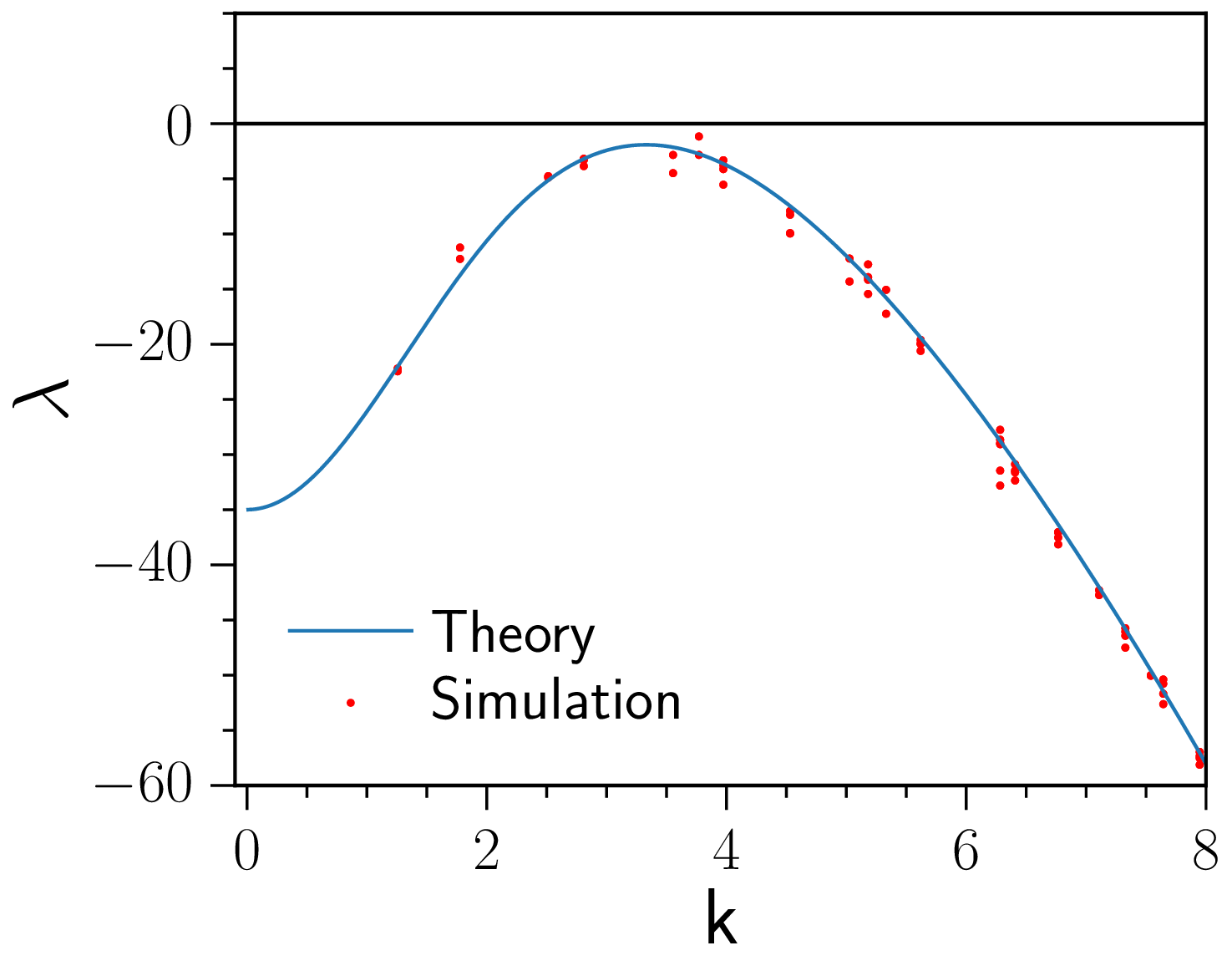}}  
    \quad
    \subfloat[Stable Homogeneous State]{    \label{fig:multi_stability_homogeneous}
    \includegraphics[height=0.25\textwidth]{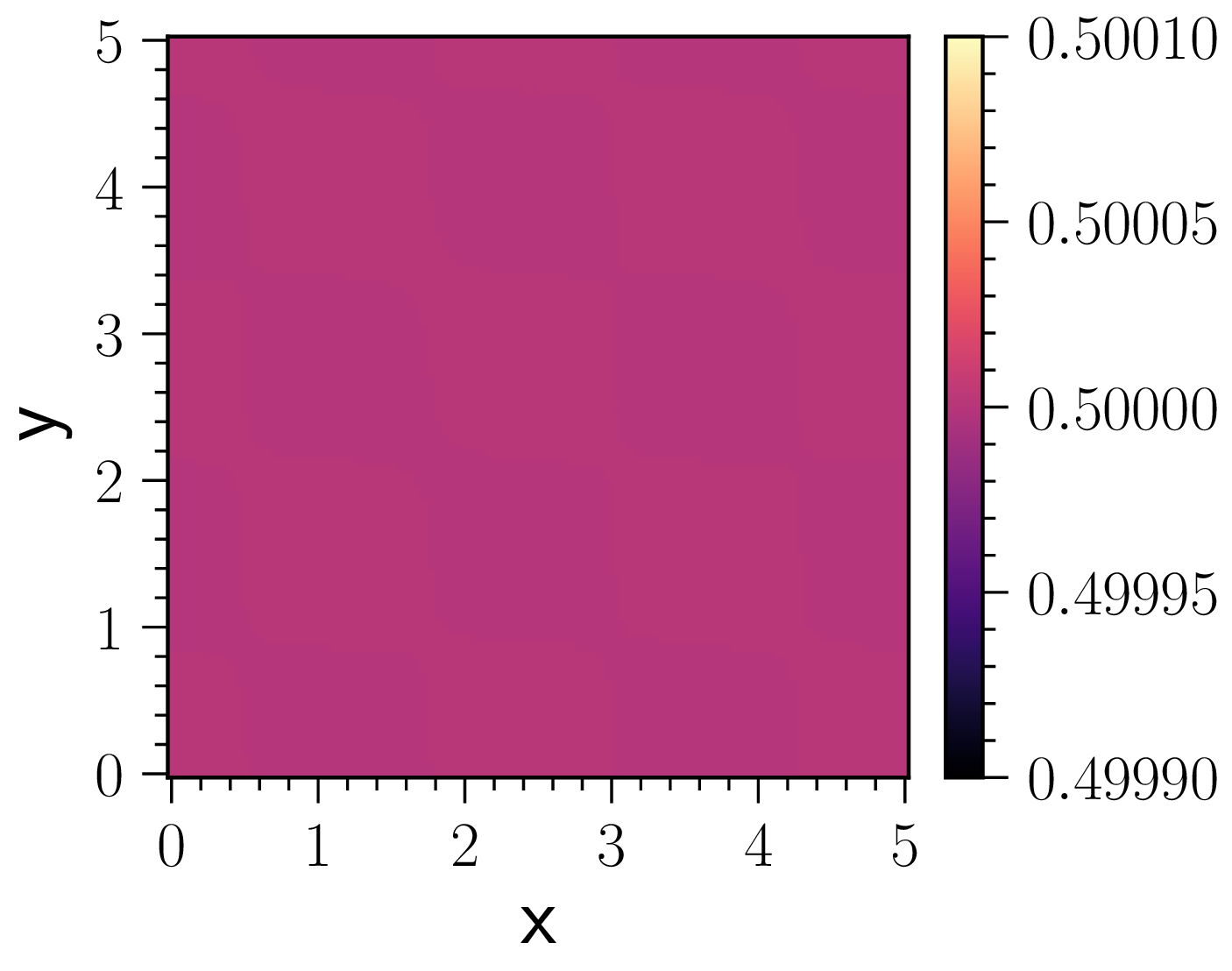}}    
    \quad
    \subfloat[Large Perturbation $t=0$]{    \label{fig:multi_stability_IC}
    \includegraphics[height=0.25\textwidth]{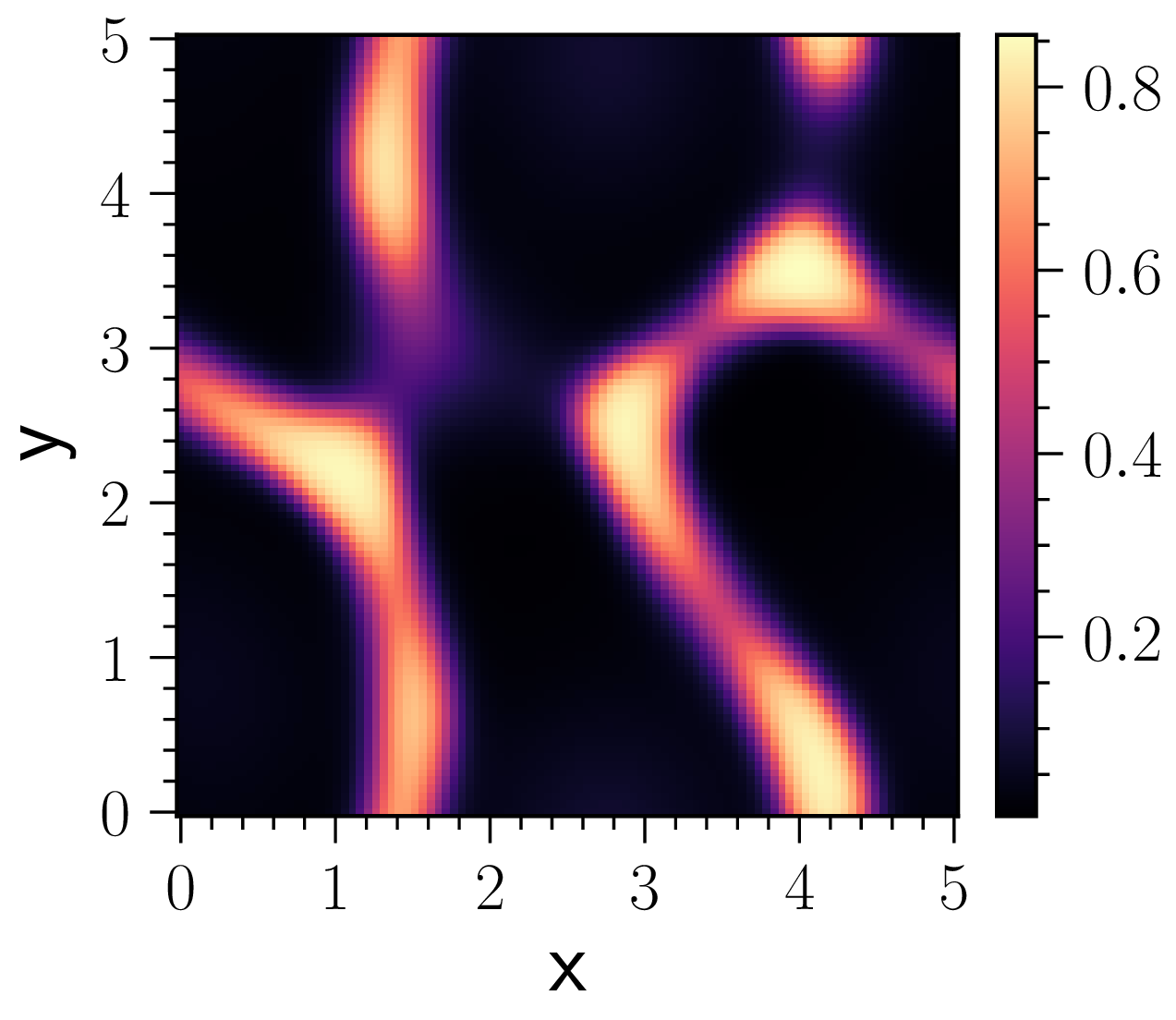}}   
  \\
    \subfloat[$t=0.15$]{    \label{fig:multi_stability_2}
    \includegraphics[height=0.25\textwidth]{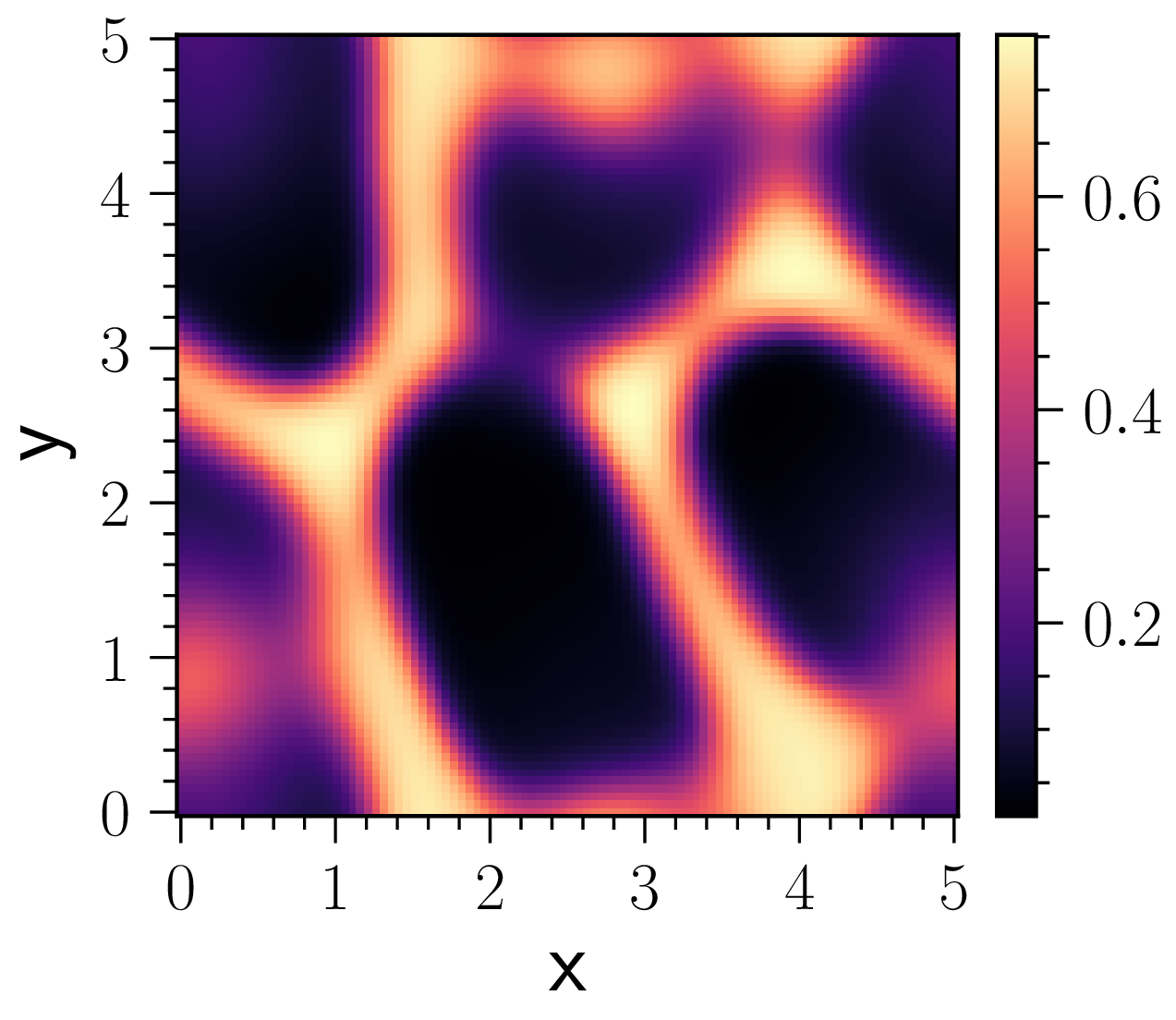}}    
    \quad
    \subfloat[$t=0.54$]{    \label{fig:multi_stability_3}
    \includegraphics[height=0.25\textwidth]{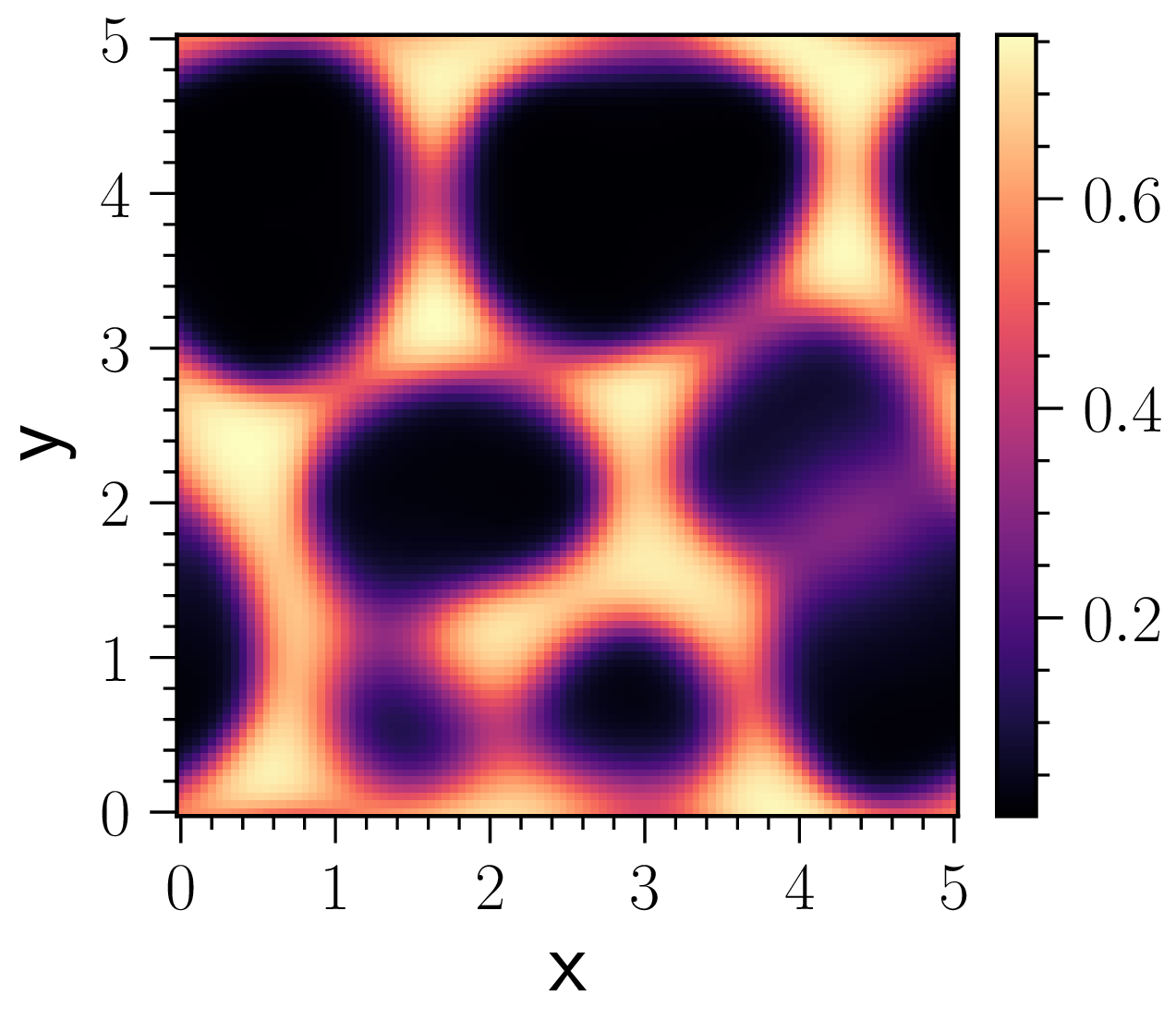}}    
    \quad
    \subfloat[$t=10.00$]{    \label{fig:multi_stability_4}
    \includegraphics[height=0.25\textwidth]{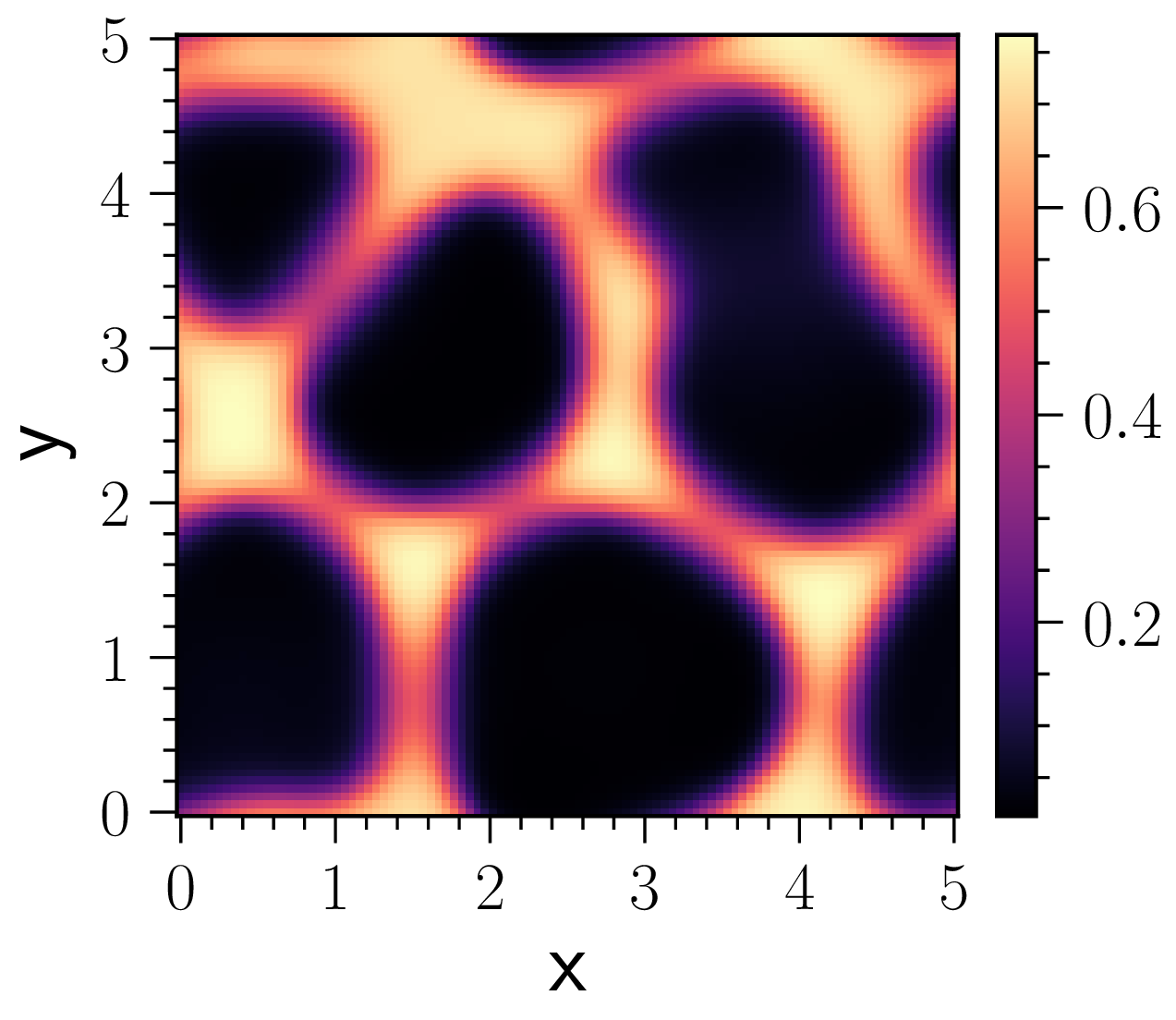}}  
  \end{center}

    \caption{Example of multi-stability in the 2D system. For a fixed set of parameters with \textbf{attractive interactions} and high proliferation: \subref{fig:multi_stability_dispersion}, the dispersion relation showing that all modes linearly decay to the homogeneous state; \subref{fig:multi_stability_homogeneous}, the heatmap of cell density for a simulation which uses a small perturbation and hence converges to the spatially homogeneous steady state; \subref{fig:multi_stability_IC}, the heatmap of a large perturbation of cell density used as an initial condition for a simulation; \subref{fig:multi_stability_2}-\subref{fig:multi_stability_4} heatmaps of cell density at subsequent points in time after the large perturbation. Parameters: \textbf{O2} kernel, $U=0.5$, $\mu=150$, $\rho=35$, $\xi=0.3$, $L=5$. An animation can be found at the repository \citep{My_GitHub}.} 
    \label{fig:multi_stability} 
\end{figure}

Further increases in the proliferation rate, $\rho$, eventually lead to a stable homogeneous steady state, as seen from Eq. \eqref{eq:2D_dispersion_relation}. Physically, this corresponds to cells that proliferate and die at such a high rate that all regions on the domain rapidly return to the carrying capacity density, $U$, regardless of transport effects. However, from simulations such as in Fig. \ref{fig:multi_stability}, we observe some parameter regimes in which large, nonlinear, perturbations may lead to a stable evolving pattern, even when the positive homogeneous steady state is linearly stable. Once again, the character of this pattern can be seen as a continuation of the evolving labyrinthine type (\ref{fig:spatio_temporal}\subref{fig:spatio_temporal_B_1}-\subref{fig:spatio_temporal_B_4}) where the proliferation is now even higher, causing all aggregates to be connected, creating an evolving structure of low density holes.

We have observed these multi-stable regimes for systems with attractive interactions and a sufficiently high proliferation rate for linear stability of the homogeneous state, but not so high as to induce a single stable homogeneous steady state. Such multi-stability is indicative of subcriticality in the Turing bifurcation, as observed in a similar nonlocal advection-diffusion model in \citet{giunta_weakly_nonlinear}, who use weakly nonlinear analysis to investigate the criticality of the Turing bifurcation.

\subsubsection{Summary}
In this section we have numerically verified the predictions of the linear stability analysis for our single species model in 2D, including the fact that repulsive nonlocal interactions can form patterns, which was observed to not be possible in 1D by \citet{Painter2015_nonlocal_rd_developmental_biology}. Beyond the linear regime, we have evidence that attractive interactions lead to steady states of spots, whereas repulsive interactions lead to steady states of stripes, perforated stripes, or a regular lattice of spots. As observed in \citet{Painter2015_nonlocal_rd_developmental_biology} for 1D, we also see spatio-temporal patterns for the 2D system for attractive interactions with high proliferation. These spatio-temporal patterns can also be multi-stable alongside the homogeneous steady state for select parameter regimes, which suggests subcriticality of the Turing bifurcation, as observed for a similar model by \citet{giunta_weakly_nonlinear}. While the above conclusions apply to our model with a single cell species, many developmental processes in biology, such as zebrafish stripe formation, are driven by the interactions of at least two different species of cell.

\section{Two Species}
\label{section:two_species}

Having examined the single species system, we now extend to two species, showing how the linear stability analysis is easily adapted from the single species case, and how there are some fundamental differences in pattern formation going from 1D to higher dimensions.

In the two species model, nonlocal \textit{homotypic} interactions can occur between members of the same species and nonlocal \textit{heterotypic} interactions can occur between members of different species. \cite{Painter2015_nonlocal_rd_developmental_biology} showed that certain combinations of attractive and repulsive homotypic and heterotypic interactions do not permit pattern formation in the 1D case. Here we show that, in higher spatial dimensions, any combination can lead to pattern formation.

Noting that kinetics can be readily incorporated in the framework, for definiteness, we immediately specialise to the transport-only case, with no proliferation and death kinetics. Thus, the time evolution of the population densities for two types of cell in this model, $u(\boldsymbol{x},t)$ and $v(\boldsymbol{x},t)$ is governed by the non-dimensional equations

\begin{equation}
\begin{split}
    \frac{\partial u(\boldsymbol{x},t)}{\partial t} = \nabla^2 u -\boldsymbol{\nabla}\cdot\Biggr( up_u(u,v)\Biggr[\frac{\mu_{uu}}{\xi_{uu}^{N}} \int_{\mathbb{R}^N} \boldsymbol{\hat{s}}\,\tilde{\Omega}_{uu}\left(\frac{s}{\xi_{uu}}\right)g_{uu}(u(\boldsymbol{x}+\boldsymbol{s},t))\boldsymbol{\diff s^N}& \\+ \frac{\mu_{uv}}{\xi_{uv}^{N}} \int_{\mathbb{R}^N} \boldsymbol{\hat{s}}\,\tilde{\Omega}_{uv}\left(\frac{s}{\xi_{uv}}\right)g_{uv}(v(\boldsymbol{x}+\boldsymbol{s},t))\boldsymbol{\diff s^N}\Biggr]\Biggr) \\
    \frac{\partial v(\boldsymbol{x},t)}{\partial t} = D\nabla^2 v -\boldsymbol{\nabla}\cdot\Biggr( vp_v(u,v)\Biggr[\frac{\mu_{vv}}{\xi_{vv}^{N}} \int_{\mathbb{R}^N} \boldsymbol{\hat{s}}\,\tilde{\Omega}_{vv}\left(\frac{s}{\xi_{vv}}\right)g_{vv}(v(\boldsymbol{x}+\boldsymbol{s},t))\boldsymbol{\diff s^N}& \\+ \frac{\mu_{vu}}{\xi_{vu}^{N}} \int_{\mathbb{R}^N} \boldsymbol{\hat{s}}\,\tilde{\Omega}_{vu}\left(\frac{s}{\xi_{vu}}\right)g_{vu}(u(\boldsymbol{x}+\boldsymbol{s},t))\boldsymbol{\diff s^N}\Biggr]\Biggr),
\end{split}
\label{eq:PDE_2_species}
\end{equation}
where $D$ is the relative diffusion coefficient. Each nonlocal interaction between a species and itself or another species has its own interaction strength $\mu_{ab}$, signalling range $\xi_{ab}$, interaction kernel $\tilde{\Omega}_{ab}$, and source function $g_{ab}$, where $a,b \in \{u,v\}$. Additionally, each species can have its own packing function $p_a(u,v)$, where $a\in \{u,v\}$. 

\subsection{Linear Stability Analysis}

The linear stability analysis of Eq. \eqref{eq:PDE_2_species} simply follows from the single species case, in that the nonlocal term becomes a hyperspherical Hankel transform of the interaction kernel in the dispersion relation. We linearise about some homogeneous steady state $(U, V)$ such that $u(\boldsymbol{x},t)=U+\tilde{u}(\boldsymbol{x},t)$ and $v(\boldsymbol{x},t)=V+\tilde{v}(\boldsymbol{x},t)$, where $(U,V)$ are dictated by the initial conditions. The dispersion relation is then given by

\begin{equation}
    \lambda^2 + \mathcal{C}(k)\lambda+\mathcal{D}(k) =0,
\label{eq:disp_relation_2_species}
\end{equation}
where
\begin{equation}
\begin{split}
    \mathcal{C}(k)&= k^2(1+D)-k(\Lambda_{uu}+\Lambda_{vv})\\
    \mathcal{D}(k)&=Dk^4 - k^3(D\Lambda_{uu}+\Lambda_{vv})+k^2(\Lambda_{uu}\Lambda_{vv}-\Lambda_{uv}\Lambda_{vu})
\end{split}
\end{equation}
in which each $\Lambda_{ab}$ is the nonlocal term for each interaction and given by
\begin{equation}
    \Lambda_{ab} = \frac{2\pi^{\frac{N}{2}}}{\Gamma\left(\frac{N}{2}\right)}\,Ap_a(U,V)\frac{\partial g_{ab}}{\partial b}\Big\rvert_{B} \frac{\mu_{ab}}{\xi_{ab}^N}\,\int_{0}^{\infty}\diff s\,\tilde{\Omega}_{ab}\left(\frac{s}{\xi_{ab}}\right)s^{N-1}\,j_1^{(N)}(ks),
\end{equation}
where $a,b \in \{u,v\}$ with \textit{corresponding} homogeneous states $A,B \in \{U,V\}$. 

Eq. \eqref{eq:disp_relation_2_species} is solved to give
\begin{equation}
    \lambda_{\pm} = \frac{-\mathcal{C}(k)\pm \sqrt{\mathcal{C}(k)^2 -4\mathcal{D}(k)}}{2}.
\label{eq:lambda_equals_two_species}
\end{equation}

 Similarly to the single species case, this dispersion relation takes the same form in the $N$D case as it does in the 1D case derived in \citet{Painter2015_nonlocal_rd_developmental_biology}, but with hyperspherical Hankel transforms instead of just the Fourier sine transform.

Pattern formation requires $\text{Re}(\lambda_+)>0$ for some $k>0$, which occurs if and only if $\mathcal{C}(k)<0$ or $\mathcal{D}(k)<0$, for some $k>0$. \citet{Painter2015_nonlocal_rd_developmental_biology} show that this can happen in 1D (assuming all $p_a(U,V)\frac{\partial g_{ab}}{\partial b}\Big\rvert_{B}>0$) for all combinations of attractive/repulsive interactions, except for repulsive-repulsive homotypic and attractive-repulsive heterotypic interactions. This exception corresponds to $\mu_{uu}, \mu_{vv}<0$ and $\mu_{uv}<0$, $\mu_{vu}>0$ (or $\mu_{uv}>0$, $\mu_{vu}<0$). 

Their conclusion relies on the fact that the integral transforms of the interaction kernels are always positive in 1D (for non-increasing kernels). However, in 2D and above, this is no longer the case. Instead, these integral transforms can be negative and so pattern formation can occur for all combinations of interactions. This is the same mechanism by which pattern formation in single species systems occurs for repulsive interactions only in 2D or above.

It is perhaps unsurprising that patterning can be driven by repulsive-repulsive homotypic interactions, given that we have shown that repulsive interactions for an isolated single species can lead to patterns. More notable is that attractive-repulsive heterotypic interactions, or equivalently run-and-chase dynamics, \textit{alone} may drive pattern formation, contrary to results from \citet{Painter2015_nonlocal_rd_developmental_biology, WoolleyShort2014, Woolley_Chiral_Chasing}. However, this does require the function $\tilde{\Omega}_{uv}\left(\frac{s}{\xi_{uv}}\right)$ to be sufficiently different from $\tilde{\Omega}_{vu}\left(\frac{s}{\xi_{vu}}\right)$ such that one of their hyperspherical Hankel transforms is positive and the other is negative for some wavenumber. Biologically, this requirement is equivalent to some asymmetry in the cross-species signalling/sensing process.

\subsection{Numerical Validation of Instability}

\begin{figure}
    \begin{center}
    \subfloat[]{    \label{fig:2_species_disp_C}
    \includegraphics[height=0.23\textwidth]{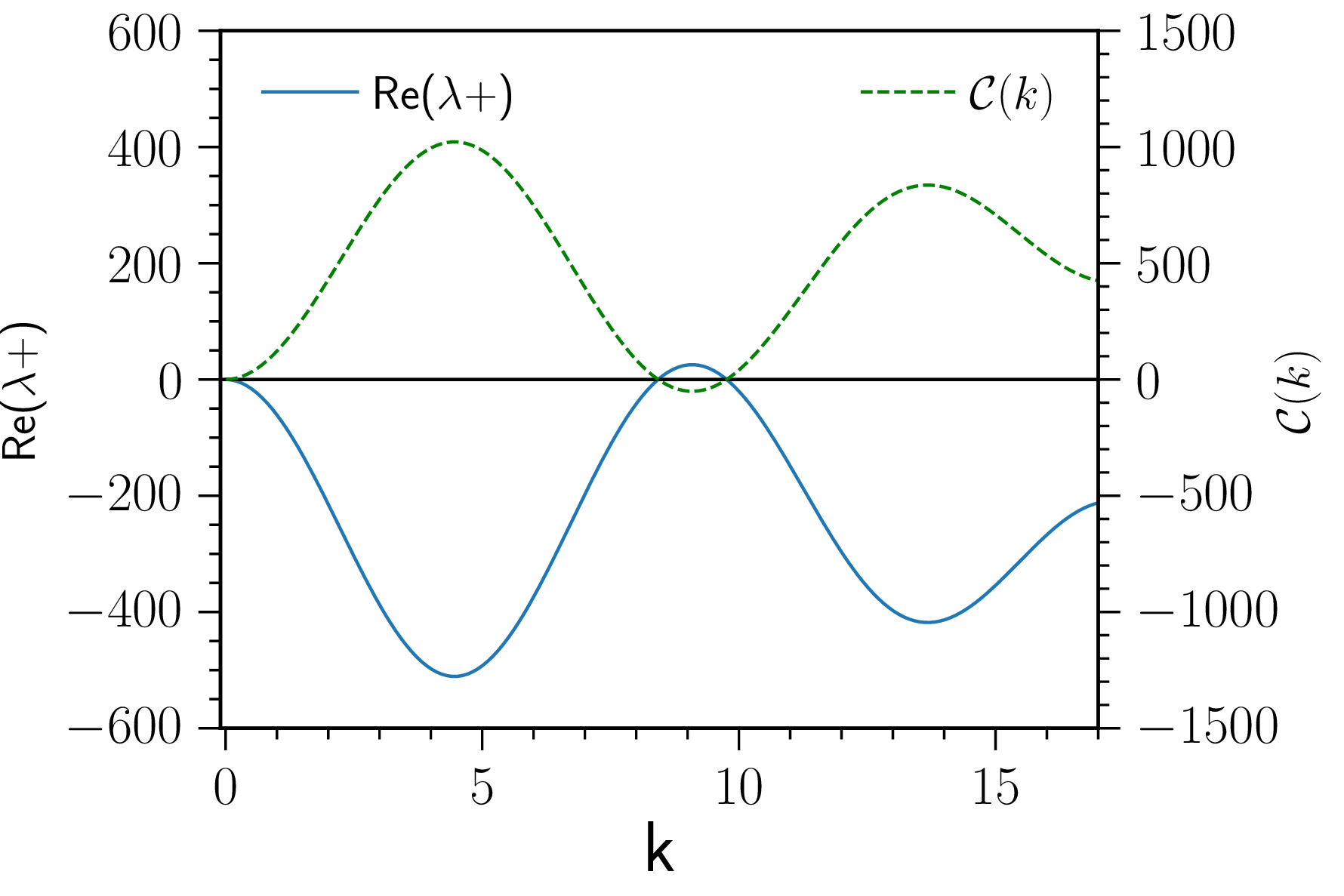}} 
    \quad
    \subfloat[$t=1.38$]{    \label{fig:2_species_f1}
    \includegraphics[height=0.23\textwidth]{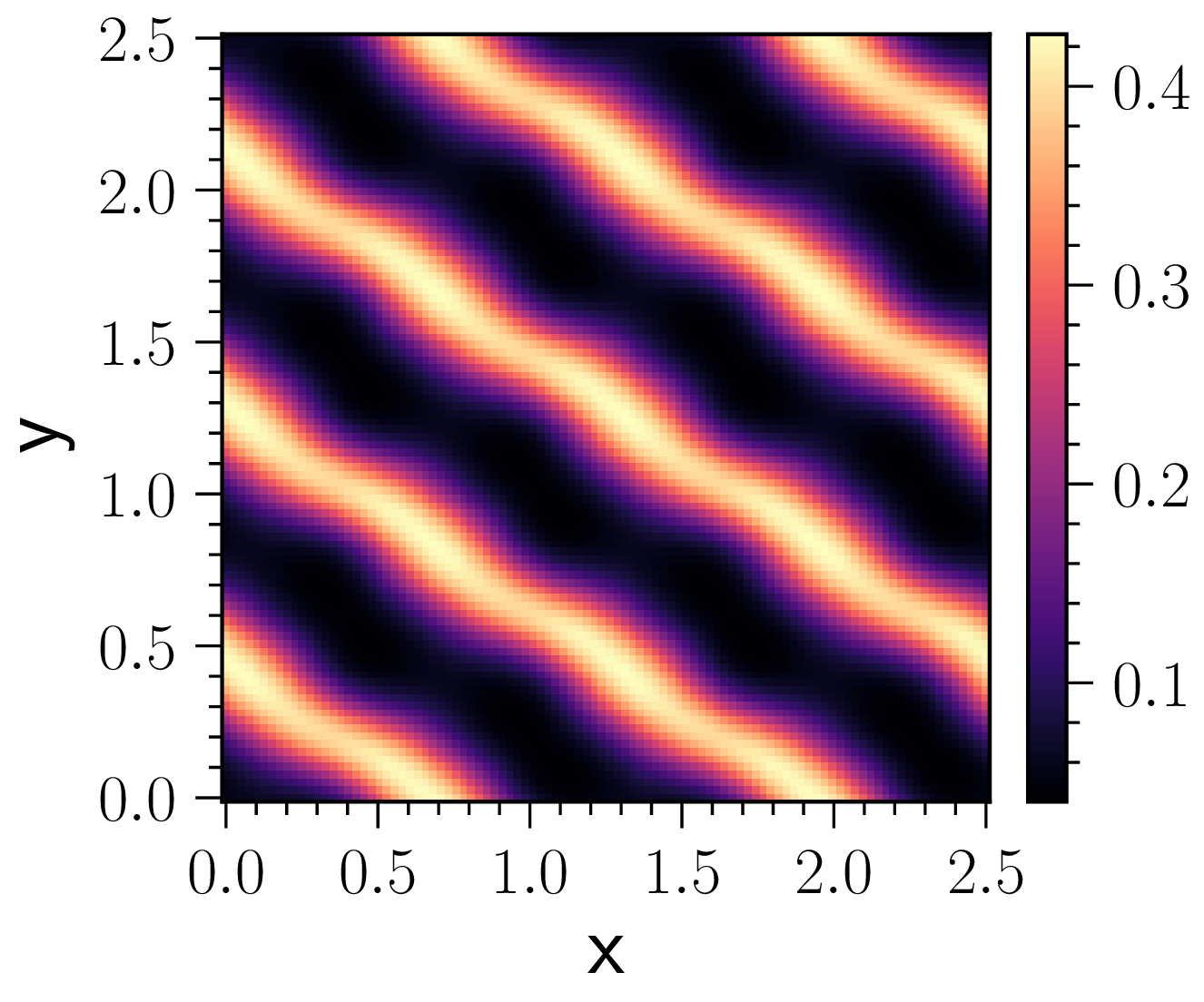}}
    \subfloat[$t=1.39$]{    \label{fig:2_species_f2}
    \includegraphics[height=0.23\textwidth]{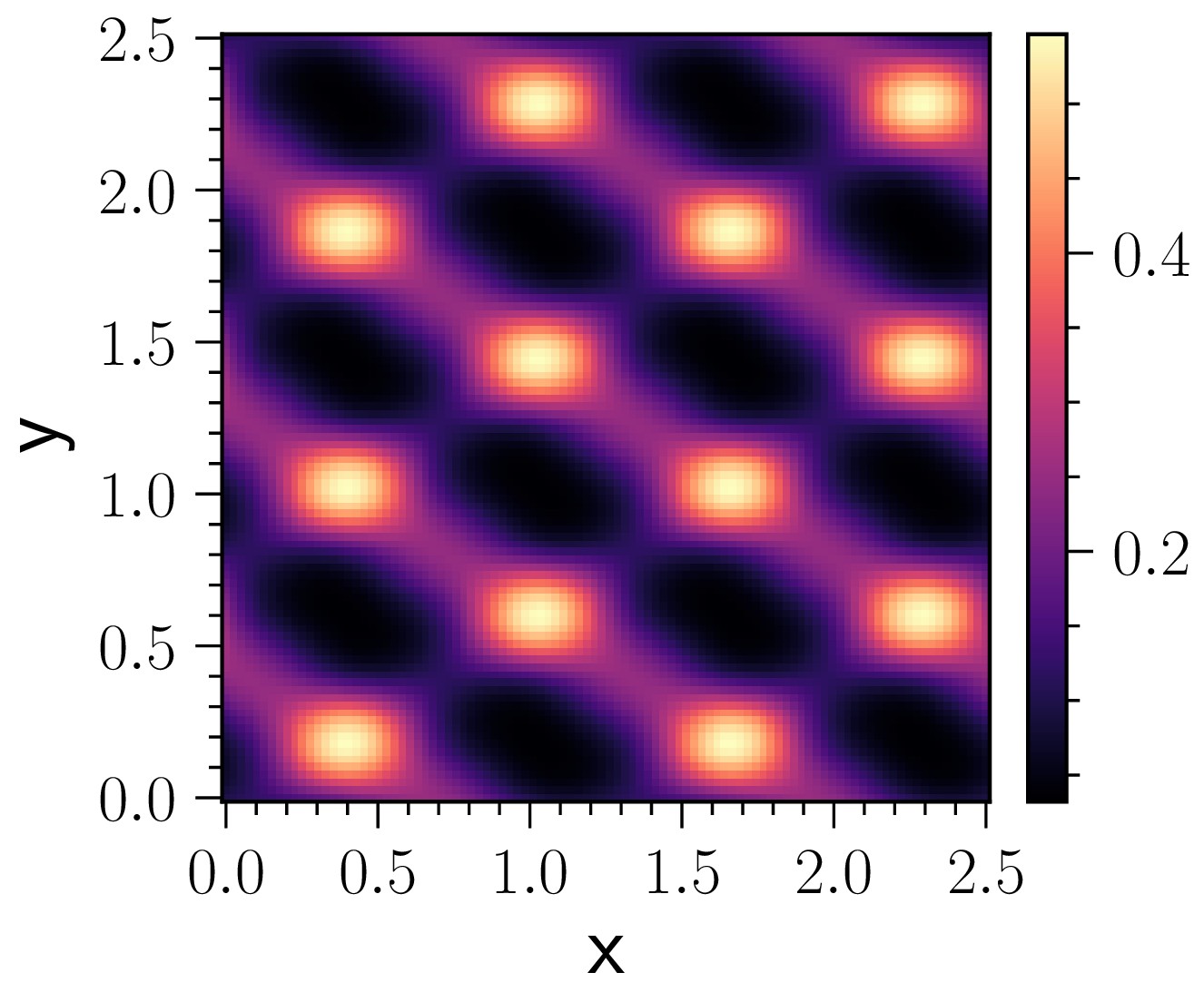}}    
    \\
    \subfloat[$t=1.42$]{    \label{fig:2_species_f3}
    \includegraphics[height=0.23\textwidth]{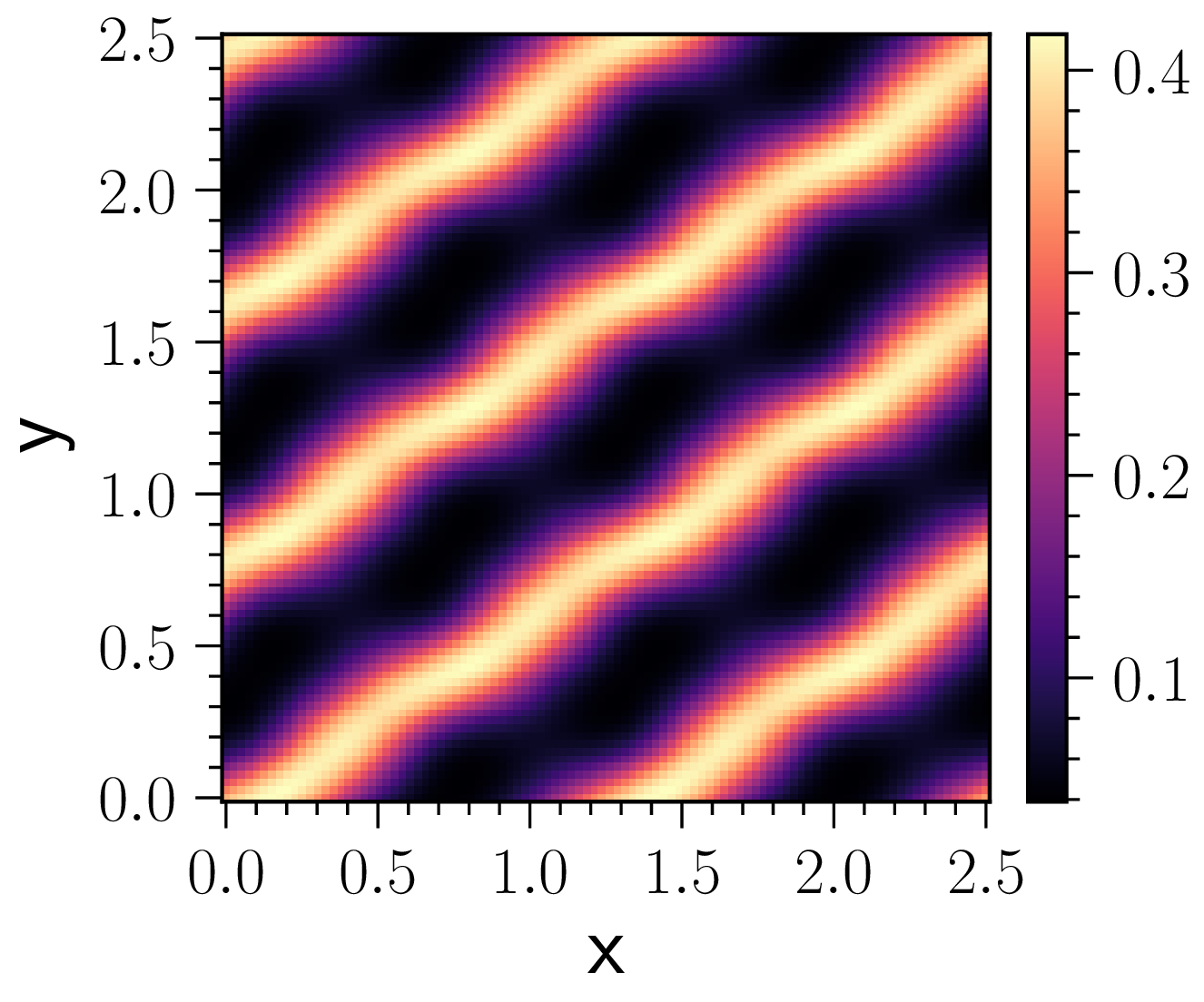}}   
    \quad
    \subfloat[$t=1.43$]{    \label{fig:2_species_f4}
    \includegraphics[height=0.23\textwidth]{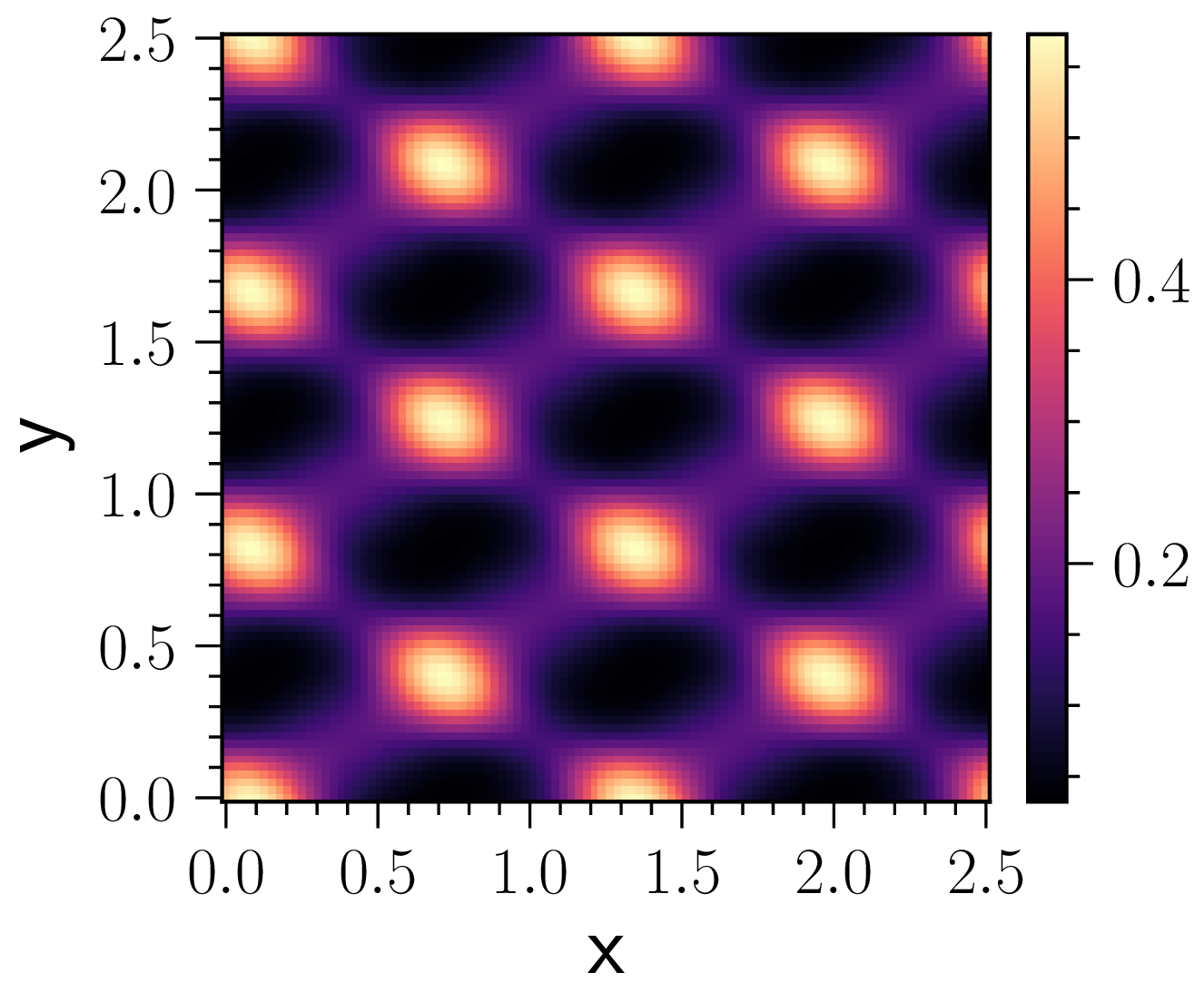}}    
    \quad
    \subfloat[$t=1.45$]{    \label{fig:2_species_f5}
    \includegraphics[height=0.23\textwidth]{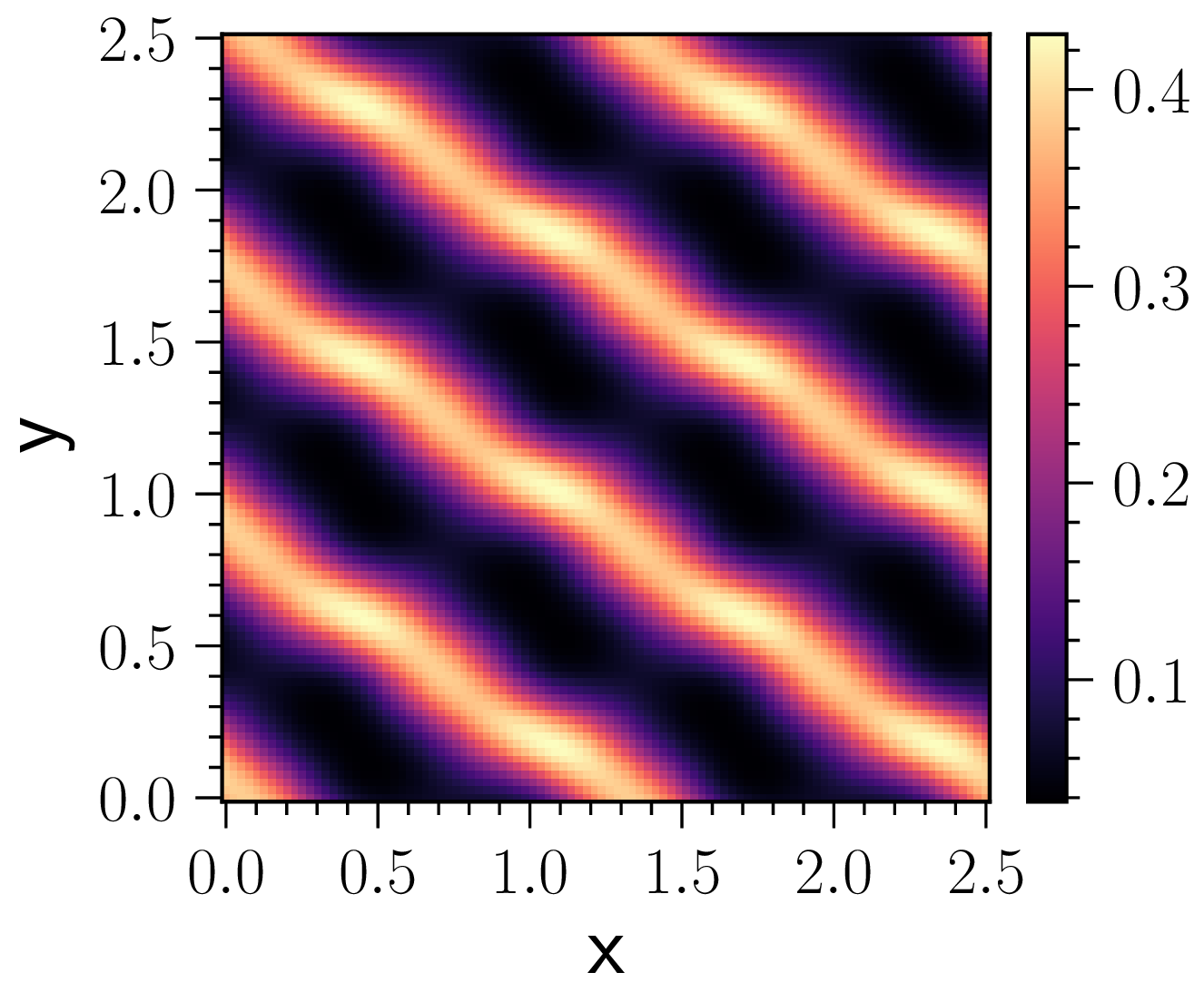}}    
  \end{center}

    \caption{Simulation results for the 2D two species model, Eq. \eqref{eq:PDE_2_species}, with repulsive-repulsive homotypic and attractive-repulsive heterotypic interactions. \subref{fig:2_species_f1}-\subref{fig:2_species_f5} heatmaps of the cell density $u$ at subsequent times, illustrating the periodic oscillatory behaviour. Note that this shows half a period of oscillation, as the maxima and minima in \subref{fig:2_species_f1} and \subref{fig:2_species_f5} are swapped; after the second half period, the pattern returns to \subref{fig:2_species_f1}. $v$ (not shown) similarly oscillates, but one quarter period behind $u$; for example, when $u$ looks like \subref{fig:2_species_f5}, $v$ looks like \subref{fig:2_species_f3}. \subref{fig:2_species_disp_C} the dispersion relation (solid line), which has one region of instability corresponding to the negative region of $\mathcal{C}(k)$ (dashed line). Parameters: $U=0.25$, $V=0.25$, $D=1$, $\mu_{uu}=-2000$, $\mu_{uv}=-1000$, $\mu_{vu}=1000$, $\mu_{vv}=-2000$, $\xi_{uu}=0.75$,  $\xi_{uv}=1$, $\xi_{vu}=1$, $\xi_{vv}=0.75$, $L=2.5$. An animation can be found at the repository \citep{My_GitHub}}. 
    \label{fig:2_species_oscillation} 
\end{figure}

\begin{figure}
    \begin{center}
    \subfloat[$u(\boldsymbol{x},t)$]{    \label{fig:2_species_u_D}
    \includegraphics[height=0.23\textwidth]{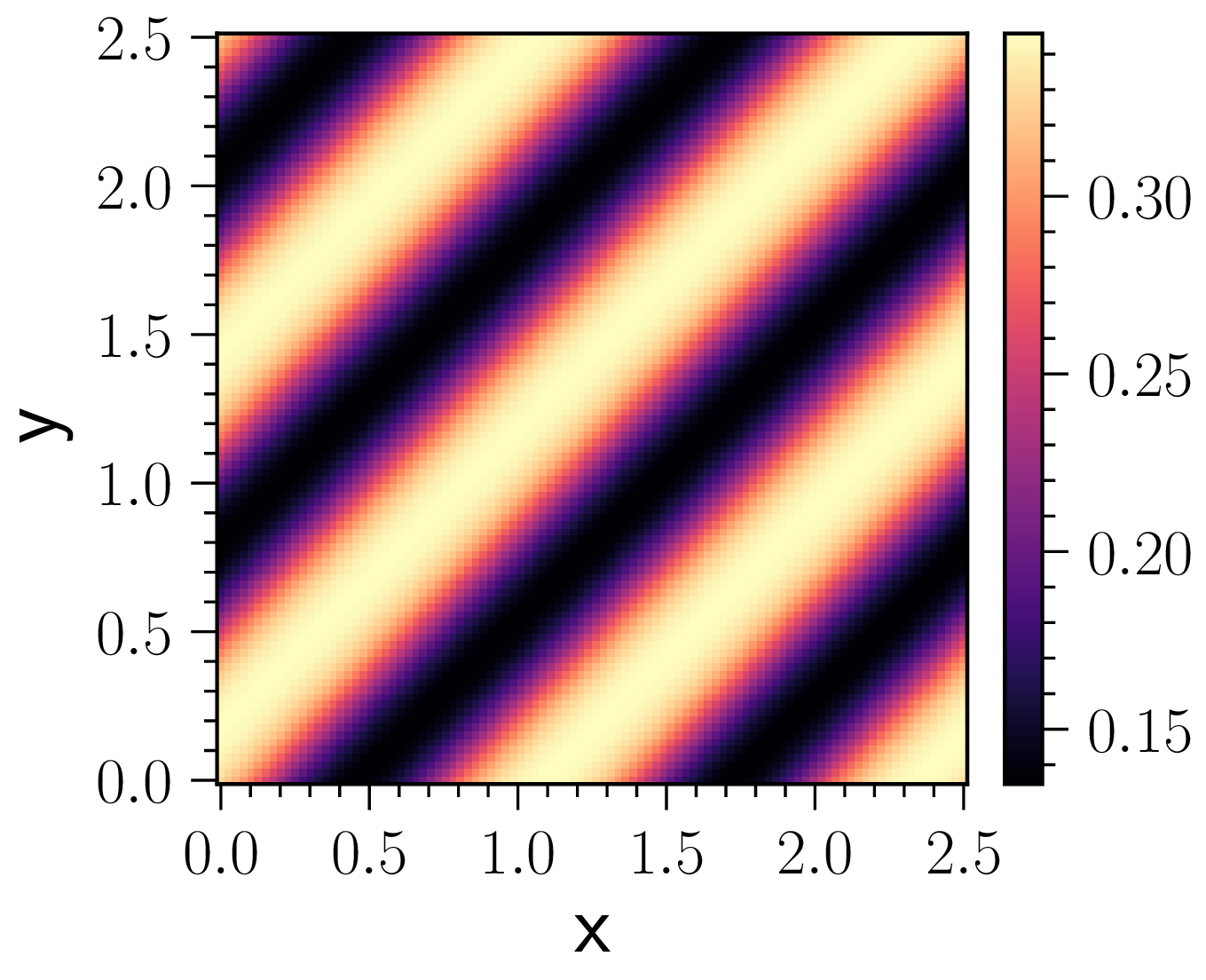}}    
    \quad
    \subfloat[$v(\boldsymbol{x},t)$]{    \label{fig:2_species_v_D}
    \includegraphics[height=0.23\textwidth]{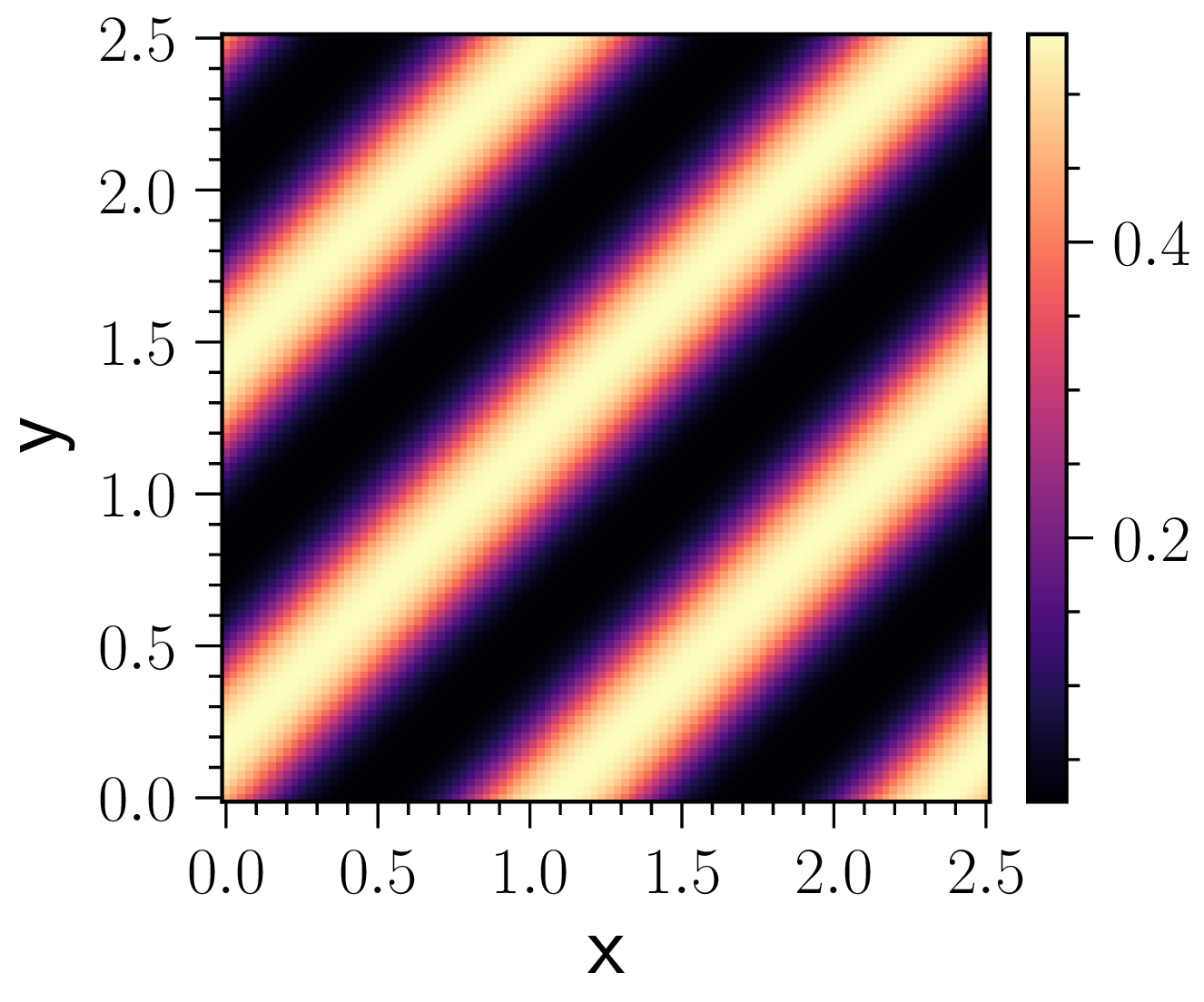}}    
    \quad
    \subfloat[]{    \label{fig:2_species_disp_D}
    \includegraphics[height=0.23\textwidth]{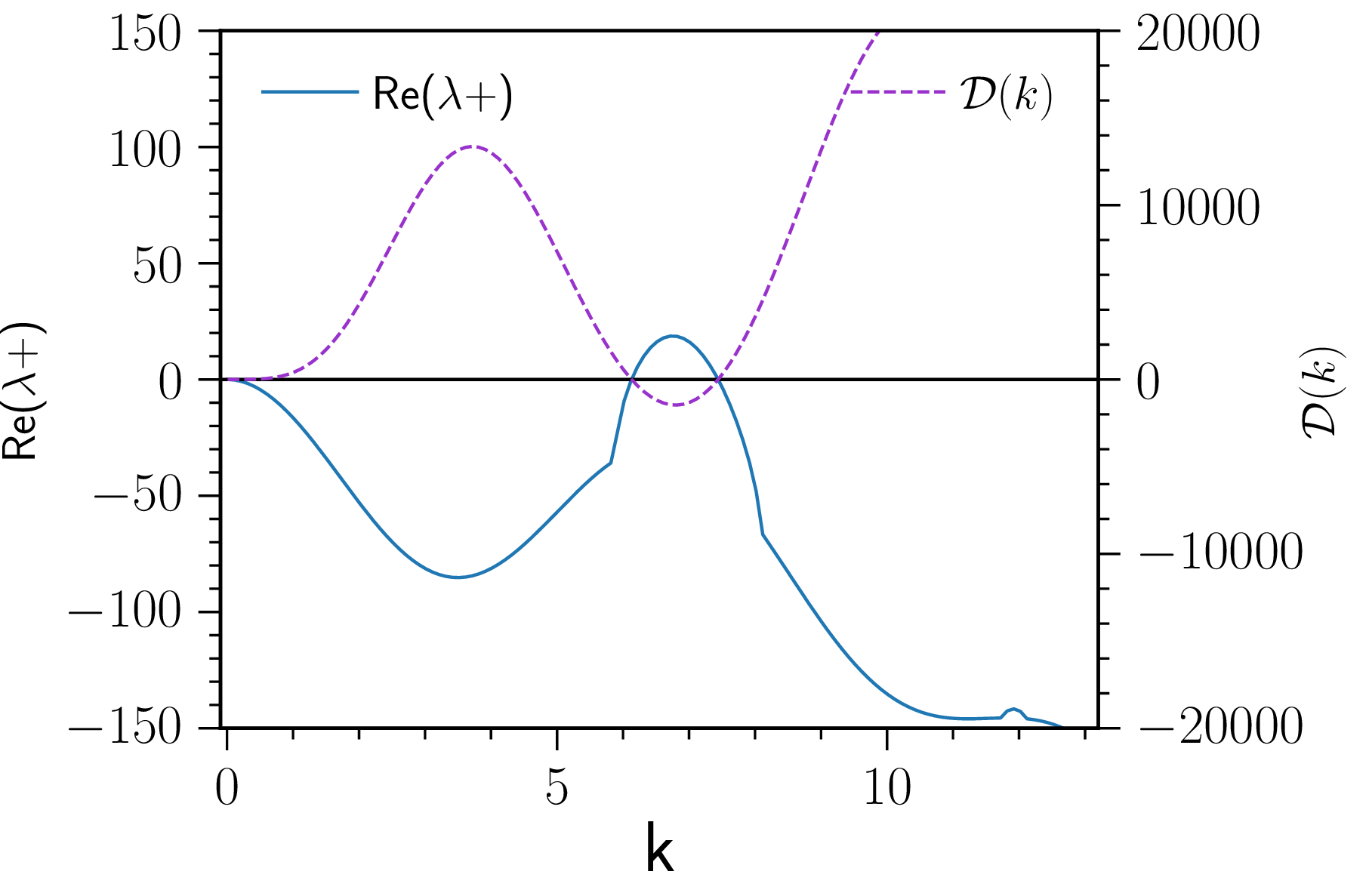}}   
  \end{center}

    \caption{Simulation results for the 2D two species model, Eq. \eqref{eq:PDE_2_species}, with repulsive-repulsive homotypic and attractive-repulsive heterotypic interactions, with different parameter values to Fig. \ref{fig:2_species_oscillation}. Heatmaps for the final stable pattern for \subref{fig:2_species_u_D} $u$, and \subref{fig:2_species_v_D} $v$. \subref{fig:2_species_disp_D} the dispersion relation (solid line), which has one region of instability corresponding to the negative region of $\mathcal{D}(k)$ (dashed line). Simulations ran to $t=20$ and used parameters:  $U=0.25$, $V=0.25$, $D=1$, $\mu_{uu}=-400$, $\mu_{uv}=-400$, $\mu_{vu}=400$, $\mu_{vv}=-400$, $\xi_{uu}=1$,  $\xi_{uv}=1$, $\xi_{vu}=0.5$, $\xi_{vv}=1$, $L=2.5$.} 
    \label{fig:2_species} 
\end{figure}

\begin{figure}
    \begin{center}
    \subfloat[$u(\boldsymbol{x},t)$]{    \label{fig:2_species_u_run_chase}
    \includegraphics[height=0.23\textwidth]{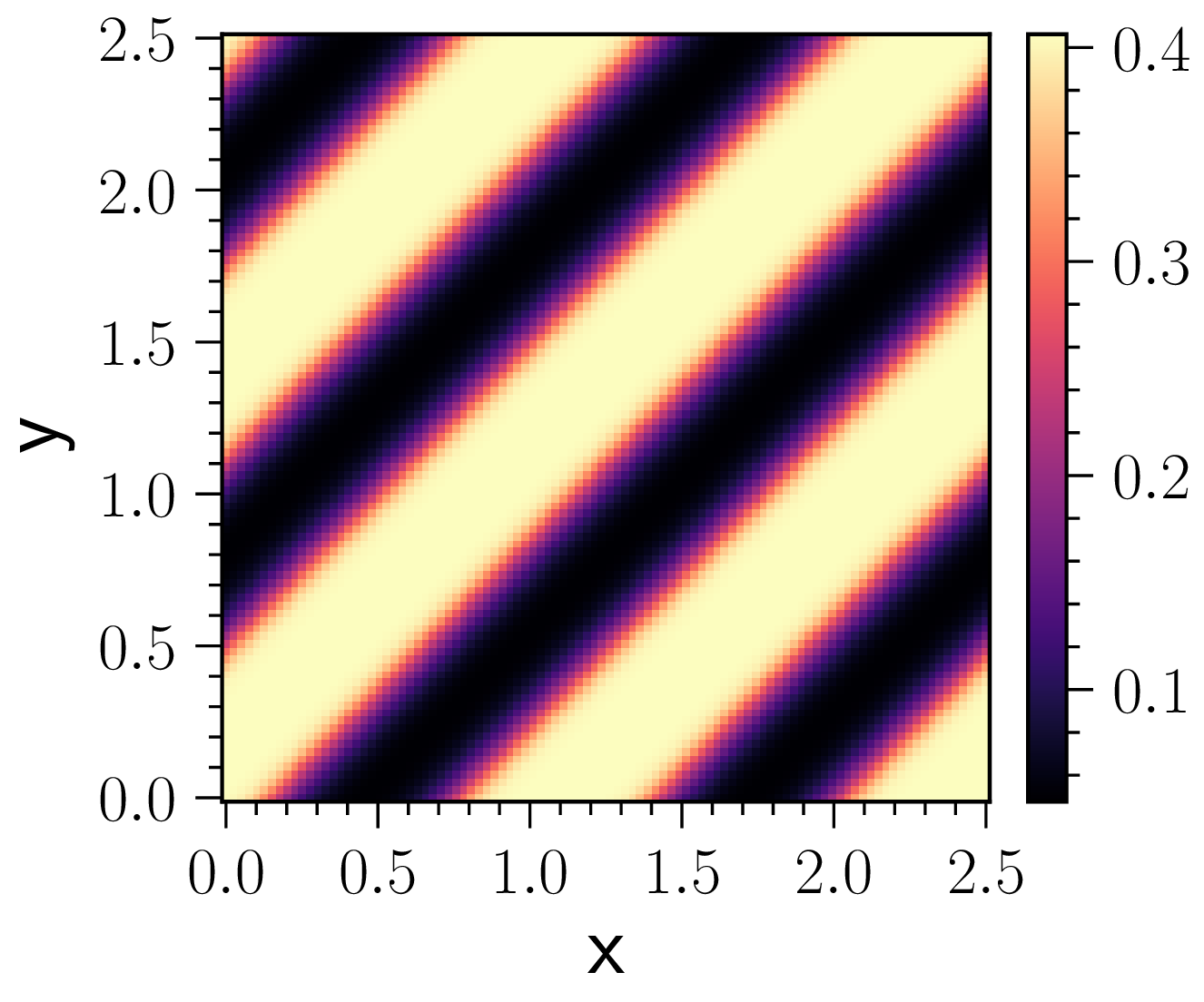}}    
    \quad
    \subfloat[$v(\boldsymbol{x},t)$]{    \label{fig:2_species_v_run_chase}
    \includegraphics[height=0.23\textwidth]{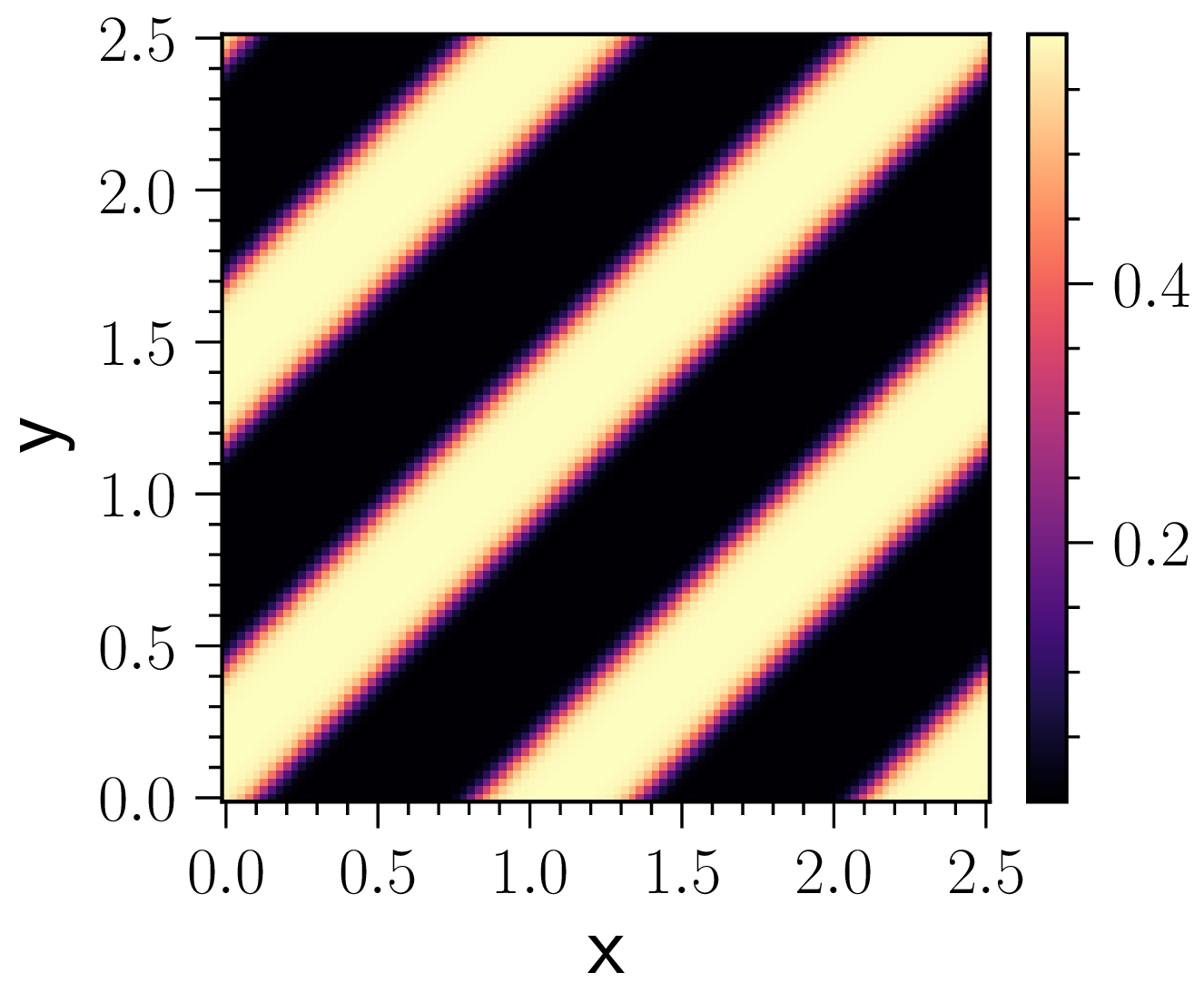}}    
    \quad
    \subfloat[]{    \label{fig:2_species_disp_run_chase}
    \includegraphics[height=0.23\textwidth]{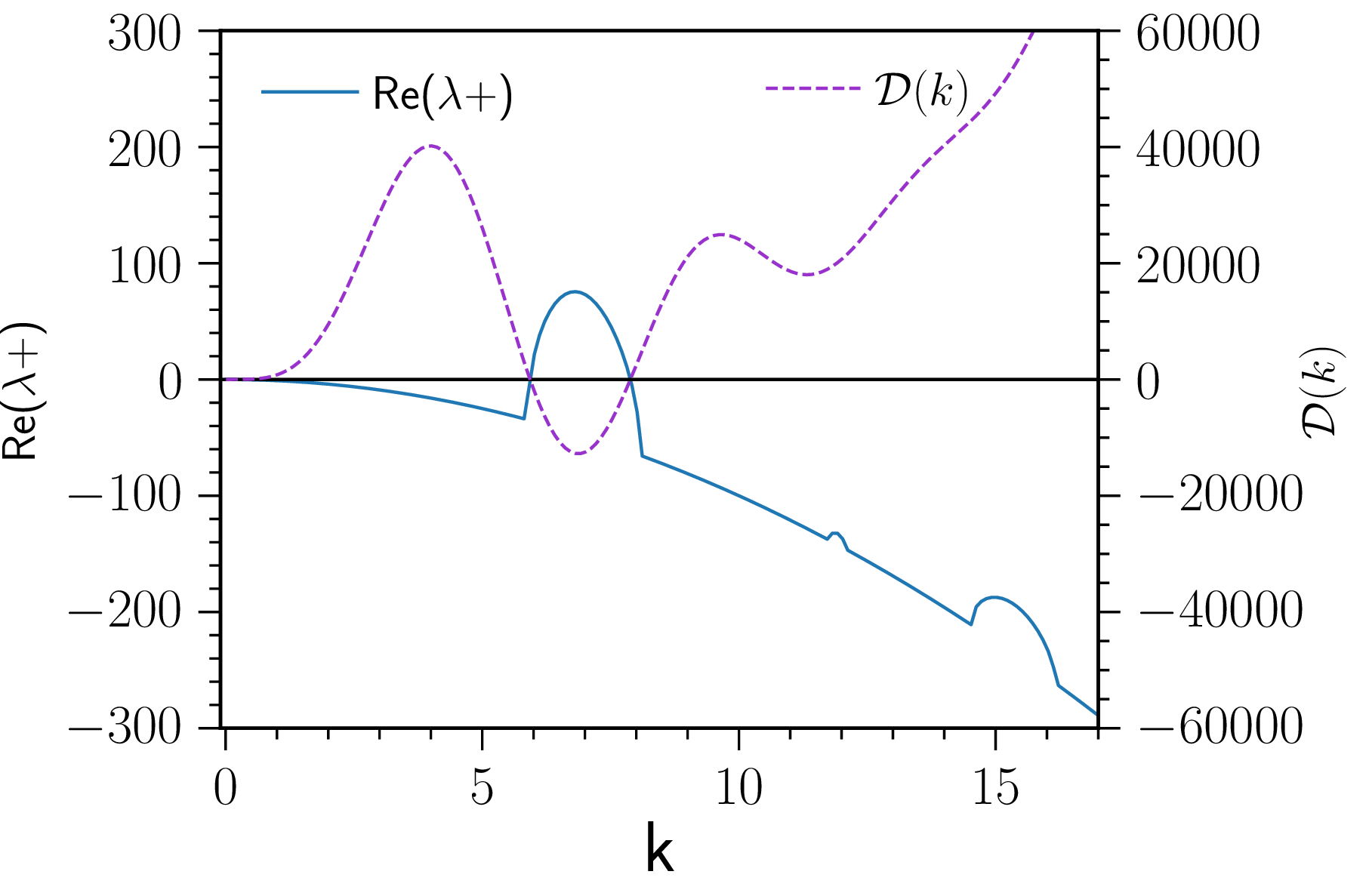}}   
  \end{center}

    \caption{Simulation results for the 2D two species model, Eq. \eqref{eq:PDE_2_species}, with attractive-repulsive heterotypic interactions and no homotypic interactions, i.e. run-and-chase dynamics. \subref{fig:2_species_u_run_chase}, shows heatmaps of $u$ for the final stable pattern, whilst \subref{fig:2_species_v_run_chase}, shows the same for $v$. \subref{fig:2_species_disp_run_chase} shows the dispersion relation (solid line), which has one region of instability corresponding to the negative region of $\mathcal{D}(k)$ (dashed line). Simulations ran to $t=20$ and used parameters: $U=0.25$, $V=0.25$, $D=1$, $\mu_{uu}=0$, $\mu_{uv}=-1000$, $\mu_{vu}=1000$, $\mu_{vv}=0$, $\xi_{uv}=1$, $\xi_{vu}=0.5$, $L=2.5$.}
    \label{fig:2_species_run_chase} 
\end{figure}

To validate our predictions numerically we look at Eq. \eqref{eq:PDE_2_species} in 2D, specifying the same packing and source functions as \citet{Painter2015_nonlocal_rd_developmental_biology}: $p_u(u,v)=p_v(u,v)=1-u-v$, $g_{uu}(u)=g_{vu}(u)=u$, $g_{vv}(v)=g_{uv}(v)=v$. We also choose \textbf{O1} (top-hat) functions for all interaction kernels. The system is then integrated using the same numerical scheme as in Section \ref{section:numerics}.

Figs. \ref{fig:2_species_oscillation}, \ref{fig:2_species}, and \ref{fig:2_species_run_chase} show examples of numerical simulations of systems with repulsive-repulsive homotypic and/or attractive-repulsive heterotypic interactions. All example systems form patterns, confirming our predictions that pattern formation is possible with such interactions in 2D. 

In contrast to the single species system, with two species the linear growth rate $\lambda$ can be complex, as seen by Eq. \eqref{eq:lambda_equals_two_species}. When $\lambda$ is complex and $\text{Re}(\lambda)>0$, the homogeneous state is destabilised into a temporally oscillating spatially heterogeneous pattern. This is sometimes referred to as a 'Turing-Hopf' bifurcation \citep{Turing-Hopf-name} and sometimes as a `Turing-wave' bifurcation \citep{Turing-wave-name}, or simply a `wave' bifurcation \citep{villar2023_wave_bifurcation}. From Eq.\,\eqref{eq:lambda_equals_two_species} we see that such bifurcations occur when $\mathcal{C}(k)$ passes through zero from positive to negative with $\mathcal{D}(k)>0$, as in this case $\lambda_+$ will have a non-zero imaginary part and its real part will go from negative to positive. The model can exhibit such bifurcations in any number of spatial dimensions, with Fig. \ref{fig:2_species_oscillation} showing a 2D example. Being a linear effect, this behaviour is distinct from the spatio-temporal patterns observed for the single species system.

Static patterns are also possible for the two species model, through the standard Turing bifurcation where $\text{Im}(\lambda_+)=0$. From Eq.\,\eqref{eq:lambda_equals_two_species} we see that this bifurcation occurs when $\mathcal{D}(k)$ passes through zero from positive to negative: $\mathcal{D}(k)=0$ implies $\lambda_+=0$, and $\mathcal{D}(k)<0$ implies $\text{Re}(\lambda_+)>0$ and $\text{Im}(\lambda_+)=0$. Figs. \ref{fig:2_species} and \ref{fig:2_species_run_chase} are examples of these static patterns. In particular, Fig. \ref{fig:2_species_run_chase} demonstrates static stripe patterns that do not require any homotypic interactions and are formed purely through attractive-repulsive heterotypic interactions, i.e. through run-and-chase dynamics alone.

\section{Discussion}
\label{section:discussion}

To gain a better understanding of how cells self-organise into the essential structures for biological development, it is crucial to identify the key factors that influence cell aggregation. This paper's primary objective has been to explore the influence of spatial dimension on pattern formation of cell aggregates within the nonlocal attraction and repulsion model proposed by \citet{Painter2015_nonlocal_rd_developmental_biology}. In particular, does the inclusion of the full two or three spatial dimensions present in real biological systems significantly alter the conditions for pattern formation? To address this question, we employed a combination of linear stability analysis in $N$D utilizing hyperspherical Bessel functions, along with numerical simulations in 2D for both the single species and two species models.

Our findings reveal a significant feature of the \citet{Painter2015_nonlocal_rd_developmental_biology} model: in contrast to standard reaction-diffusion systems, its linear stability fundamentally changes with the number of spatial dimensions. Changing dimensionality thus not only affects the shapes of possible patterns, but changes the very capacity for pattern formation itself, as well as the capacity for linearised spatio-temporal dynamics. Our results underscore the importance of considering 2D and 3D cases when studying systems with nonlocal interactions, as simplistic 1D models may overlook essential behaviors in biological systems.

That said, we have shown that the dispersion relation derived in \citet{Painter2015_nonlocal_rd_developmental_biology} for 1D generalises to a very similar form in $N$D, for both the single species and two species models. Essentially, the only change is the generalisation of the Fourier sine transform of the interaction kernel to a hyperspherical Hankel transform. This consistency allowed us to show that patterning through a Turing bifurcation can occur for this model in any number of dimensions, given appropriate nonlocal interaction directions and strengths. For example, attractive interactions between a single species can always form patterns, given a sufficiently high interaction strength.

Additionally, for all $N$, we have illustrated the mathematical mechanism by which a system dominated by nonlocal transport produces a pattern with wavelength directly proportional to the signalling range. We have shown how the proportionality constant depends only on the dimensionality, the shape of the interaction kernel, and whether the interactions are attractive or repulsive. This insensitivity to changes in all other parameters is an essential trait for any mechanism in developmental biology, in which structures with precise size have to form in noisy environments.

However, as captured by the behaviour of hyperspherical Hankel transforms, there are some fundamental differences in pattern formation in this model for different dimensions, and we have presented a physically intuitive argument for this in terms of the flux induced by nonlocal interactions pointing at different angles depending on the location of the source of the signal. An important example of such a difference is the ability for single species systems to form patterns driven by repulsive interactions, which is only possible in 2D or higher and hence not observed in \citet{Painter2015_nonlocal_rd_developmental_biology}. Some developmental processes do indeed feature a single species of cell that move apart when contact is made between cell protrusions, such as with fibroblast cells during neural crest development \citep{neural_crest_development}, although we are not claiming that this particular example is reflected perfectly in the model.

Another significant difference between the 1D and higher dimensional cases is the behaviour of the two species model with attractive-repulsive heterotypic interactions or run-and-chase dynamics. \citet{YamanakaKondo2014} proposed that such dynamics could underlie zebrafish stripe formation. Despite this, \citet{Painter2015_nonlocal_rd_developmental_biology} found from their analysis of the 1D case that run-and-chase \textit{alone} cannot produce patterns, and this has been supported by studies of related models in 2D by \citet{WoolleyShort2014} and \citet{Woolley_Chiral_Chasing}. Nevertheless, our work demonstrates that in 2D, which more accurately represents skin, the \citet{Painter2015_nonlocal_rd_developmental_biology} model actually predicts that such dynamics alone are sufficient to drive pattern formation. The resulting patterns appropriately take the form of stripes in the simulations considered.

However, this does require some asymmetry in the cross-species signalling ranges or interaction kernels. This could explain why \citet{WoolleyShort2014} and \citet{Woolley_Chiral_Chasing} do not predict such patterning - they implicitly assume symmetric signalling distributions. In fact, it is possible to see that the inclusion of asymmetric signalling would indeed enable the similar nonlocal model of \citet{Potts2019_spatial_memory_ecosystems} to produce patterns with run-and-chase dynamics, where previously such patterning was not predicted from the restricted set of cases examined. In zebrafish, asymmetric signalling would correspond to the ratio of the response by xanthophores and the response by melanophores differing in their dependence on separation distance. In this case, `response' could be measured as a magnitude of induced velocity or flux, or a probability to move, for example.

A significant caveat to the application of the \citet{Painter2015_nonlocal_rd_developmental_biology} model to zebrafish is that the model assumes that the nonlocal interaction induces flux parallel to the separation between cells. In contrast, zebrafish  melanophores have been observed to move away from xanthophores on average in an anticlockwise direction \citep{YamanakaKondo2014}. Although we have demonstrated that run-and-chase dynamics can drive patterning in 2D (where it was previously shown to not be possible 1D), it is unclear whether this result extends to interactions that induce a cellular flux that is not parallel to separation. Incorporating anisotropy into the framework, to model flux induced at different angles, would thus be a valuable avenue for future work. \citet{Woolley_Chiral_Chasing} has already shown that, in the limit of signalling range tending to zero, the choice of angle can heavily affect the capacity for patterns to form and the shapes that they take, which include patterns that are typically not seen in previous reaction-diffusion models. Anisotropy in a more generalised model could also be used to represent some bias of direction in the domain which, biologically, could correspond to directed fibres in an extracellular matrix, for example.

Another possible direction for future work is to consider the model on different geometries, which could be more biologically realistic and also could provide insight into the generality of our current results. In this work, we consider only an infinite domain or a finite domain with periodic boundary conditions. Curved manifolds, such as spheres or prolate ellipsoids, could thus be a natural extension as these shapes automatically feature such periodicity. Beyond periodicity, recent work \citep{new_operators_no_flux, 1D_no_flux_existence} has focused on formulating \citet{Painter2015_nonlocal_rd_developmental_biology}-like models on bounded domains with zero-flux and other boundary conditions, including in 2D \citep{2D_no_flux_existence}. These works largely focus on proving existence and well-posedness. To the best of our knowledge, investigation of pattern formation in these models through linear stability analysis has not yet been implemented.

In addition to providing biological and modelling insights, the analytical work in this paper has also demonstrated how hyperspherical Bessel functions with hyperspherical Hankel transforms can be powerful tools for the linear stability analysis of integro-partial-differential equations, in any number of dimensions. Our analysis can be directly applied to simplify the analysis of 2D and 3D models with equivalent nonlocal terms, such as in \citet{nonlocal_Vasculogenesis_Villa2022}. Furthermore, our analysis is easily adapted to nonlocal models that do not feature a radial vector $\boldsymbol{\hat{s}}$, including those with nonlocal advection such as \citet{Potts2019_spatial_memory_ecosystems, potts2019directionally, murakawa2015_density_dependent}, and those with nonlocal reaction-kinetics such as \citet{nonlocal_kinetics_britton, nonlocal_kinetics_2006, da_Cunha_ecology_nonlocal_competition, piva2021_nonlocal_reaction_kinetics}. Extending the linear analysis of these models from 1D to $N$D will simply use hyperspherical Hankel transforms of order $l=0$, instead of the order $l=1$ used in our work. Integro-partial-differential-equations are also used in contexts without physical space, such as evolutionary models in `phenotype space' \citep{phenotype_models_review_Lorenzi, phenotype_space_original, phenotype_space_kernel}, and hyperspherical Hankel transforms could potentially be useful for extending these to higher dimensions.

Through numerical simulation in 2D, we have verified the results of our linear stability analysis. Beyond the linear onset of instability, these simulations have also demonstrated some complex behaviour of the \citet{Painter2015_nonlocal_rd_developmental_biology} model. For the single species system, we observe distinct differences between the patterns generated by attractive interactions and those generated by repulsive interactions. The former are observed to form spots, which do not necessarily have constant spacing, whilst the latter form more regular structures: stripes, perforated stripes, or spot lattices. Spatio-temporal patterns driven by proliferation are also observed, which resemble those found in local chemotaxis models, as previously identified by \citet{Painter2015_nonlocal_rd_developmental_biology} in 1D. Systems can also feature multi-stability, with the possibility of stable evolving patterns even whilst the homogeneous state is linearly stable. Investigating such nonlinear behaviour analytically presents a promising direction for future work.

In particular, weakly nonlinear analysis has been shown to be tractable for similar nonlocal models \citep{giunta_weakly_nonlinear}. Such analysis could potentially provide more definite insight about which structures can form through attractive or repulsive interactions. Weakly nonlinear analysis could also potentially reveal whether the Turing bifurcation is supercritical or subcritical. The latter would explain the observed multi-stability and could also have important biological implications \citep{subcriticality_champneys}. Specifically, subcritical systems enable a small change in a bifurcation parameter to result in a large (discontinuous) change in macroscopic behaviour. For example, a small change in interaction strength could result in the formation of large amplitude stripes instead of a homogeneous steady state, whereas in supercritical systems such a change could only produce small amplitude stripes. Additionally, once this large amplitude pattern has formed, reversing the change in the bifurcation parameter will not necessarily destroy the pattern, because of how subcriticality induces multi-stability. In this way, subcriticality could enable the stochastic fluctuations intrinsic to biological systems to drive the formation of patterns which are then simultaneously robust to such fluctuations.

Finally, to close, we briefly discuss the importance of understanding the specifics of how nonlocal signalling varies with separation, for each biological or ecological application.
Regarding the \citet{Painter2015_nonlocal_rd_developmental_biology} model, we emphasise that the capacity for pattern formation is dependent on the behaviour of the interaction kernel, and in particular, whether $s^{\frac{N-1}{2}}\tilde{\Omega}\left(\frac{s}{\xi}\right)$ increases over any region, as this dictates whether the hyperspherical Hankel transform can be negative. The behaviour of such interaction kernels depends on the given biological application, specifically the mechanism by which signals are transmitted. For example, we could construct a simplistic kernel for interactions caused by the extension of filopodia by assuming that the filopodium has equal probability of being any length (up to a limit) and then touches all points within the circle/sphere of this length with equal probability. In 2D a filopodium that extends to distance $s$ might then sample any point on a circle of circumference $2\pi s$, and so the rate of interaction at a point at this distance should decrease by a factor $\frac{1}{2\pi s}$. Similarly in 3D, the choice of samples scales as the area of a sphere, $4\pi s^2$, and the interaction rate should decrease by a factor of $\frac{1}{4\pi s^2}$. In general, this `splattergun' effect would correspond to a kernel that decays with a factor $\mathcal{O}\left(s^{1-N}\right)$. In this case, $s^{\frac{N-1}{2}}\tilde{\Omega}\left(\frac{s}{\xi}\right)$ is never an increasing function, and so the hyperspherical Hankel transform cannot be negative, and so pattern formation cannot occur with repulsion interactions in a single species or run-and-chase dynamics with two species. However, the above example is only a very simplistic construction. Interaction kernels could equally also not decay in the above way, such as if the filopodia have higher probability of being a particular length, or if the filopodia could rotate at a faster timescale than they extend, or if signalling was due to some chemical released in all directions with activation only requiring a threshold concentration to be detected. In these cases, the model could plausibly predict pattern formation with repulsion interactions in a single species or run-and-chase dynamics with two species. This illustrates how the predictions of pattern formation in the model are dependent on the specific details of the signalling mechanism.

 Ultimately, we have shown in this work how pattern formation in a nonlocal reaction-diffusion-advection model is intrinsically dependent on the number of spatial dimensions, in a way which is not the case for local models. This motivates the need for future nonlocal models to reflect the dimensionality of their subject system. However, we should also consider the biological detail of the interaction mechanism's potentially complicated dependence on distance. Such details can only be properly understood through experiment or observation for each specific biological application. This suggests that pattern formation modelling should not merely focus on complex reaction kinetics with simple diffusion, but also on the details of more complex transport dynamics.

\section*{Acknowledgements}
We gratefully acknowledge Dr Dan J. Hill for valuable insight on hyperspherical Bessel functions.
\\ \\
Thomas Jun Jewell is supported by funding from the Edmund J. Crampin Scholarship at Linacre College, University of Oxford; and the Engineering and Physical Sciences Research Council [EP/W524311/1].

\section*{Declarations of Interest}
The authors have no competing interests to declare.

\bibliographystyle{MYplainnat}
\bibliography{refs}

\end{document}